\documentclass[12pt]{article}
\usepackage{amssymb, amsthm, amsmath}
\usepackage{bm}
\usepackage{graphicx,psfrag,epsf}
\usepackage{enumerate}
\usepackage{natbib}
\usepackage{url} % not crucial - just used below for the URL 
\usepackage{subfig}
\usepackage{color}
\usepackage{hyperref}

\newcommand{\blind}{1}

\begin{document}

	\def\spacingset#1{\renewcommand{\baselinestretch}%
		{#1}\small\normalsize} \spacingset{1}

	%%%%%%%%%%%%%%%%%%%%%%%%%%%%%%%%%%%%%%%%%%%%%%%%%%%%%%%%%%%%%%%%%%%%%%%%%%%%%%
	
	\if1\blind
	{
		\title{
			\bf A Bayesian Mark Interaction Model\\ for Analysis of Tumor Pathology Images
		}
		\author{
			Qiwei Li\\ %\thanks{The authors gratefully acknowledge \textit{please remember to list all relevant funding sources in the unblinded version}}\hspace{.2cm}\\
			Department of Clinical Sciences\\ University of Texas Southwestern Medical Center, Dallas, Texas\\
			and \\
			Xinlei Wang\\
			Department of Statistics\\ Southern Methodist University, Dallas Texas\\
			and\\
			Faming Liang\\
			Department of Statistics\\ Purdue University, West Lafayette, Indiana\\
			and\\
			Guanghua Xiao\thanks{To whom correspondence should be addressed.}\\
			Department of Clinical Sciences\\ University of Texas Southwestern Medical Center, Dallas, Texas
		}
		\maketitle
	} \fi
	
	\if0\blind
	{
		\bigskip
		\bigskip
		\bigskip
		\begin{center}
			{\LARGE\bf A Bayesian Mark Interaction Model for Analysis of Tumor Pathology Images}
		\end{center}
		\medskip
	} \fi
	
	\newpage
	\bigskip
	\begin{abstract}
		With the advance of imaging technology, digital pathology imaging of tumor tissue slides is becoming a routine clinical procedure for cancer diagnosis. This process produces massive imaging data that capture histological details in high resolution. Recent developments in deep-learning methods have enabled us to identify and classify individual cells from digital pathology images at large scale. The randomly distributed cells can be considered from a marked point process, where each point is defined by its position and cell type. Reliable statistical approaches to model such marked spatial point patterns can provide new insight into tumor progression and shed light on the biological mechanisms of cancer. In this paper, we consider the problem of modeling spatial correlations among three commonly seen cells (i.e. lymphocyte, stromal, and tumor) observed in tumor pathology images. A novel marking model of marked point processes, with interpretable underlying parameters (some of which are clinically meaningful), is proposed in a Bayesian framework. We use Markov chain Monte Carlo (MCMC) sampling techniques, combined with the double Metropolis-Hastings (DMH) algorithm, to sample from the posterior distribution with an intractable normalizing constant. On the benchmark datasets, we demonstrate how this model-based analysis can lead to sharper inferences than ordinary exploratory analyses. Lastly, we conduct a case study on the pathology images of $188$ lung cancer patients from the National Lung Screening Trial. The results show that the spatial correlation between tumor and stromal cells predicts patient prognosis. This statistical methodology not only presents a new model for characterizing spatial correlations in a multi-type spatial point pattern, but also provides a new perspective for understanding the role of cell-cell interactions in cancer progression. %We have also developed a web application (\href{https://qiwei.shinyapps.io/bayesmif/}{https://qiwei.shinyapps.io/bayesmif/}) to demonstrate our method on relatively small datasets.
	\end{abstract}
	
	\noindent%
	\vfill
	
	\newpage
	\spacingset{1.45} % DON'T change the spacing!
	\section{Introduction}\label{introduction}
	
	Cancer is a complex disease characterized by uncontrolled tumor cell growth. Pathological examination of H\&E-stained tissue slides is an essential step in cancer diagnosis. It has been reported that cell growth patterns are associated with the survival outcome \citep{Gleason2002, Amin2002, Borczuk2009, Barletta2010} and treatment response \citep{Tsao2015} of cancer patients. In addition, the interactions between tumor cells and other types of cells (e.g. immune cells) play vital roles in the progression and metastasis of cancer \citep{Mantovani2002,Orimo2005,Merlo2006,Polyak2009,Hanahan2011,Gillies2012,Junttila2013}. Spatial variations among cell types and their association with patient prognosis have been previously reported in breast cancer \citep{Mattfeldt2009}. Pathological examination of tissue slides requires a pathologist to match the observed image slides with his/her memory for certain patterns and features (such as tumor content, nuclei counts and tumor boundary). This process is laborious, tedious and subject to errors. More importantly, due to the limitations of the human brain in interpreting highly-complex pathology images, it is extremely hard for pathologists to systematically explore those subtle but essential patterns, such as tumor cell distribution and interaction with the surrounding micro-environment. Pathological examination by the human eyes is insufficient to decipher the large amount of complex and comprehensive information harbored in the high resolution pathology images.  
	
	With the advance of imaging technology, H\&E-stained pathology imaging is becoming a routine clinical procedure, which produces massive digital pathology images on a daily basis. Recent studies \citep{Beck2011, Yuan2012, Luo2016, Yu2016} have demonstrated the feasibility of using digital pathology image analysis to assist pathologists in clinical diagnosis and prognosis. However, current studies of pathology image analysis mainly focus on the morphology features, such as tissue texture and granularity. These imaging data, which capture histological details in high resolution, still leave unexplored more undiscovered knowledge. Computer vision and machine learning algorithms have enabled us to automatically identify and classify individual cells from digital pathology images at large scale \citep[e.g.][]{Yuan2012}. Recent developments in deep-learning methods have greatly facilitated this process. We have developed a convolutional neural network (CNN) to identify individual cells and classify the cell types into three categories: lymphocyte (a type of immune cell), stromal, and tumor. 
	
	Consequently, a pathology image is abstracted into a spatial map of marked points, where each cell (i.e. points in the spatial map) belongs to one of the three distinct types (i.e. qualitative marks), and the spatial location of each cell is known. The analysis of pathology images thus becomes an investigation of those spatial maps, which will provide a new perspective for the role of cell-cell interactions in cancer progression. Currently, a patient cohort usually contains hundreds of patients, and each patient has one or more pathology images. These rich datasets provide a great opportunity to study the cell-cell interactions in cancer. Recently, \cite{Li2017} developed a modified Potts model to study the spatial patterns observed in tumor pathology images, by projecting irregularly distributed cells into a $2$-dimensional lattice. However, this approximate method relies on selection of an \textit{ad hoc} lattice. More importantly, this method models the interaction among different regions (small squares defined by the lattice), but not those among individual cells.
	
	The study of interactions between objects, which results in the spatial correlation of marks, has been a primary focus in spatial statistics. It is a key aspect in population forestry \citep{Stoyan2000} and ecology \citep{Dale2000} theory, but receives little attention in biology. \cite{Illian2008} discussed in detail a large variety of numerical, functional, and second-order summary characteristics, which can be used to describe the spatial dependency between different types of points in a planar region. The most common approaches are based on generalizing the standard distance-dependent G-, K-, J-, and L-functions to their ``cross-type" versions \citep[see e.g.][]{Ripley1977_2,Besag1977,Diggle1981,Lotwick1982,Diggle1983,Vincent1989,Lieshout1996,Van1999}. Mark connection functions (MCFs) are another well recognized tool for qualitative marks, which are more suitable for the detection of mark correlation in an exploratory analysis \citep{Wiegand2004}. The \textit{ad hoc} testing of hypotheses, such as spatial independences of the marks, based on some suitable summary characteristics (e.g. K-functions) has also been discussed in the literature \citep{Grabarnik2011}. However, model-based analysis, which may sharpen inferences about the spatial pattern, is lagging. \cite{Diggle2006} formulated a pairwise interaction model for a spatial pattern of bivariate marked points and argued that model-based inference is statistically more efficient.
	
	In this paper, motivated by the emerging needs of tumor pathology images analysis, we develop a novel marking model, which aims to study the mark formulation in a spatial pattern through a Bayesian framework. A local energy function of three groups of parameters, i.e. first- and second-order intensities, and an exponential decay rate to the inter-point distance, is carefully defined, as is the related Gibbs distribution. The proposed model can serve as a novel model-based approach to characterize the spatial pattern/correlation among marks. We use the double Metropolis-Hastings (DMH) algorithm \citep{Liang2010} to sample from the posterior distribution with an intractable normalizing constant in the Gibbs distribution. The model performs well in simulated studies and three benchmark datasets. We also conduct a case study on a large cohort of lung cancer pathology images. The result shows that the spatial correlation between tumor and stromal cells is significantly associated with patient prognosis (\textit{P}-value=$0.0021$). Although the morphological features of stroma in tumor regions have been discovered to be associated with patient survival, there is no strong statistical evidence to support this, due to a lack of rigorous statistical methodology. In the study, the proposed statistical methodology not only delivers a new perspective for understanding how marks (i.e. cell types in pathology images) formulate in marked point processes, but also provides a refined statistical tool to characterize spatial interactions, which the existing approaches (e.g. MCF) may lack sufficient power to do so.
	
	The remainder of the paper is organized as follows: Section \ref{model} introduces the proposed modeling framework, including the local energy function and its related Gibbs distribution (i.e. the model likelihood), the choices of priors, and the model interpretation. Section \ref{model fitting} describes the Markov chain Monte Carlo (MCMC) algorithm and discusses the resulting posterior inference. Section \ref{simulation} assesses performance of the proposed model on simulated data. Section \ref{application} investigates the results of the data analyses from three benchmark datasets and a large cohort of lung cancer pathology images from the National Lung Screening Trial (NLST). Section \ref{conclusion} concludes the paper with some remarks on future research directions.

	\section{Model}\label{model}
	We describe a spatial map of cells in a Cartesian coordinate system, with $n$ observed cells indexed by $i$. We use $(x_i,y_i)\in\mathbb{R}^2$ to denote the $x$- and $y$- coordinates and $z_i\in\{1,\ldots,Q\},Q\ge2$ to denote the type of cell $i$. In spatial point pattern analysis, such data are considered as multi-type point pattern data, where $(x_1,y_1),\ldots,(x_n,y_n)$ are the point locations in a compact subset of the $2$-dimensional Euclidean space $\mathbb{R}^2$ (note that the proposed model can be easily extend to the general case of $\mathbb{R}^k,k\ge3$) and $z_1,\ldots,z_n$ are their associated qualitative (i.e. categorical or discrete) univariate marks. The mark attached to each point indicates which type/class it is (e.g. on/off, case/control, species, colors, etc.). Without loss of generality, we assume that the data points are restricted within the unit square $[0,1]^2$. This can be done by rescaling each pair of coordinates $(x_i,y_i)$ to $(x_i',y_i')$, with $x_i'=(x_i-\min\{\bm{x}\})/L$ and $y_i'=(y_i-\min\{\bm{y}\})/L$. Usually, $L$ is known, defined as the maximum possible Manhattan distance between any two points (both observed and unobserved) in the space. When $L$ is unknown, it can be estimated from the data itself by: 1) roughly setting $L=\max\{\max\{\bm{x}\}-\min\{\bm{x}\},\max\{\bm{y}\}-\min\{\bm{y}\}\}$; or 2) computing the Ripley-Rasson estimator \citep{Ripley1977} of a rectangle window, given the points, and then setting $L$ equal to the maximum side length of the window.

	\subsection{Energy Functions}\label{energy function}
	In the analysis of tumor pathology images, cell distribution and cell-cell interaction may reveal important messages about the tumor cell growth and its micro-environment. Therefore, it is of great interest to study the arrangements of cell types associated with the observed cells, given their locations. In spatial point pattern analysis, such a problem is called \textit{marking modeling}, which is to study the formulation of the marks $\bm{z}$ in a pattern, given the points $(\bm{x},\bm{y})$. In this subsection, we explore the formulation of energy functions, accounting for both of the first- and second-order properties of the point data.
	
	At the initial stage, we assume that each point interacts with all other points in the space. A complete undirected graph $G=(V,E)$ can be used to depict their relationships, with $V$ denoting the set of points (i.e. the $n$ observed cells) and $E$ denoting the set of direct interactions (i.e. the $(n-1)n/2$ cell-cell pairs). We define $G$ as the interaction network and define its potential energy as
	\begin{equation}\label{energy}
		V(\bm{z}|\bm{\omega},\bm{\Theta})=\sum_{q}\omega_q\sum_iI(z_i=q)+\sum_{q}\sum_{q'}\theta_{qq'}\sum_{(i\sim i')\in E}I(z_i=q,z_{i'}=q'),
	\end{equation}
	where the notation $(i\sim i')$ denotes that points $i$ and $i'$ are the interacting pair in $G$ (i.e. they are connected by an edge in $G$), and $I$ denotes the indicator function. Note that $\theta_{qq'}=\theta_{q'q}$ as the edge between any pairs of points has no orientation. On the right-hand side of Equation (\ref{energy}), the first term can be viewed as the weighted average of the numbers of points with different marks, while the second term can be viewed as the weighted average of the numbers of pairs connecting two points with the same or different marks. In the context of spatial point pattern analysis, the first and second terms are referred to the first- and second-order potentials/characteristics, respectively. Their corresponding parameters $\bm{\omega}=(\omega_1,\ldots,\omega_Q)$ and $\bm{\Theta}=\left(\begin{array}{ccc} \theta_{11} & \cdots & \theta_{1Q}\\ & \ddots & \vdots\\ & & \theta_{QQ}\end{array}\right)$ are defined as the first- and second-order intensities. These two groups of parameters control the enrichment of different marks and the spatial correlations among them simultaneously. A detailed interpretation of $\bm{\omega}$ and $\bm{\Theta}$ is discussed in Section \ref{interpretation}. 
	
\begin{figure}
		\begin{center}
			\includegraphics[width=0.6\linewidth]{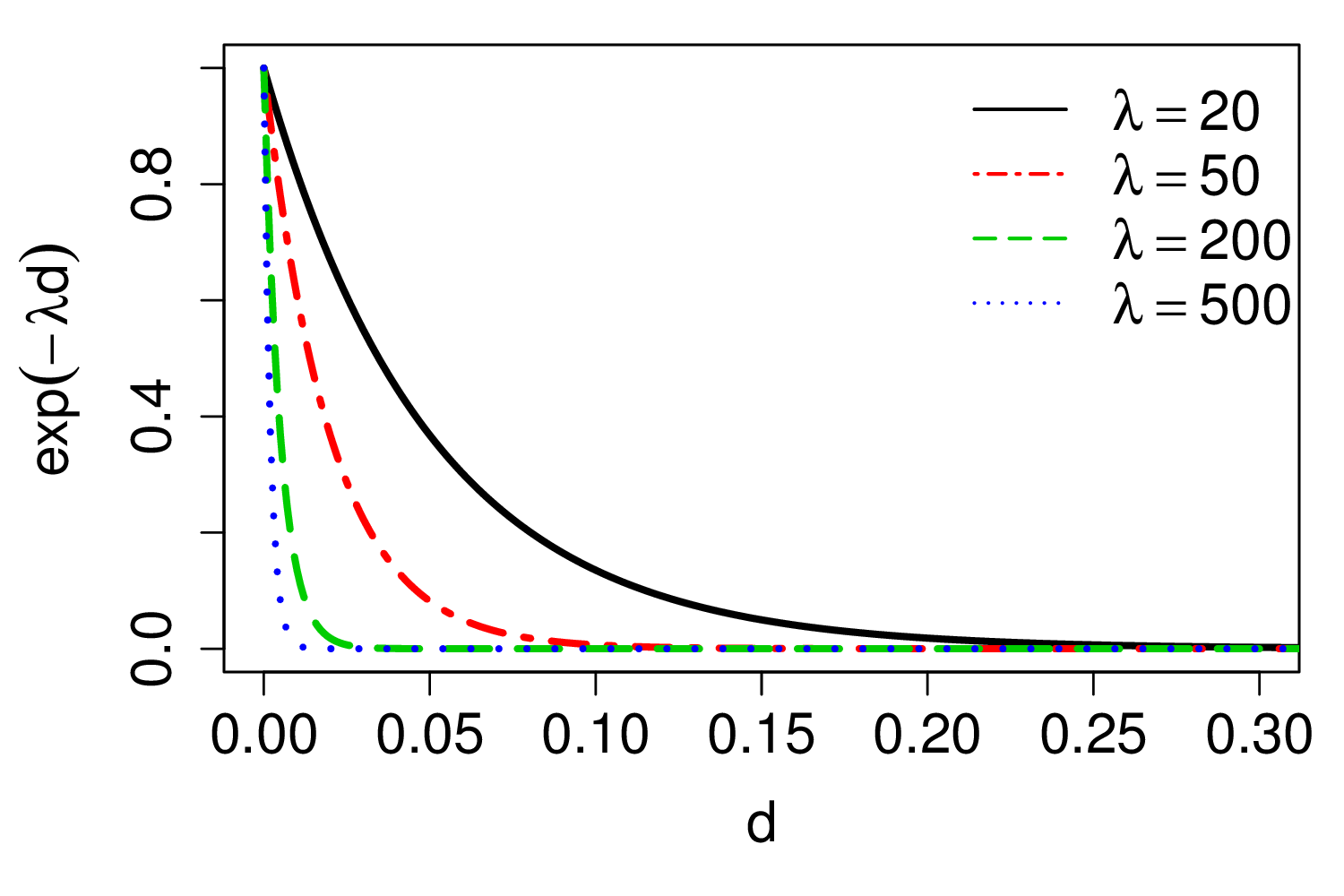}
		\end{center}
		\caption{Examples of exponential decay functions under different choices of $\lambda$}\label{Figure 21}
	\end{figure}

	In mathematical physics and statistical thermodynamics, the interaction energy between two points (i.e. particles and cells) is usually an exponential decay function with respect to the distance between the two points \citep[see e.g.][]{Penrose1974,Kashima2010,Avalos2014,Chulaevsky2014,Rincon2015}. Similarly, exponential decay has also been observed in biological systems, such as cell-cell interactions \citep{Segal1984,Hui2007} and gene-gene correlations \citep{Xiao2009,Xiao2011}. In this study, we assume the interaction energy between a pair of points decreases exponentially at a rate $\lambda$ proportional to the distance,
	\begin{equation}\label{energy2}
		V(\bm{z}|\bm{\omega},\bm{\Theta},\lambda)=\sum_{q}\omega_q\sum_iI(z_i=q)+\sum_{q}\sum_{q'}\theta_{qq'}\sum_{(i\sim i')\in E}e^{-\lambda d_{ii'}}I(z_i=q,z_{i'}=q'),
	\end{equation}
	where $d_{ii'}=\sqrt{(x_i-x_{i'})^2+(y_i-y_{i'})^2}$ is the Euclidean distance between points $i$ and $i'$. A larger value of the decay parameter $\lambda$ makes the interaction energy vanish much more rapidly with the distance, while a smaller value leads to $e^{-\lambda d_{ii'}}\approx 1$ and Equation (\ref{energy2}) $\rightarrow$ Equation (\ref{energy}). See Figure \ref{Figure 21} for examples of exponential decay functions with different values of parameter $\lambda$. 
	
\begin{figure}
		\centering
		\subfloat[][$c=0.3$]{\includegraphics[width=0.33\linewidth]{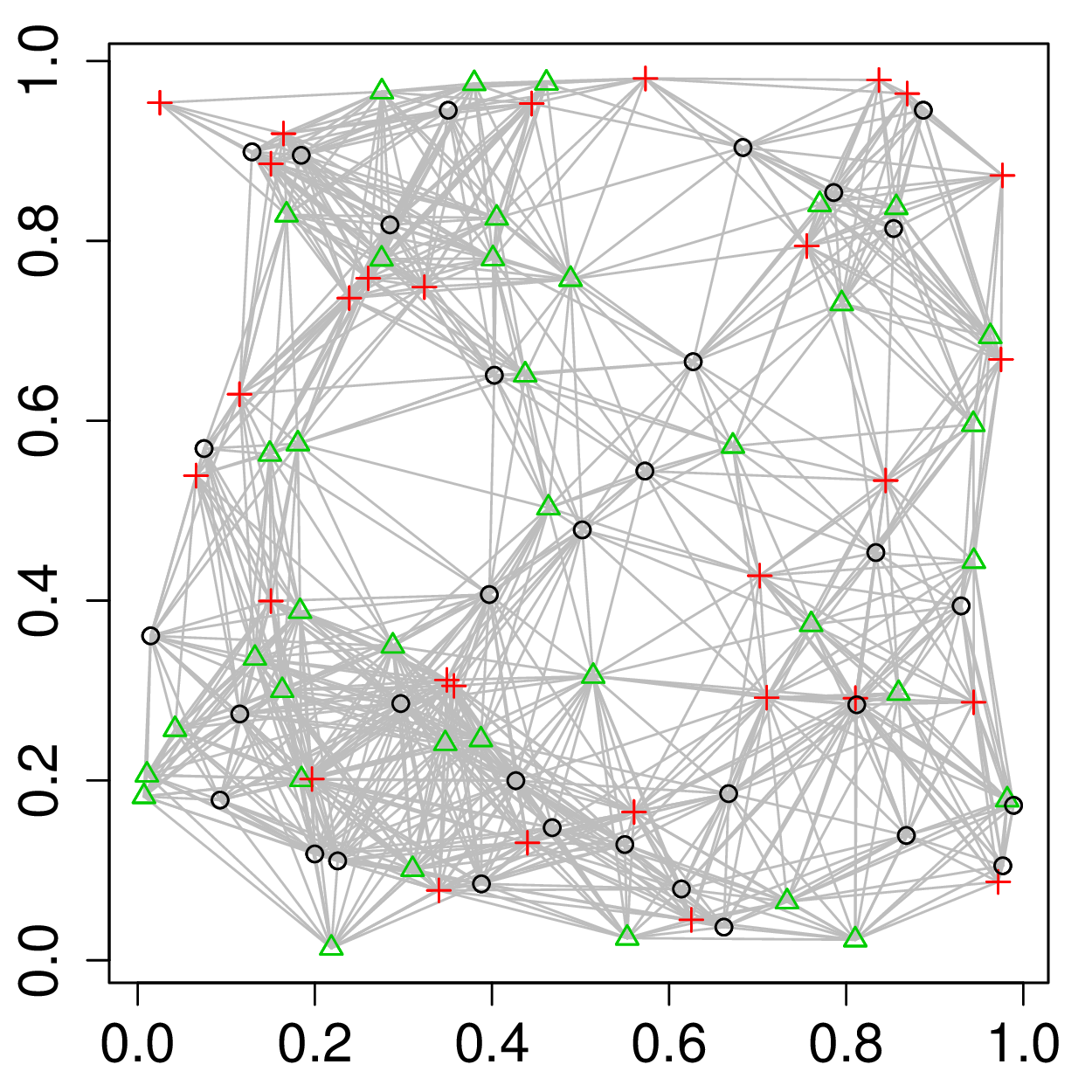}}
		\subfloat[][$c=0.2$]{\includegraphics[width=0.33\linewidth]{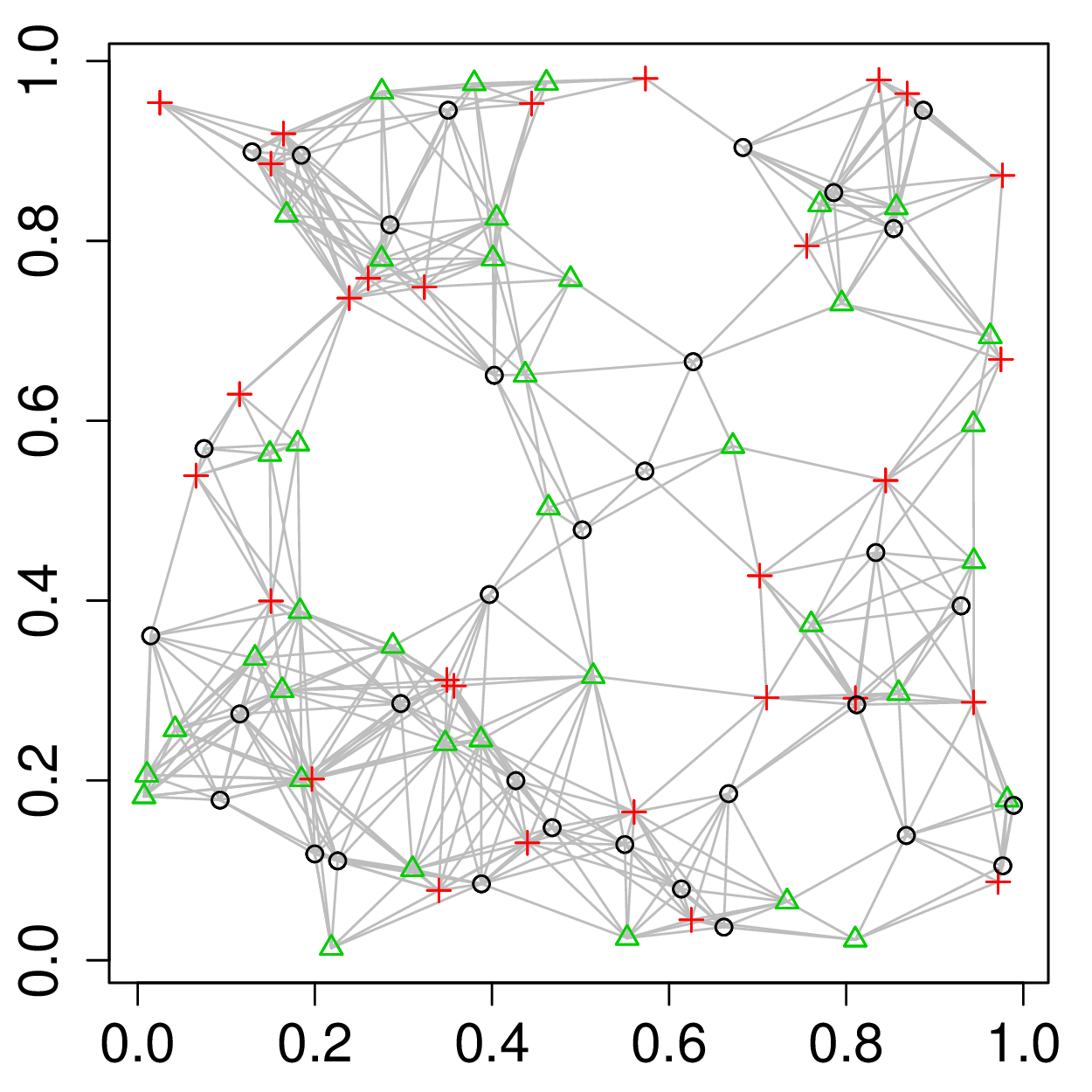}}
		\subfloat[][$c=0.1$]{\includegraphics[width=0.33\linewidth]{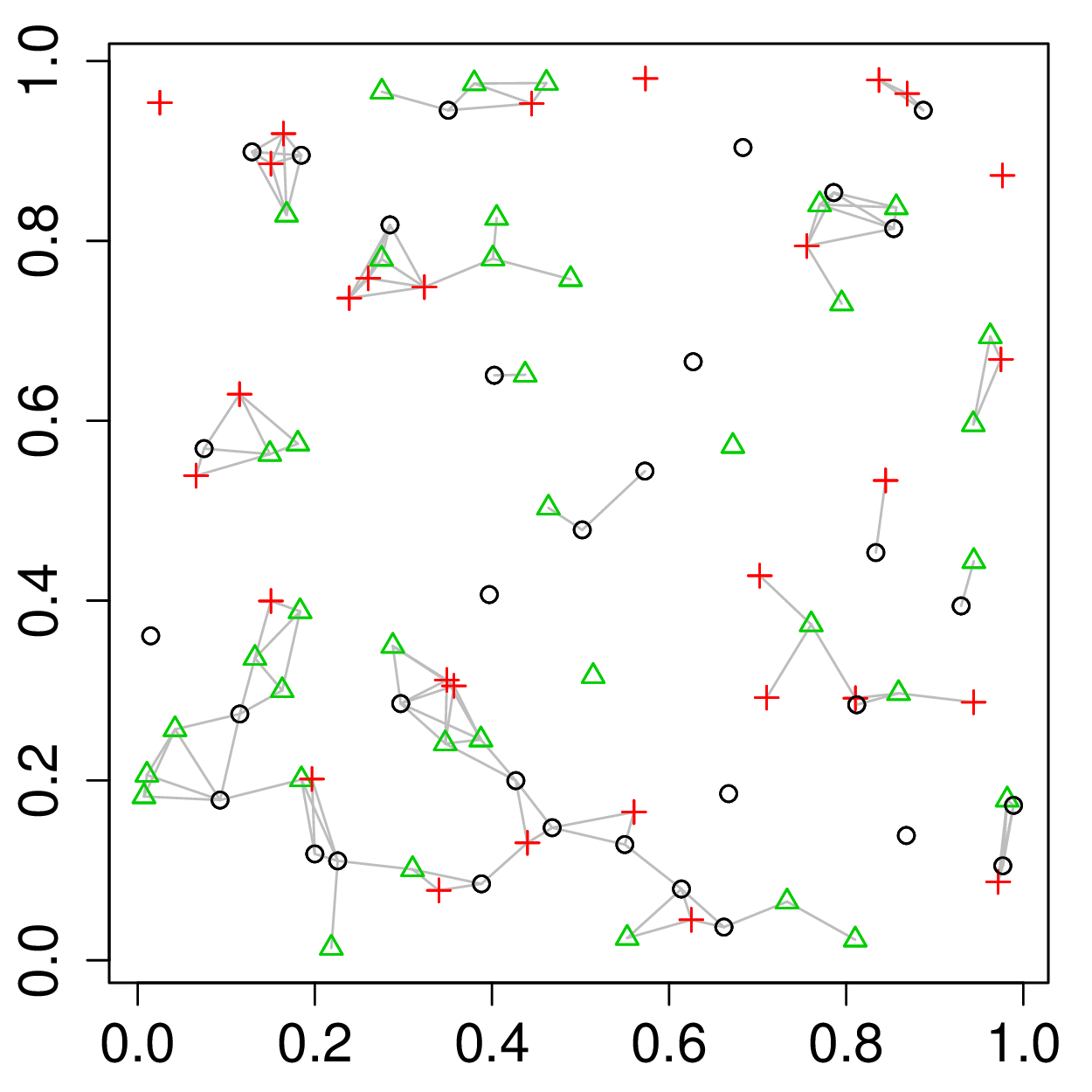}}
		\caption{An example of three-type point pattern data ($n=100$) and its corresponding network $G'$ under different choices of (a) $c=0.3$, (b) $c=0.2$, and (c) $c=0.1$.}\label{Figure 22}
	\end{figure}

	As shown in Equation (\ref{energy2}),  it needs to sum over $n$ data points and $(n-1)n/2$ pairs of data points to compute the potential energy, resulting in an extremely tedious computation, especially when $n$ is large. An alternative way is to obtain an approximate value of $V(\bm{z}|\bm{\omega},\bm{\Theta},\lambda)$ by neglecting those pairs with distance beyond a certain threshold $c,c\in[0,1]$. This is feasible as long as the decay function $e^{-\lambda d}$ causes exponentially decreasing weights for those pairs being placed on the potential energy.  It can be illustrated that a point (i.e. a cell) can only interact with its nearby points within a certain range $c$. Therefore, the complete network $G$ reduces to a sparse network $G'=(V,E')$, with $E'\subseteq E$ denoting the set of edges joining pairs of points $i$ and $i'$ in $G'$, if their distance  $d_{ii'}$ is smaller than a threshold $c$. We write the potential energy of the interaction network $G'$ as 
	\begin{equation}\label{energy3}
		V(\bm{z}|\bm{\omega},\bm{\Theta},\lambda)=\sum_{q}\omega_q\sum_iI(z_i=q)+\sum_{q}\sum_{q'}\theta_{qq'}\sum_{(i\sim i')\in E'}e^{-\lambda d_{ii'}}I(z_i=q,z_{i'}=q')
	\end{equation}
	Note that $c$ is not a model parameter, but a user-defined value. We may determine its value from a mark connection function analysis (discussed in Section \ref{simulation}) or from the subjective assessment of an experienced expert in the related field. The choice of a large $c$ causes an extremely complex network, while a too small value results in a sparse network that may neglect some important spatial information. See Figure \ref{Figure 22} for an example of three-type point pattern data ($n=100$) and its corresponding mark interaction networks $G'$ under different choices of $c$. By introducing the sparse network $G'$, we not only reduce the computational cost in calculating the potential energy, but also define a local spatial structure. %Such an approximation, which was proposed in physics \citep{Yi1989}, makes a local energy function.

	\subsection{Data Likelihood}
	According to the fundamental Hammersley-Clifford theorem \citep{Hammersley1971}, if we have a locally defined energy, such as Equation (\ref{energy3}), then a probability measure with a Markov property exists. Specifically, this frequently seen measure in many problems of probability theory and statistical mechanics is called a \textit{Gibbs measure}. It gives the probability of observing the marks associated with their locations in a particular state,
	\begin{equation}\label{globle_prob}
		\begin{split}
			& p(\bm{z}|\bm{\omega},\bm{\Theta},\lambda)=\frac{\exp(-V(\bm{z}|\bm{\omega},\bm{\Theta},\lambda))}{\sum_{z'}\exp(-V(\bm{z}'|\bm{\omega},\bm{\Theta},\lambda))}\\
			&=\frac{1}{C(\bm{\omega},\bm{\Theta},\lambda)}\exp\left(-\sum_{q}\omega_q\sum_iI(z_i=q)+\sum_{q}\sum_{q'}\theta_{qq'}\sum_{(i\sim i')\in E'}e^{-\lambda d_{ii'}}I(z_i=q,z_{i'}=q')\right).
		\end{split}
	\end{equation}
	The normalizing constant $C(\bm{\omega},\bm{\Theta},\lambda)=\sum_{z'}\exp(-V(\bm{z}'|\bm{\omega},\bm{\Theta},\lambda))$ is also called a partition function. An exact evaluation of $C(\bm{\omega},\bm{\Theta},\lambda)$ needs to sum over the entire space of $\bm{z}$, which consists of $Q^{n}$ states. Thus, it is intractable even for a small size model. Take $Q=2$ and $n=100$ for example, it needs to sum over $2^{100}\approx1.268\times10^{30}$ elements. To address this issue, we employ the double Metropolis-Hastings (DMH) algorithm \citep{Liang2010} to make inference on the model parameters $\bm{\omega}$, $\bm{\Theta}$, and $\lambda$. DMH is an auxiliary variable MCMC algorithm, which can make the normalizing constant ratio canceled by augmenting appropriate auxiliary variables through a short run of the ordinary Metropolis-Hastings (MH) algorithm. More details are given in Section \ref{algorithm}. 
	
	Equation (\ref{globle_prob}) serves as the full data likelihood of the proposed model. Since the model satisfies the local Markov property, we can also write the probability of observing point $i$ belonging to class $q$ conditional on its neighborhood configuration(s),
	\begin{equation}\label{local_prob}
		p(z_i=q|\bm{z}_{-i},\bm{\omega},\bm{\Theta},\lambda)\propto\exp\left(-\omega_q-\sum_{q'}\theta_{qq'}\sum_{\{i':(i\sim i')\in E'\}}e^{-\lambda d_{ii'}}I(z_{i'}=q')\right),
	\end{equation}
	where $\bm{z}_{-i}$ denotes the the collection of all marks excluding the $i$-th one. According to Equation (\ref{local_prob}), the conditional probability depends on the first-order intensity $\omega_q$, the second-order intensities $\theta_{qq'},q'=1,\ldots,Q$, the decay parameter $\lambda$, and the neighborhood of the points defined by $c$. Although it is not easy to describe how the parameters affect the conditional probability, we can still draw the following conclusions: 1) the smaller the value of $\omega_{q}$ or $\theta_{qq}$, the more likely that point $i$ belongs to class $q$; and 2) the smaller the value of $\lambda$, the more impact from the neighborhood configuration(s).
	
	\subsection{Parameter Priors}
	The proposed model in the Bayesian framework requires the specification of prior distributions for the unknown parameters. In this subsection, we specify the priors for all three groups of parameters: $\bm{\omega}$, $\bm{\Theta}$, and $\lambda$. For the first- and second-order intensities $\bm{\omega}$ and $\bm{\Theta}$, we notice that an identifiability problem arises from Equation (\ref{globle_prob}) or (\ref{local_prob}). For instance, adding a non-zero constant, say $s$, into $\omega_q,q=1,\ldots,Q$ does not change the probability of observing point $i$ belonging to class $q$. Similarly, the settings of $\bm{\Theta}$ and $\bm{\Theta}+s\bm{I}$ lead to the same conditional probability. Therefore, imposing an appropriate constraint is necessary. Without loss of generality, suppose the points with mark $Q$ have the largest population and we set $\omega_Q=1$ and $\theta_{QQ}=1$. For the other parameters in $\bm{\omega}$ and $\bm{\Theta}$, we consider normal priors and set $\omega_q\sim\text{N}(\mu_\omega,\sigma_\omega^2),q=1,\ldots,Q-1$ and $\theta_{qq'}\sim\text{N}(\mu_\theta,\sigma_\theta^2),q=1,\ldots,Q-1,q'=q,\ldots,Q$. We suggest users choose the standard normal distribution; that is, $\mu_\omega=\mu_\theta=0$ and $\sigma_\omega=\sigma_\theta=1$. For the decay parameter $\lambda$, we specify a gamma prior $\lambda\sim\text{Ga}(a_\lambda,b_\lambda)$. One standard way of setting a weakly informative gamma prior is to choose small values for the two parameters, such as $a_\lambda=b_\lambda=0.001$\citep{Gelman2006}.
	
	\subsection{Interpretation}\label{interpretation}
	In this subsection, we aim to interpret the meanings of the model parameters $\bm{\omega}$ and $\bm{\Theta}$, because it is crucial for describing the observed spatial pattern as well as studying their associations with any other measurements of interest. 
	
	Suppose there is only one point in the space. Then Equation (\ref{local_prob}) reduces to $p(z_1=q|\cdot)\propto\exp\left(-\omega_q\right)$, which implies the probability of observing a point with mark $q$ in this single-point system is equal to
	\begin{equation}\label{pi}
		\pi_q=\exp\left(-\omega_q\right)/\sum_{q}\exp\left(-\omega_q\right).
	\end{equation}
	Note that the vector $\bm{\pi}=(\pi_1,\ldots,\pi_Q)$ has a natural constraint; that is, $\sum_q\pi_q=1$. Furthermore, suppose there are $n$ points in the space and there are almost no mark interactions. This can be fulfilled by any one of the following conditions: 1) the distance between any pairs of two points is beyond the given value $c$, i.e. $d_{ii'}>c,\forall (i\sim i')\in E$; 2) the second-order intensities are all equal, i.e. $\bm{\Theta}=s\bm{1},\forall s\in\mathbb{R}$; or 3) the decay parameter $\lambda$ goes to infinity, i.e. $\lambda\rightarrow\infty$. Then Equation (\ref{local_prob}) converges to $p(z_i=q|\bm{z}_{-i},\bm{\omega},\bm{\Theta},\lambda)\propto\exp\left(-\omega_q\right)=\pi_q$, implying that the expected number of points with mark $q$ is $n\pi_q$. Thus, after transforming the first-order intensities $\bm{\omega}$ to their probability measures $\bm{\pi}$, we find a clear path to describe the abundance of different marks in the above simplified situations. 
	
	Suppose there are only two points $1$ and $2$ in the space, with the type of the second point known; say $z_{2}=q'$. For convenience, we further assume $\omega_1=\cdots=\omega_Q$. We first consider the case of the two points being at the same location, i.e. $d_{12}=0$. Then Equation (\ref{local_prob}) turns out to be $p(z_1=q|z_2=q',\cdot)\propto\exp\left(-\theta_{qq'}\right)$, which implies the probability of observing the point with unknown mark belonging to type $q$, given the one with the known mark $q'$ (at the same location), is
	\begin{equation}\label{phi}
		\phi_{qq'}=\exp\left(-\theta_{qq'}\right)/\sum_{q}\exp\left(-\theta_{qq'}\right).
	\end{equation}
	We use a $Q$-by-$Q$ matrix $\bm{\phi}$ to denote the collection of $\phi_{qq'},q=1,\ldots,Q,q'=1,\ldots,Q$. Note that each column in $\bm{\phi}$ should be summed to $1$ and $\bm{\phi}$ is not necessary to be a symmetric matrix as $\bm{\Theta}$. In this duo-point system (and more complex cases therein), the larger the value of $\phi_{qq'}$, the more likely the points with mark $q$ get attracted to the nearby points with mark $q'$. Thus, the spatial correlations among marks can be easily interpreted by the probability matrix $\bm{\phi}$. 
	
	In the aforementioned duo-point model with known parameters, if the assumption of equivalent first-order intensities is relaxed, then the probability of assigning mark $q$ to point $1$ conditional on the mark of point $2$ is $q'$ is a strictly monotonic function of their distance $d$,
	\begin{equation}\label{bmif}
	\text{MIF}_{q|q'}(d)=\frac{\exp\left(-\omega_q-\theta_{qq'}e^{-\lambda d}\right)}{\sum_{q''}\exp\left(-\omega_{q''}-\theta_{q''q'}e^{-\lambda d}\right)}.
	\end{equation}
	We call the above equation the \textit{mark interaction function} (MIF) of mark $q$ given mark $q'$. As the distance increases, its value ultimately converges to $\pi_q$. The plot of MIF is a more comprehensive way to describe the spatial correlation/interaction between marks.
	
	In conclusion, $\bm{\pi}$, $\bm{\phi}$, and MIF directly characterize a single point behavior (i.e. the assignment of its mark) in a model with small size, such as $n=1$ and $2$. However, the observed spatial marked point pattern is a reflection of how each individual point reacts with its neighbors. Note that the mappings from $\bm{\omega}$ to $\bm{\pi}$ and from $\bm{\Theta}$ to $\bm{\phi}$ are one-to-one/unique, so we can implement this step after obtaining the estimates of $\bm{\omega}$ and $\bm{\Theta}$.

	\section{Model Fitting}\label{model fitting}
	In this section, we describe the MCMC algorithm for posterior inference. Our inferential strategy allows for simultaneously estimating 1) the first-order intensities $\bm{\omega}$, which reveal the abundance of different marks; 2) the second-order intensities $\bm{\Theta}$, which capture the spatial correlation among marks; and 3) the decay parameter $\lambda$. We first give the full details of our MCMC algorithm and then discuss the resulting posterior inference.
	
	\subsection{MCMC Algorithm}\label{algorithm}
	We are interested in estimating $\bm{\omega}$, $\bm{\Theta}$, and $\lambda$, which define the Gibbs measure based on the local energy function. However, the data likelihood, as shown in Equation (\ref{globle_prob}), includes an intractable normalizing constant $C(\bm{\omega},\bm{\Theta},\lambda)$, making the Metropolis-Hastings algorithm infeasible in practice. To address this issue, we use the double Metropolis-Hastings algorithm (DMH) proposed by \cite{Liang2010}. The DMH is an asymptotic algorithm, which has been shown to produce accurate results by various spatial models. Unlike other auxiliary variable MCMC algorithms \citep{Moller2006,Murray2012} that also aim to have the normalizing constant ratio canceled, the DMH sampler is more efficient because: 1) it removes the need for exact sampling; and 2) it does not require drawing the auxiliary variables from a perfect sampler. \cite{Liang2016} also proposed an adaptive exchange algorithm, which generates auxiliary variables via an importance sampling procedure from a Markov chain running in parallel. However, this exact algorithm is more computationally intensive than the DMH.
	
	\textbf{Update of $\bm{\omega}$:} We update each of ${\omega}_{q},q=1,\ldots,Q-1$ by using the DMH algorithm. We first propose a new $\omega_{q}^*$ from $\text{N}(\omega_{q},\tau_\omega^2)$. Next, according to Equation (\ref{local_prob}), we implement the Gibbs sampler to simulate an auxiliary variable $\bm{z}^*$ starting from $\bm{z}$ based on the new $\bm{\omega}^*$, where all the elements are the same as $\bm{\omega}$ excluding the $q$-th one. The proposed value $\omega_{q}^*$ is then accepted to replace the old value with probability $\min(1,r)$. The Hastings ratio $r$ is given as below,
	\begin{align*}
	r=\frac{p(\bm{z}^*|\bm{\omega},\bm{\Theta},\lambda)}{p(\bm{z}|\bm{\omega},\bm{\Theta},\lambda)}\frac{p(\bm{z}|\bm{\omega}^*,\bm{\Theta},\lambda)}{p(\bm{z}^*|\bm{\omega}^*,\bm{\Theta},\lambda)}\frac{\text{N}(\omega_{q}^*;\mu_\omega,\sigma_\omega^2)}{\text{N}(\omega_{q};\mu_\omega,\sigma_\omega^2)}\frac{J(\omega_q;\omega_q^*)}{J(\omega_q^*;\omega_q)},
	\end{align*}
	where the form of $p(\bm{z}|\bm{\omega},\bm{\Theta},\lambda)$ is given by Equation (\ref{globle_prob}). As a result, the normalizing constant in Equation (\ref{globle_prob}) can be canceled out. Note that the last fraction term, which is the proposal density ratio, equals $1$ for this random walk Metropolis update on $\omega_{q}$.
	
	\textbf{Update of $\bm{\Theta}$:} We update each of ${\theta}_{qq'},q=1,\ldots,Q-1,q'=q,\ldots,Q$ by using the DMH algorithm. We first propose a new $\theta_{qq'}^*$ from $\text{N}(\theta_{qq'},\tau_\theta^2)$ and set $\theta_{q'q}^*=\theta_{qq'}^*$ as the matrix is symmetric. Next, according to Equation (\ref{local_prob}), an auxiliary variable $\bm{z}^*$ is simulated via the Gibbs sampler with $\bm{z}$ as the starting point. This simulation should be based on the new $\bm{\Theta}^*$, where all the elements are the same as $\bm{\Theta}$ except the two elements corresponding to $\theta_{qq'}$ and $\theta_{q'q}$. The proposed value $\theta_{qq'}^*$ as well as $\theta_{q'q}^*$ is then accepted to replace the old values with probability $\min(1,r)$. The Hastings ratio $r$ is given as below:
	\begin{align*}
	r=\frac{p(\bm{z}^*|\bm{\omega},\bm{\Theta},\lambda)}{p(\bm{z}|\bm{\omega},\bm{\Theta},\lambda)}\frac{p(\bm{z}|\bm{\omega},\bm{\Theta}^*,\lambda)}{p(\bm{z}^*|\bm{\omega},\bm{\Theta}^*,\lambda)}\frac{\text{N}(\theta_{qq'}^*;\mu_\theta,\sigma_\theta^2)}{\text{N}(\theta_{qq'};\mu_\theta,\sigma_\theta^2)}\frac{J(\theta_{qq'};\theta_{qq'}^*)}{J(\theta_{qq'}^*;\theta_{qq'})},
	\end{align*}
	where the form of $\text{Pr}(\bm{z}|\bm{\theta},\lambda)$ is given by Equation (\ref{globle_prob}). As a result, the normalizing constant in Equation (\ref{globle_prob}) can be canceled out. Note that the last fraction term, which is the proposal density ratio, equals $1$ for this random walk Metropolis update on $\theta_{qq'}$.
	
	\textbf{Update of $\lambda$:} We update the decay parameter $\lambda$ by using the DMH algorithm. We first propose a new $\lambda^*$ from a gamma distribution $\text{Ga}(\lambda^2/\tau_\lambda,\lambda/\tau_\lambda)$, where the mean is $\lambda$ and the variance is $\tau_\lambda$. Next, according to Equation (\ref{local_prob}), we implement the Gibbs sampler to simulate an auxiliary variable $\bm{z}^*$ starting from $\bm{z}$ based on the new $\lambda^*$. The proposed value $\lambda^*$ is then accepted to replace the old value with probability $\min(1,r)$. The Hastings ratio $r$ is given as below:
	\begin{align*}
	r=\frac{p(\bm{z}^*|\bm{\omega},\bm{\Theta},\lambda)}{p(\bm{z}|\bm{\omega},\bm{\Theta},\lambda)}\frac{p(\bm{z}|\bm{\omega},\bm{\Theta},\lambda^*)}{p(\bm{z}^*|\bm{\omega},\bm{\Theta},\lambda^*)}\frac{\text{Ga}(\lambda^*;a,b)}{\text{Ga}(\lambda;a,b)}\frac{J(\lambda;\lambda^*)}{J(\lambda^*;\lambda)},
	\end{align*}
	where the form of $\text{Pr}(\bm{z}|\bm{\theta},\lambda)$ is given by Equation (\ref{globle_prob}). As a result, the normalizing constant in Equation (\ref{globle_prob}) can be canceled out. Note that the last fraction term, which is the proposal density ratio, equals $1$ for this random walk Metropolis update on $\lambda$.

	\subsection{Posterior Estimation}
	We obtain posterior inference by post-processing the MCMC samples after burn-in. Suppose that multiple sequences of MCMC samples,
	\begin{align*}
	\omega_{q}^{(1)},\ldots,\omega_{q}^{(U)}&,q=1,\ldots,Q-1,\\
	\theta_{qq'}^{(1)},\ldots,\theta_{qq'}^{(U)}&,q=1,\ldots,Q-1,q'=q,\ldots,Q,\\
	\lambda^{(1)},\ldots,\lambda^{(U)}&,
	\end{align*}
	have been collected, where $u,u=1,\ldots,U$ indexes the iteration after burn-in. An approximate Bayesian estimator of each parameter can be simply obtained by averaging over the samples, $\hat{\omega}_{q}=\sum_{u=1}^U{\omega}_{q}^{(u)}/U$, $\hat{\theta}_{qq'}=\sum_{u=1}^U{\theta}_{qq'}^{(u)}/U$, and $\hat{\lambda}=\sum_{u=1}^U{\lambda}^{(u)}/U$. For a better understanding of the model, we suggest to project the parameters $(\bm{\omega},\bm{\Theta})$ to $(\bm{\pi},\bm{\Phi})$ according to Equations (\ref{pi}) and (\ref{phi}), or plot the mark interaction functions as given in Equation (\ref{bmif}).
	
	\section{Simulation}\label{simulation}
	In this section, we use simulated data generated from the proposed model to assess performance of our strategy for posterior inference on the model parameters, $\bm{\omega}$, $\bm{\Theta}$, and $\lambda$. In addition, we discuss how to choose the tunable parameter $c$ based on the mark connection function plots and investigate the sensitivity of the proposed model to the choices of $c$.
	
\begin{figure}[!p]
		\centering
		\subfloat[][]{\includegraphics[width=0.25\linewidth]{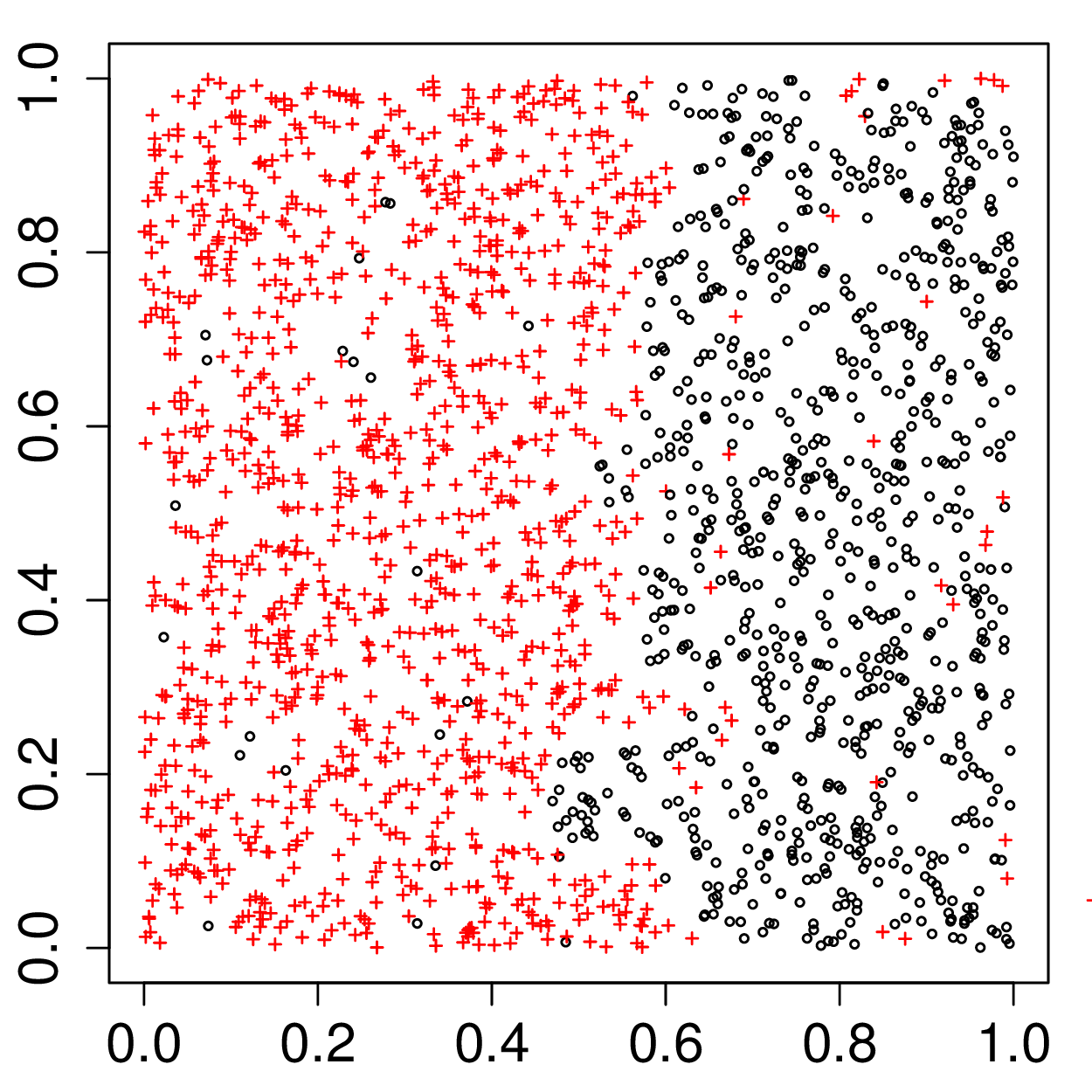}}
		\subfloat[][]{\includegraphics[width=0.25\linewidth]{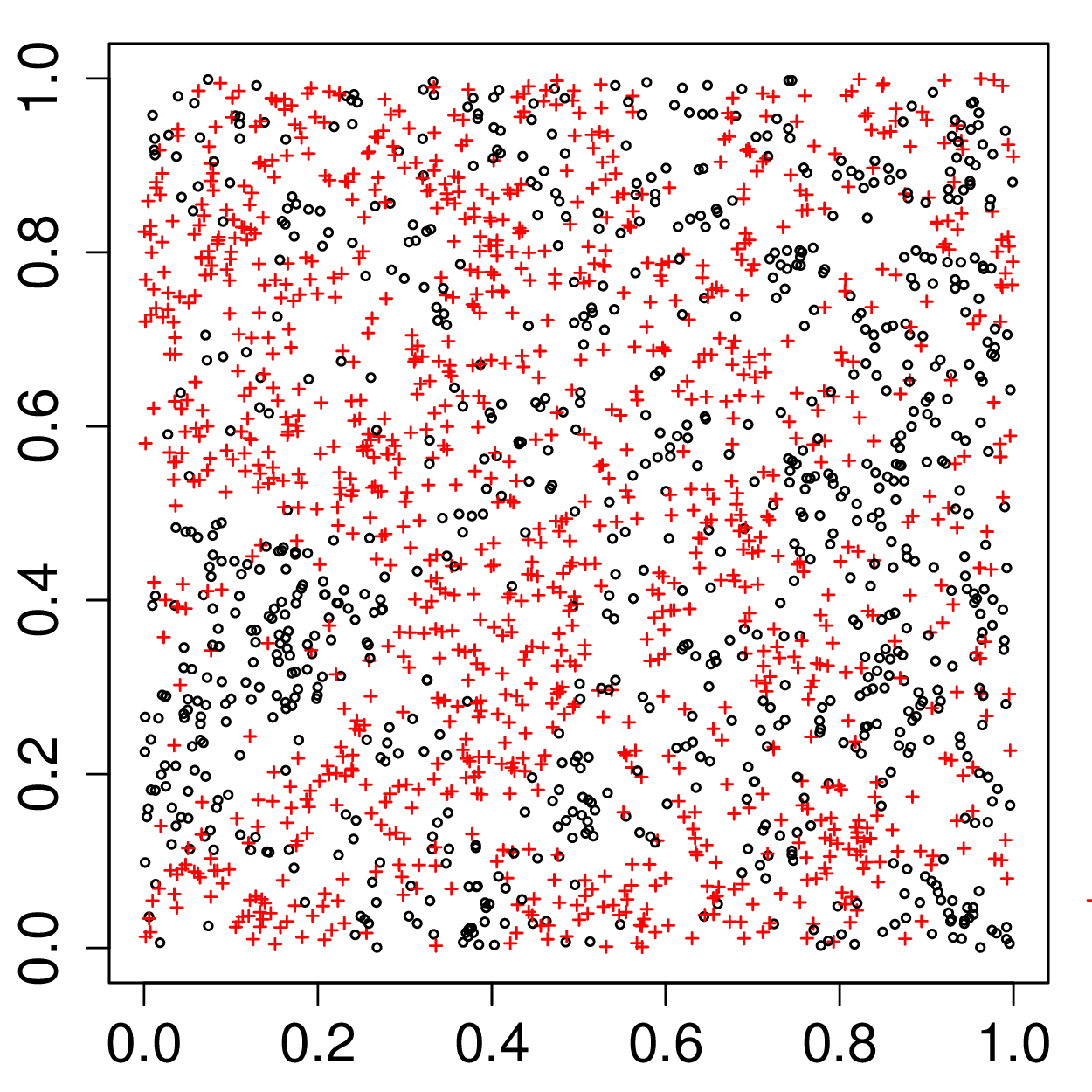}}
		\subfloat[][]{\includegraphics[width=0.25\linewidth]{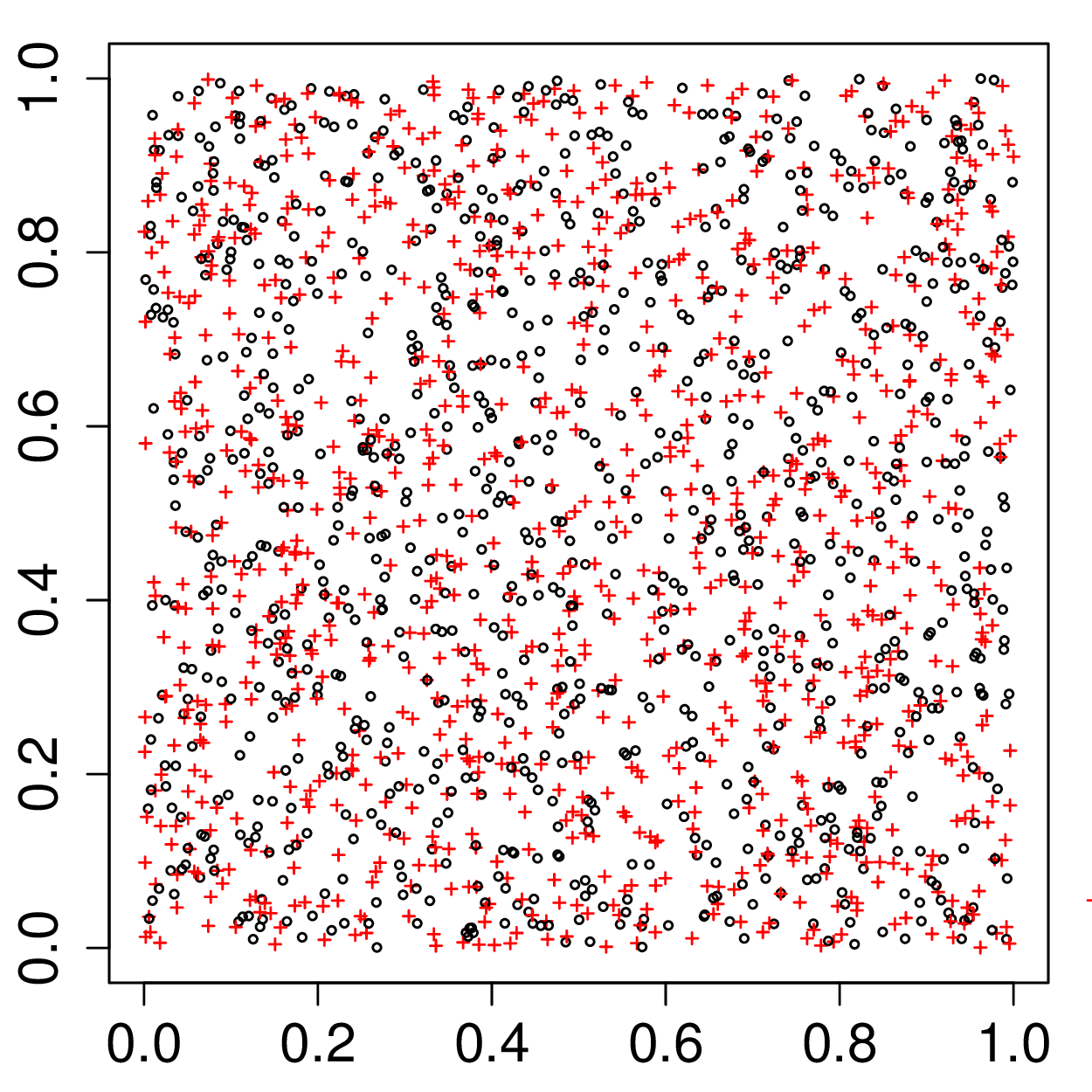}}
		\subfloat[][]{\includegraphics[width=0.25\linewidth]{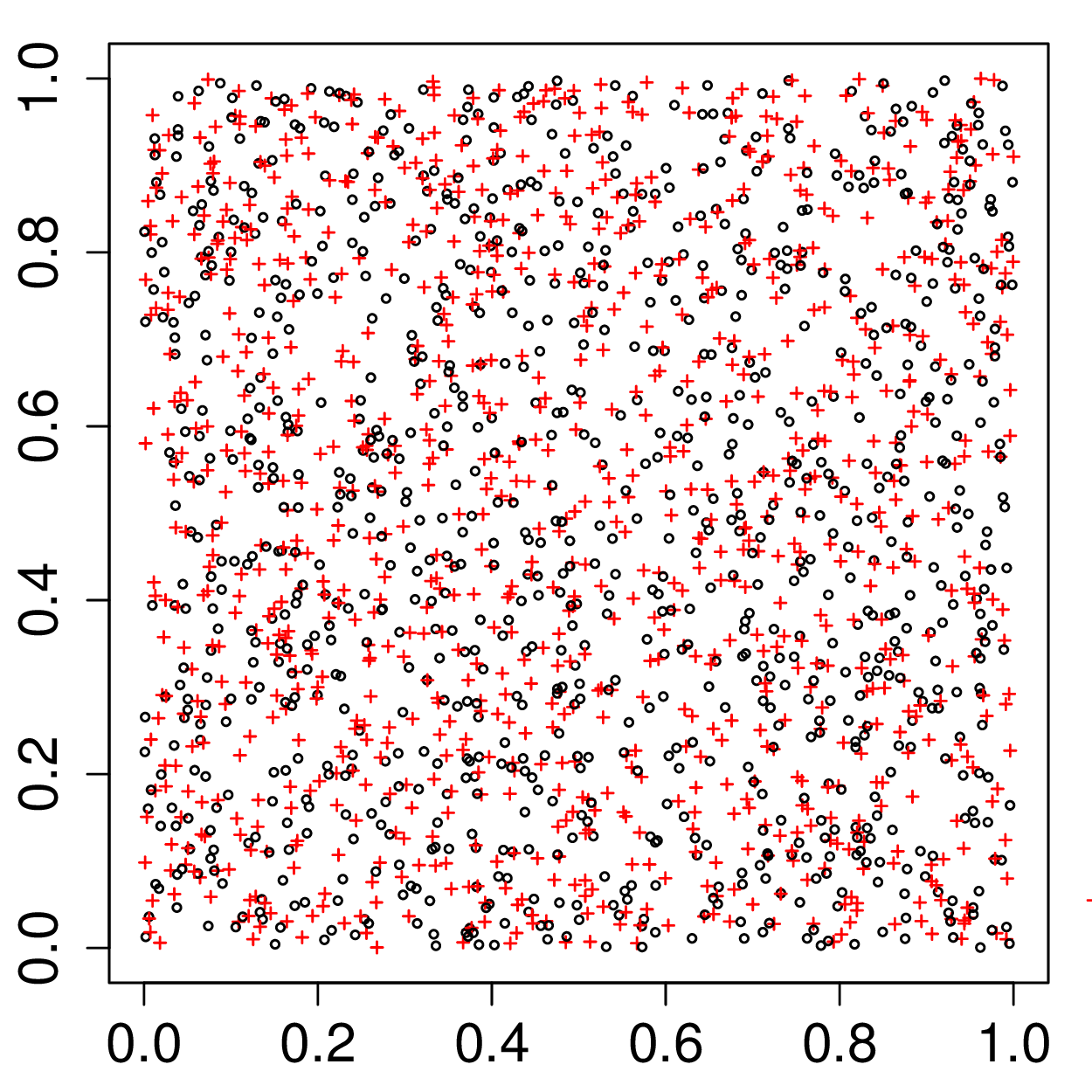}}\\\vspace{-12pt}
		\subfloat[][]{\includegraphics[width=0.25\linewidth]{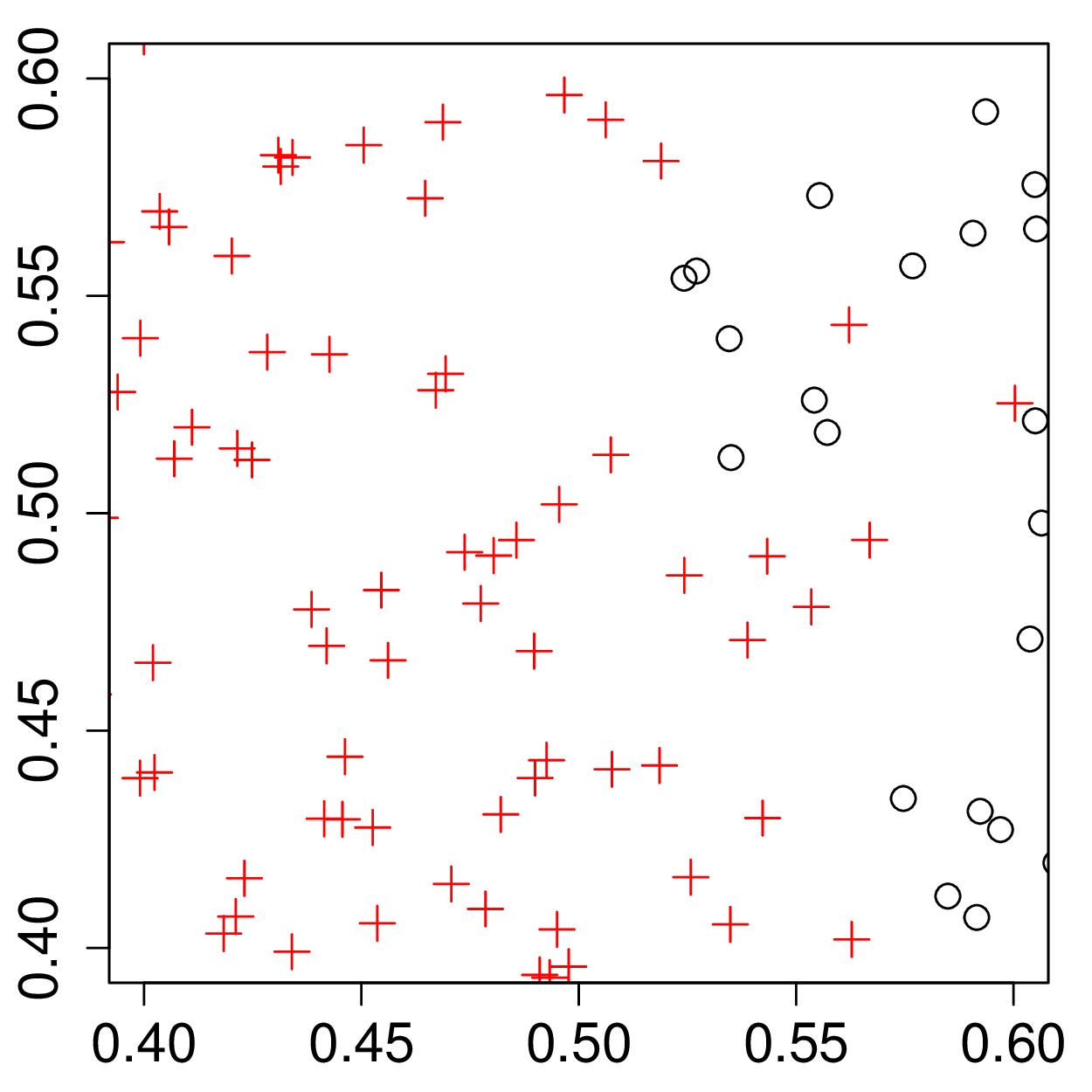}}
		\subfloat[][]{\includegraphics[width=0.25\linewidth]{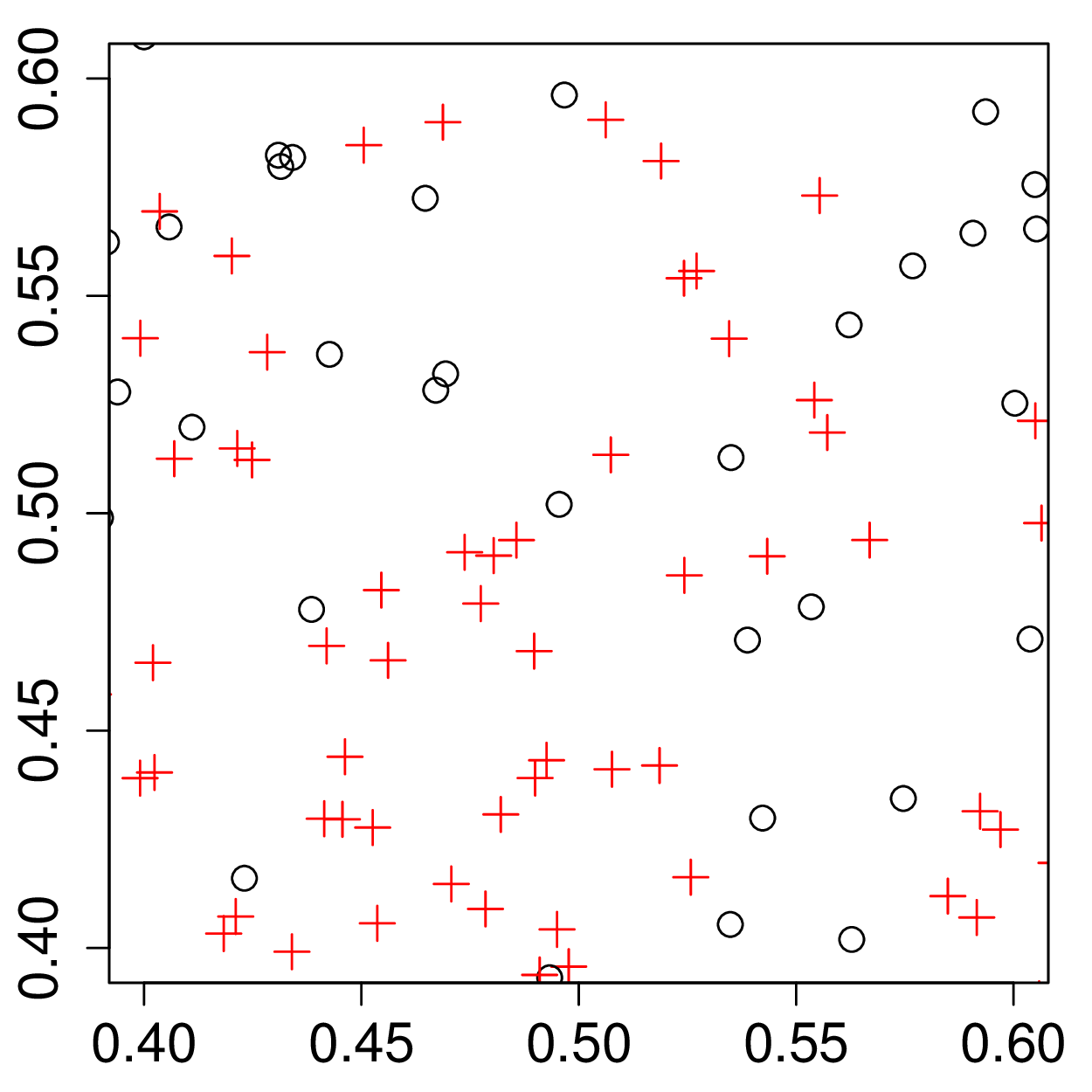}}
		\subfloat[][]{\includegraphics[width=0.25\linewidth]{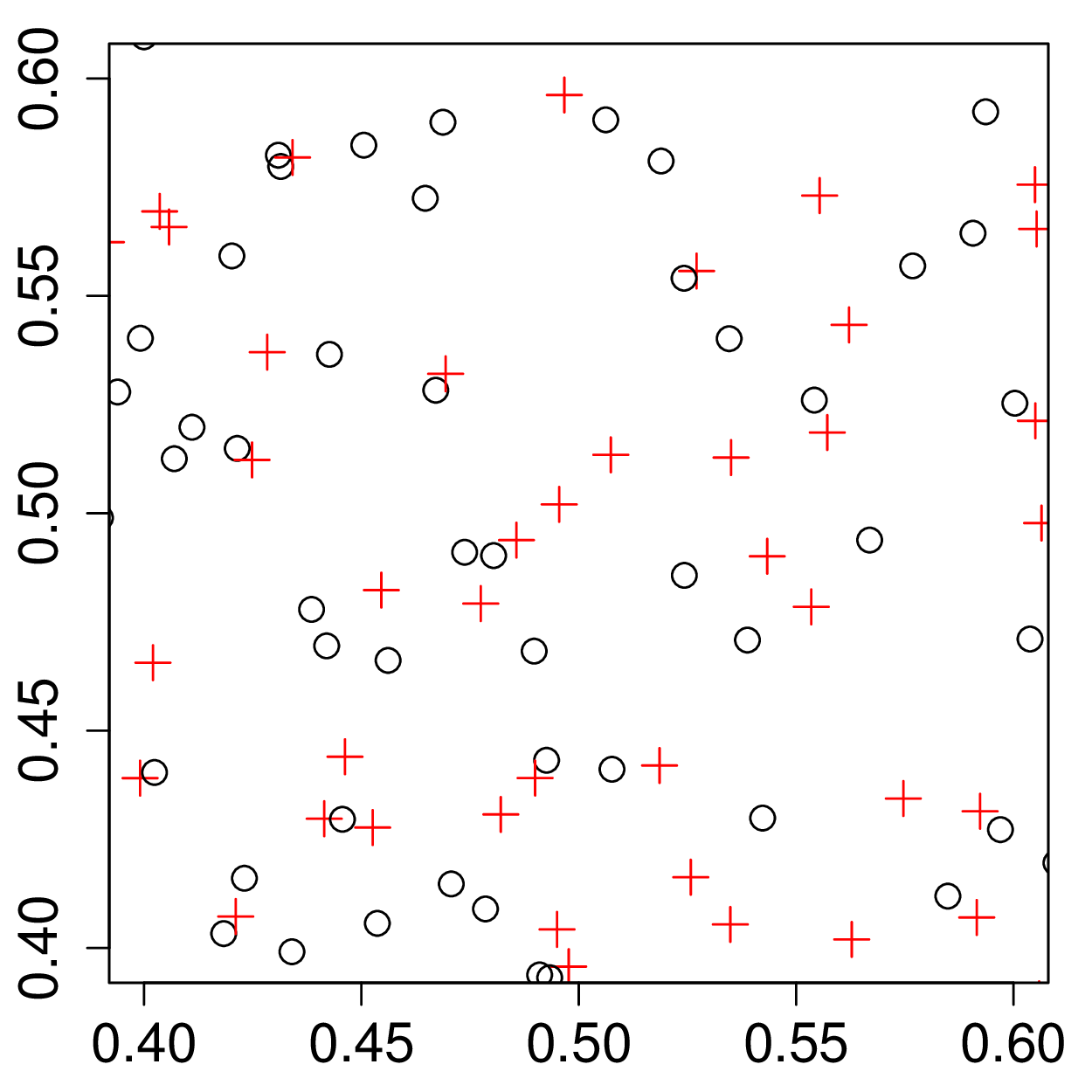}}
		\subfloat[][]{\includegraphics[width=0.25\linewidth]{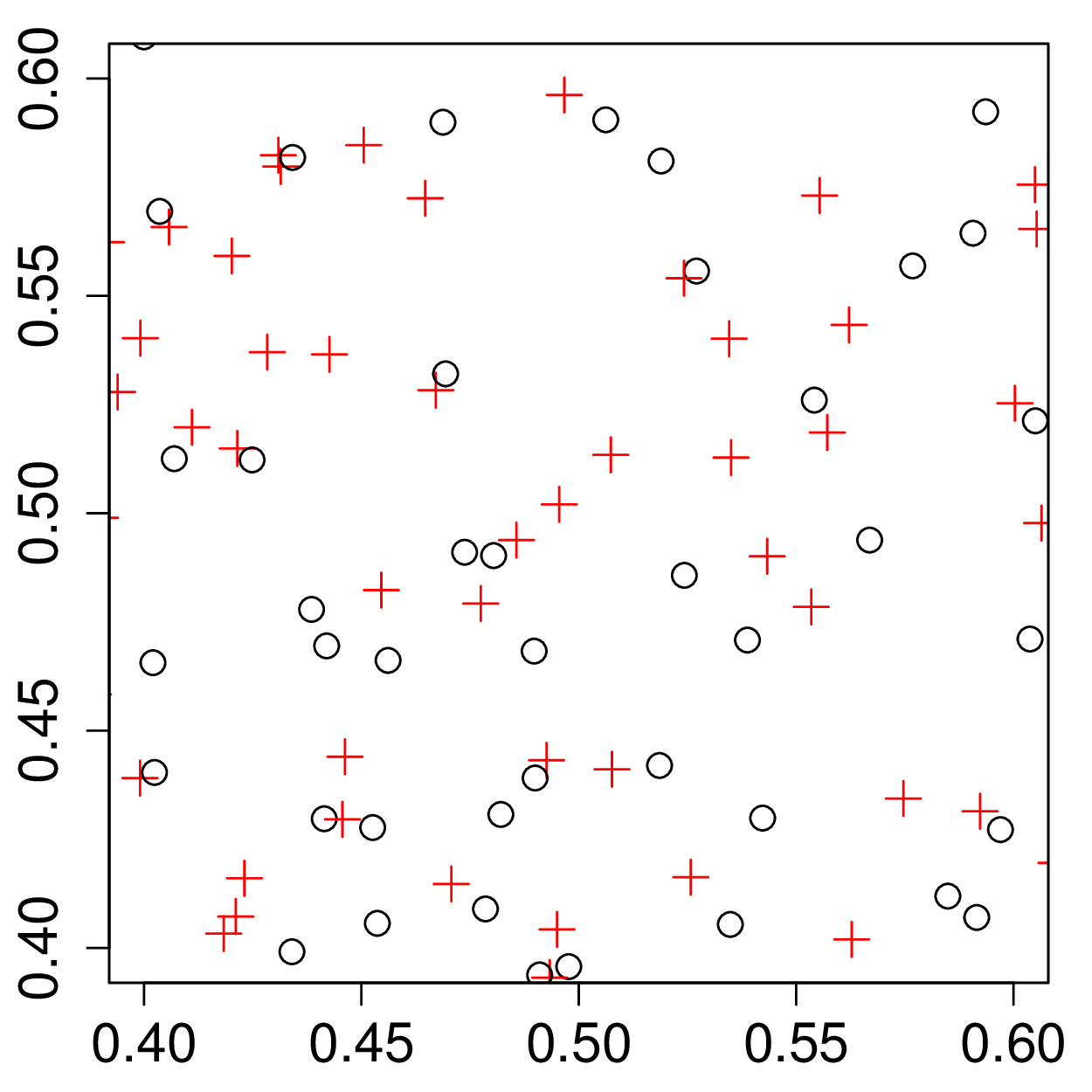}}\\\vspace{-12pt}
		\subfloat[][\\High attraction]{\includegraphics[width=0.25\linewidth]{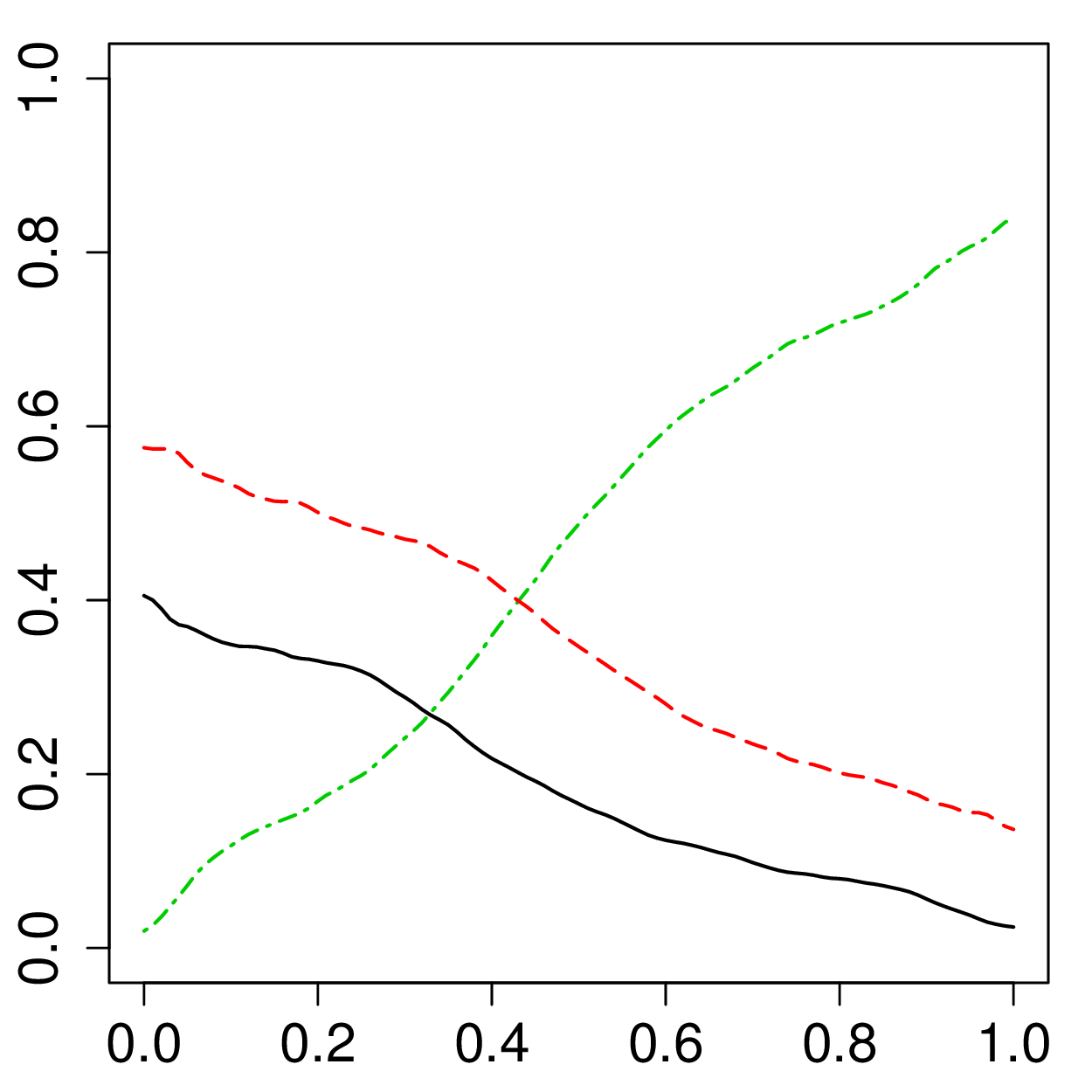}}
		\subfloat[][\\Low attraction]{\includegraphics[width=0.25\linewidth]{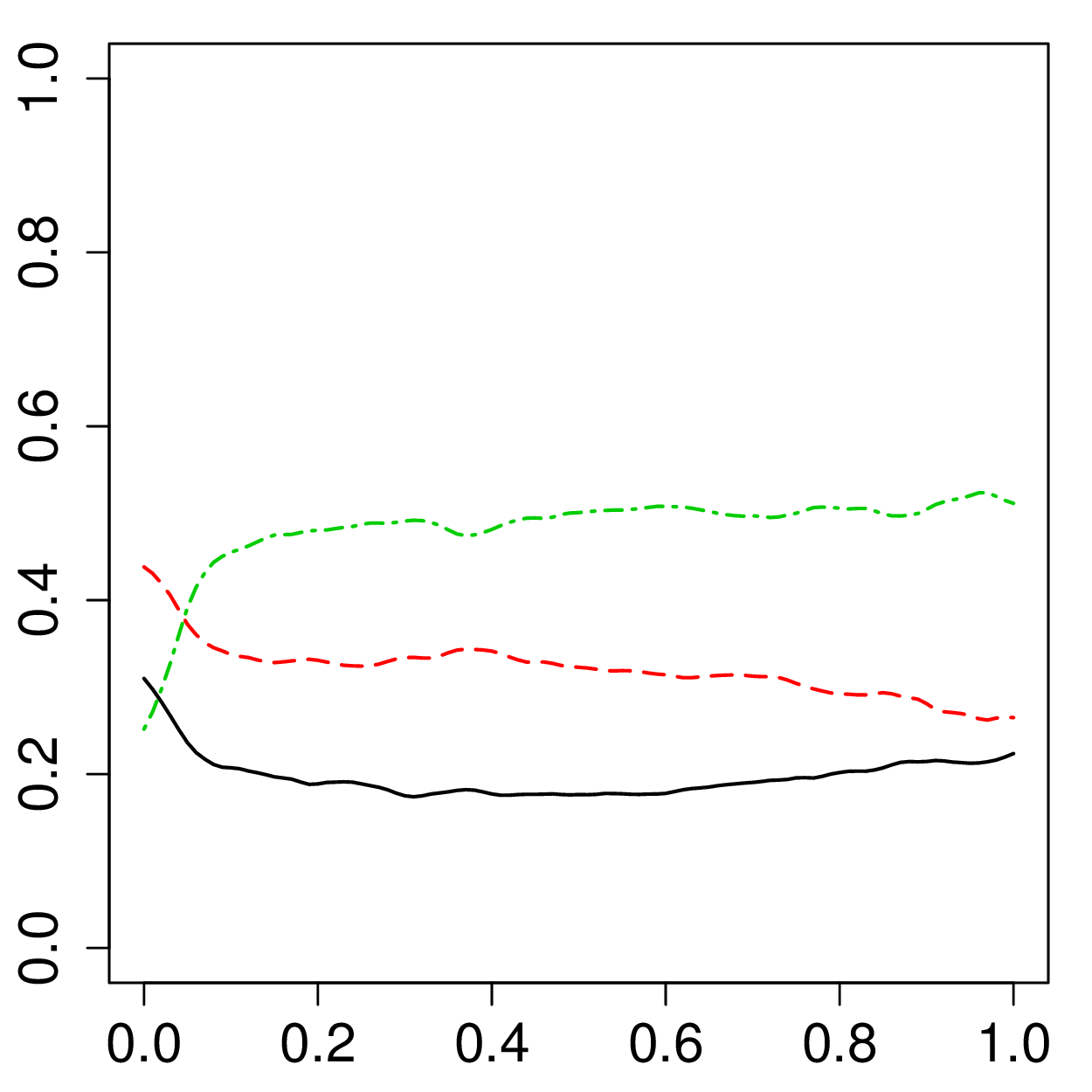}}
		\subfloat[][\\Low repulsion]{\includegraphics[width=0.25\linewidth]{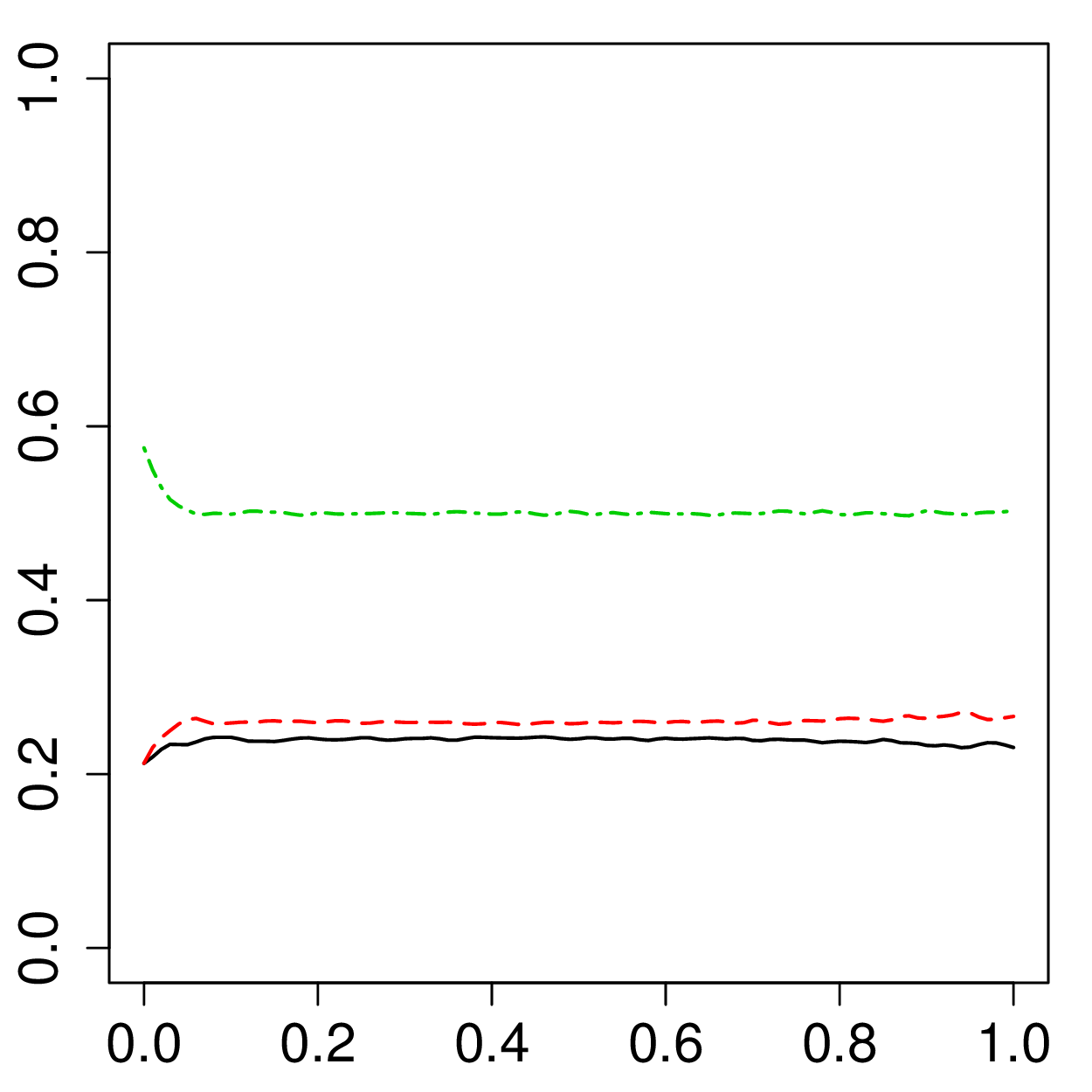}}
		\subfloat[][\\High repulsion]{\includegraphics[width=0.25\linewidth]{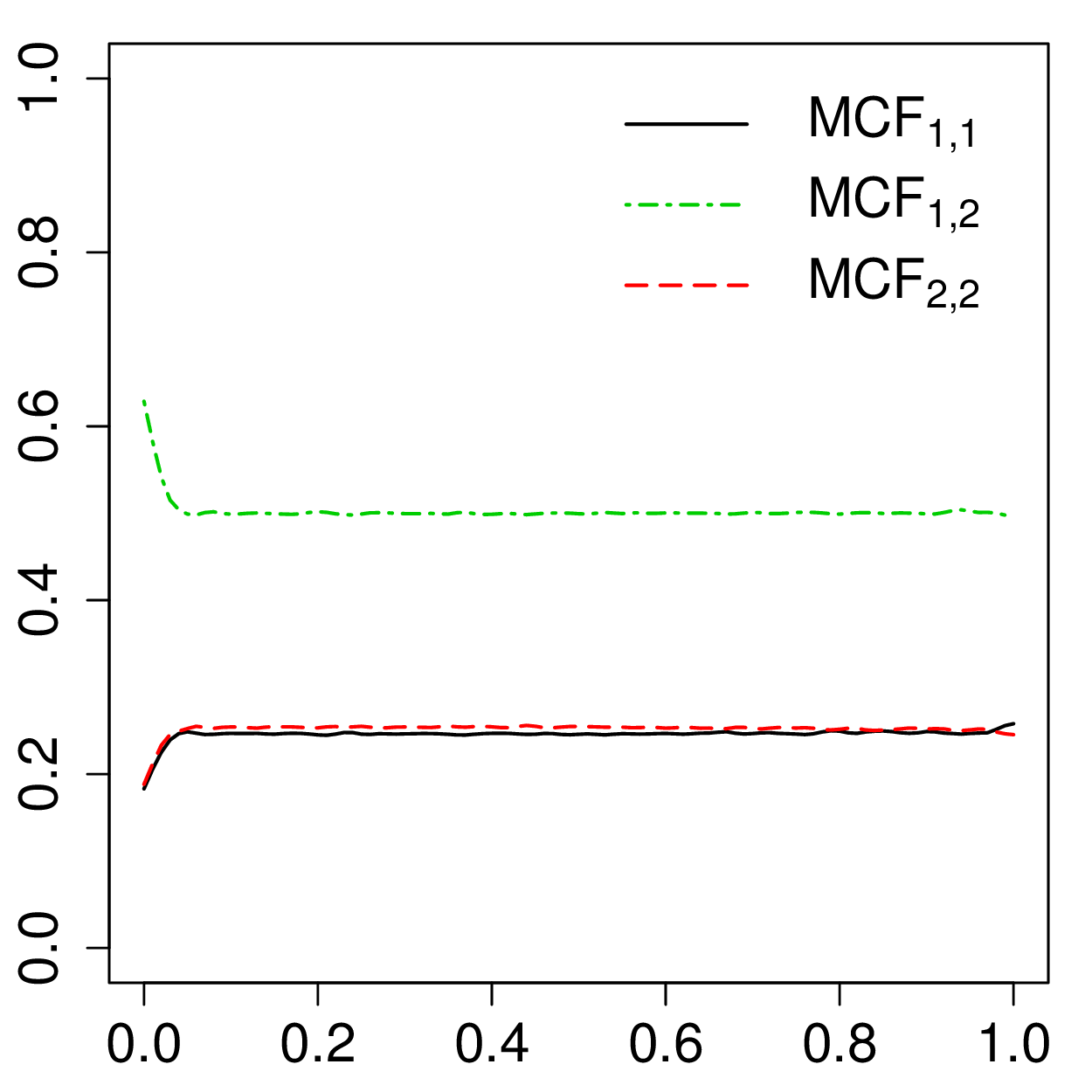}}
		\caption{Simulated datasets: (a)-(d): Examples of the generated data from the homogeneous Poisson process with the four scenarios: high/low attraction and high/low repulsion, with $\lambda=60$. (e)-(h): The center sub-regions of the simulated data as shown in (a)-(d). (i)-(l): The corresponding empirical mark connection function plots.}
		\label{Figure 1}
	\end{figure}

	We considered to generate the points by using two different point processes: 1) a homogeneous Poisson point process with a constant intensity $\eta=2000$ over the space $[0,1]^2$; and 2) a log Gaussian Cox process (LGCP) with an inhomogeneous intensity $\eta(x,y)=\exp(6+|x-0.3|+|y-0.3|+\mathcal{GP}(x,y)),x\in[0,1],y\in[0,1]$ and $\mathcal{GP}$ denotes a zero-mean Gaussian process with variance equal to $1$ and scale equal to $1$ (The LGCP setting was also used in \cite{Shirota2016}). We assumed that there are $Q=2$ different types of points. The mark of each point, $z_i$, was simulated by using a Gibbs sampler based on Equation (\ref{local_prob}). We ran $100,000$ iterations with a completely random starting configuration of $\bm{z}$. The true parameters were set as follows: 1) the decay parameter $\lambda=60$ or $\lambda=0$, and the threshold $c=0.05$, which implies that any pair of points with distance large than $0.05$ were not considered in the model construction; 2) the first-order intensities $\bm{\omega}=(\omega_1,\omega_2)=(1,1)$, which correspond to $\bm{\pi}=(0.5,0.5)$; and 3) the second-order intensities $\bm{\Theta}$ were set according to each of the five scenarios, as shown in Table \ref{Table 2}. They are high/low attraction, complete randomness, and high/low repulsion. \textit{Attraction} is defined as the clustering of points with the same type, while \textit{repulsion} (also known as inhibition or suppression) is defined as the clustering of points with different marks. We repeated the above steps to generate $30$ independent datasets for each point process and each setting of $\lambda$ and $\bm{\Theta}$. See Figure \ref{Figure 1} (a)-(d) for examples of simulated data generated by the homogeneous Poisson process under settings of $\bm{\Theta}$ and $\lambda=60$. Their corresponding mark connection function (MCF) plots are shown in \ref{Figure 1} (i)-(l). MCF is used to describe the spatial correlations of marks, where its quantity $\text{MCF}_{qq'}(d)$ is interpreted as the empirical probability that two points at distance $d$ have marks $q$ and $q'$. An upward trend in $\text{MCF}_{qq}(d)$ with a downward trend in $\text{MCF}_{qq'}(d)$ indicates attraction, while the opposite case suggests repulsion.
	\begin{table}[h]
		\renewcommand*{\arraystretch}{1.0}
		\begin{center}
			\caption{Simulated datasets: The five settings of the second-order intensities $\bm{\Theta}$ and their corresponding $\bm{\Phi}$.}\label{Table 2}
			{\begin{tabular}{@{}cccccc@{}}
					& High & Low & Complete & Low & High \\
					& attraction & attraction & randomness & repulsion & repulsion \\\hline
					$\bm{\Theta}$ & $\left(\begin{array}{cc}1.0 & 3.2\\3.2 & 1.0\end{array}\right)$ & $\left(\begin{array}{cc}1.0 & 1.9\\1.9 & 1.0\end{array}\right)$& $\left(\begin{array}{cc}1.0 & 1.0\\1.0 & 1.0\end{array}\right)$& $\left(\begin{array}{cc}1.0 & 0.2\\0.2 & 1.0\end{array}\right)$& $\left(\begin{array}{cc}1.0 & -1.2\\-1.2 & 1.0\end{array}\right)$\\\hline
					$\bm{\Phi}$ & $\left(\begin{array}{cc}0.9 & 0.1\\0.1 & 0.9\end{array}\right)$ & $\left(\begin{array}{cc}0.7 & 0.3\\0.3 & 0.7\end{array}\right)$& $\left(\begin{array}{cc}0.5 & 0.5\\0.5 & 0.5\end{array}\right)$& $\left(\begin{array}{cc}0.3 & 0.7\\0.7 & 0.3\end{array}\right)$& $\left(\begin{array}{cc}0.1 & 0.9\\0.9 & 0.1\end{array}\right)$\\
			\end{tabular}}
		\end{center}
	\end{table}
	
	For the prior on $\omega_1$, we used a normal distribution $\text{N}(1,1)$, corresponding that $\pi_1\in[0.125,0.878]$ with $95\%$ probability \textit{a priori}. For the priors on $\theta_{11}$ and $\theta_{12}$, we used a standard normal distribution $\text{N}(0,1)$. Note that we set the constraints $\omega_2=1$ and $\theta_{22}=1$ to avoid the identifiability problem. We set the hyperparameters that control the gamma prior on the exponential decay to $a_\lambda=b_\lambda=0.001$, which leads to a vague prior with variance equal to $1,000$. This is one of the most commonly used weak gamma priors \citep{Gelman2006}. For the tunable parameter $c$, we chose its true value $c=0.05$. Results we report below were obtained by running the MCMC chain with $50,000$ iterations, discarding the first $50\%$ sweeps as burn in. We started the chain from a model by randomly drawing $\omega_1$, $\theta_{11}$, $\theta_{12}$, and $\lambda$ from their prior distributions and assigning a random mark to each $z_i$. All experiments were implemented in \texttt{R} with \texttt{Rcpp} package to accelerate computations on a Mac PC with $2.60$GHz CPU and $16$GB memory. In our implementation, the MCMC algorithm ran about $20$ minutes for each dataset. We also assessed convergence by using the Raftery-Lewis diagnostic \citep{Raftery1992}, as included in \texttt{coda} package.
	
	Tables \ref{Table 3} - \ref{Table 44} summarize the results of posterior inference on the model parameters, under the $20$ scenarios (two point process and two settings of $\lambda$, and five settings of $\bm{\Theta}$). Each estimate was obtained by averaging over $30$ independent datasets. Overall, the tables indicate that our model fitting strategy based on the DMH algorithm works well, whichever point process is given. However, we notice that the decay parameter $\lambda$ was greatly overestimated in the complete randomness scenarios. This is not surprising because all $z_i$'s are completely irrelevant to each other (i.e. $p(z_i=q|\cdot)\propto\exp(-\omega_q)$) under this scenario. Therefore, $\lambda$ is ill-defined in this situation. The observed large values of $\lambda$ also indicate the weights, associated with the second-order intensities, decrease faster and thus explain why each mark is dominated by the first-order intensities only. It is also found that the high attraction scenarios had the worst performance on $\theta_{12}$, which measures the interaction strength between different types of points. The reason is that we can only observe a small number of the interacting pairs between type 1 and 2 points. Take Figure \ref{Figure 1} (a) for example, such interacting pairs can be only seen near the border between the two clumps. Therefore, we may expect a biased estimation on $\theta_{12}$.
	\begin{table}[htbp]
		\begin{center}
			\caption{Simulated datasets from the homogeneous Poisson process with $\lambda=60$: Results of posterior inference on the model parameters. Values are averaged over $30$ simulated datasets for each scenario, with standard deviations indicated in parentheses.}\label{Table 3}
			{\begin{tabular}{@{}cccccc@{}}
					\hline
					& High & Low & Complete & Low & High \\
					& attraction & attraction & randomness & repulsion & repulsion \\\hline
					$\omega_1$ & $1.0$ & $1.0$ & $1.0$ & $1.0$& $1.0$\\
					$\hat{\omega}_1$ & $1.30(0.39)$ & $1.05(0.12)$ & $1.04(0.09)$ & $1.08(0.18)$& $1.05(0.19)$\\\hline
					$\theta_{11}$ & $1.0$ & $1.0$ & $1.0$& $1.0$& $1.0$\\
					$\hat{\theta}_{11}$ &$0.71(0.37)$ & $0.97(0.09)$ & $0.84(0.33)$ & $0.94(0.15)$ & $0.95(0.14)$\\\hline
					$\theta_{12}$ & $3.2$ & $1.9$ & $1.0$ & $0.2$ & $-1.2$\\
					$\hat{\theta}_{12}$ &$2.52(0.26)$ & $1.82(0.17)$ & $0.81(0.30)$& $0.05(0.19)$ & $-1.14(0.20)$\\\hline
					$\lambda$ & $60$ & $60$& $60$& $60$& $60$\\
					$\hat{\lambda}$ &$48.36(6.79)$ & $58.77(7.49)$ & $186.76(118.09)$ & $65.58(11.82)$ & $58.75(4.81)$\\
			\end{tabular}}
		\end{center}
	\end{table}
	\begin{table}[htbp]
		\begin{center}
			\caption{Simulated datasets from the homogeneous Poisson process with $\lambda=0$: Results of posterior inference on the model parameters. Values are averaged over $30$ simulated datasets for each scenario, with standard deviations indicated in parentheses.}\label{Table 4}
			{\begin{tabular}{@{}cccccc@{}}
					& High & Low & Complete & Low & High \\
					& attraction & attraction & randomness & repulsion & repulsion \\\hline
					$\omega_1$ & $1.0$ & $1.0$ & $1.0$ & $1.0$& $1.0$\\
					$\hat{\omega}_1$ & $1.31(0.26)$ & $1.19(0.54)$ & $1.04(0.07)$ & $1.01(0.26)$& $0.97(0.43)$\\\hline
					$\theta_{11}$ & $1.0$ & $1.0$ & $1.0$& $1.0$& $1.0$\\
					$\hat{\theta}_{11}$ &$0.89(0.10)$ & $0.91(0.14)$ & $0.77(0.38)$ & $1.00(0.04)$ & $0.99(0.15)$\\\hline
					$\theta_{12}$ & $3.2$ & $1.9$ & $1.0$ & $0.2$ & $-1.2$\\
					$\hat{\theta}_{12}$ &$2.66(0.13)$ & $2.00(0.16)$ & $0.73(0.34)$& $0.11(0.05)$ & $-1.07(1.15)$\\\hline
					$\lambda$ & $0$ & $0$& $0$& $0$& $0$\\
					$\hat{\lambda}$ &$1.74(0.58)$ & $2.74(1.15)$ & $234.21(135.37)$ & $0.57(0.42)$ & $6.93(35.36)$\\
			\end{tabular}}
		\end{center}
	\end{table}
	\begin{table}[htbp]
		\begin{center}
			\caption{Simulated datasets from the LGCP process with $\lambda=60$: Results of posterior inference on the model parameters. Values are averaged over $30$ simulated datasets for each scenario, with standard deviations indicated in parentheses.}\label{Table 33}
			{\begin{tabular}{@{}cccccc@{}}
					\hline
					& High & Low & Complete & Low & High \\
					& attraction & attraction & randomness & repulsion & repulsion \\\hline
					$\omega_1$ & $1.0$ & $1.0$ & $1.0$ & $1.0$& $1.0$\\
					$\hat{\omega}_1$ & $1.05(0.14)$ & $1.02(0.09)$ & $1.04(0.10)$ & $1.00(0.13)$& $0.98(0.14)$\\\hline
					$\theta_{11}$ & $1.0$ & $1.0$ & $1.0$& $1.0$& $1.0$\\
					$\hat{\theta}_{11}$ &$0.87(0.26)$ & $0.87(0.30)$ & $0.61(0.50)$ & $0.91(0.31)$ & $0.96(0.22)$\\\hline
					$\theta_{12}$ & $3.2$ & $1.9$ & $1.0$ & $0.2$ & $-1.2$\\
					$\hat{\theta}_{12}$ &$2.64(0.48)$ & $1.75(0.22)$ & $0.63(0.40)$& $-0.07(0.33)$ & $-1.20(0.31)$\\\hline
					$\lambda$ & $60$ & $60$& $60$& $60$& $60$\\
					$\hat{\lambda}$ &$48.29(12.63)$ & $61.11(22.04)$ & $194.14(119.57)$ & $75.70(32.85)$ & $60.758(9.23)$\\
			\end{tabular}}
		\end{center}
	\end{table}
	\begin{table}[htbp]
		\begin{center}
			\caption{Simulated datasets from the LGCP process with $\lambda=0$: Results of posterior inference on the model parameters. Values are averaged over $30$ simulated datasets for each scenario, with standard deviations indicated in parentheses.}\label{Table 44}
			{\begin{tabular}{@{}cccccc@{}}
					& High & Low & Complete & Low & High \\
					& attraction & attraction & randomness & repulsion & repulsion \\\hline
					$\omega_1$ & $1.0$ & $1.0$ & $1.0$ & $1.0$& $1.0$\\
					$\hat{\omega}_1$ & $1.12(0.26)$ & $1.05(0.28)$ & $1.04(0.09)$ & $1.00(0.20)$& $0.98(0.24)$\\\hline
					$\theta_{11}$ & $1.0$ & $1.0$ & $1.0$& $1.0$& $1.0$\\
					$\hat{\theta}_{11}$ &$0.80(0.31)$ & $0.91(0.17)$ & $0.57(0.52)$ & $0.99(0.10)$ & $1.01(0.11)$\\\hline
					$\theta_{12}$ & $3.2$ & $1.9$ & $1.0$ & $0.2$ & $-1.2$\\
					$\hat{\theta}_{12}$ &$3.01(0.25)$ & $1.91(0.12)$ & $0.60(0.43)$& $0.05(0.17)$ & $-1.35(0.22)$\\\hline
					$\lambda$ & $0$ & $0$& $0$& $0$& $0$\\
					$\hat{\lambda}$ &$1.00(0.47)$ & $2.28(1.16)$ & $208.17(107.95)$ & $1.49(1.89)$ & $0.67(0.51)$\\
			\end{tabular}}
		\end{center}
	\end{table}

\begin{figure}[!p]
		\centering
		\subfloat[][High attraction]{\includegraphics[width=0.42\linewidth]{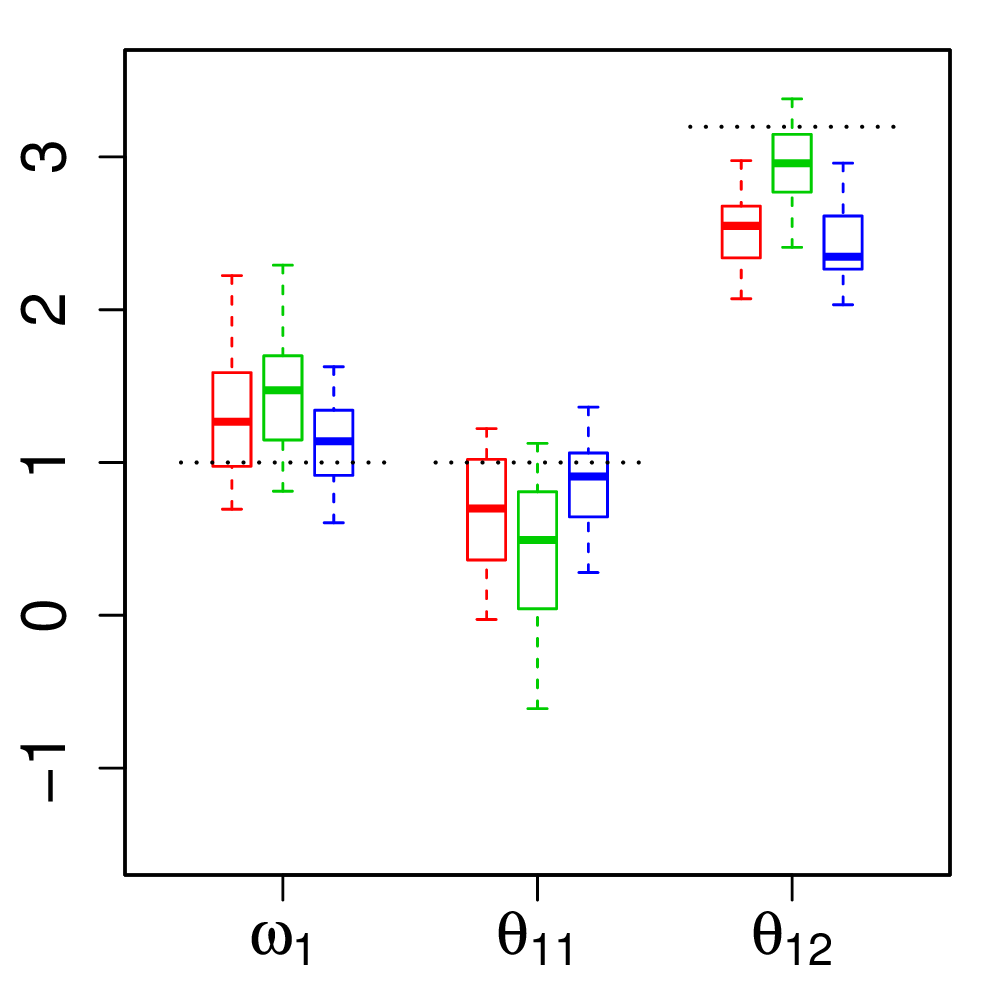}}
		\subfloat[][Low attraction]{\includegraphics[width=0.42\linewidth]{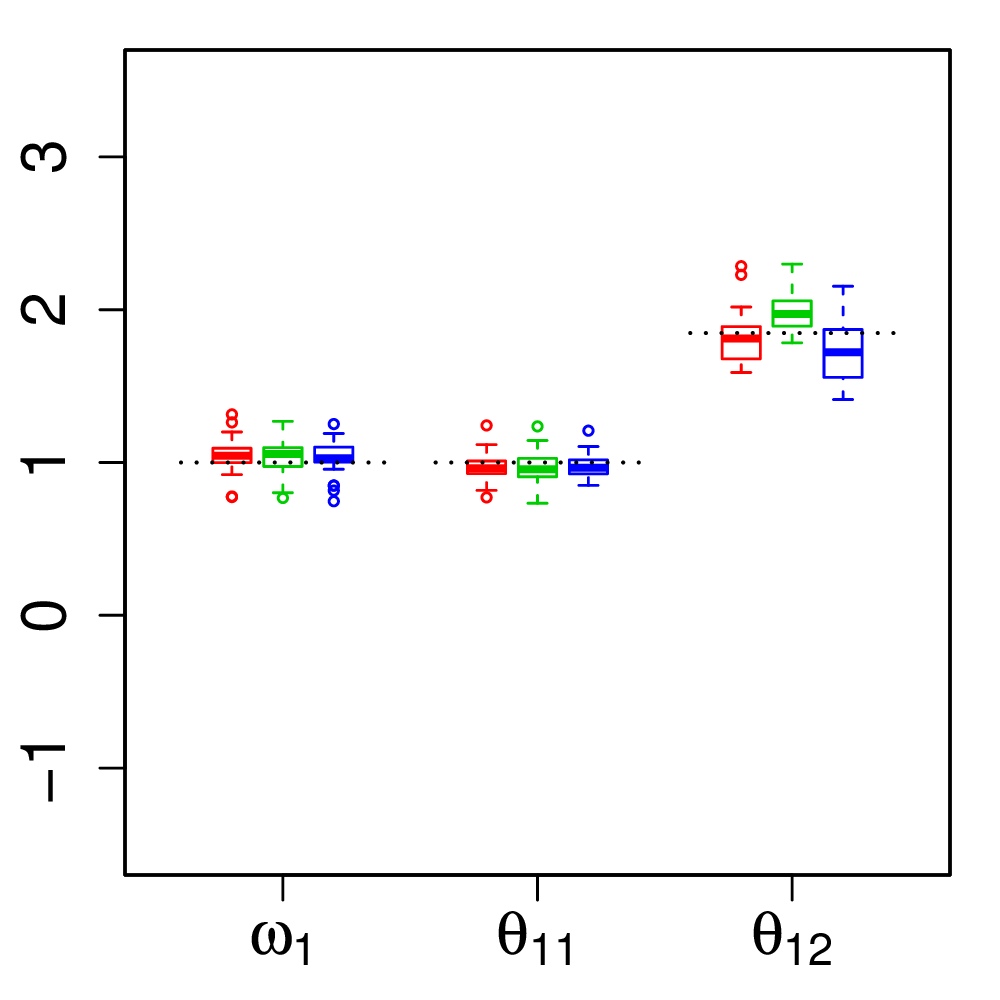}}\\\vspace{-13pt}
		\subfloat[][Low repulsion]{\includegraphics[width=0.42\linewidth]{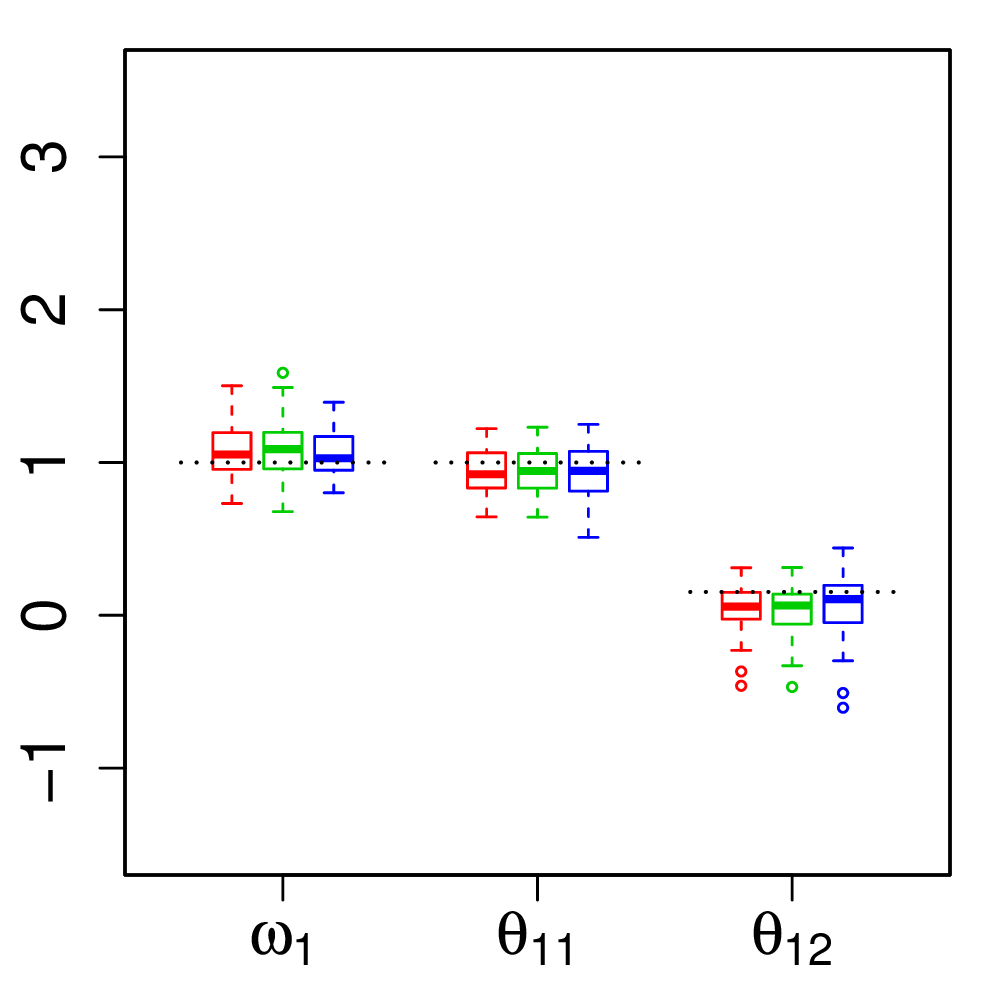}}
		\subfloat[][High repulsion]{\includegraphics[width=0.42\linewidth]{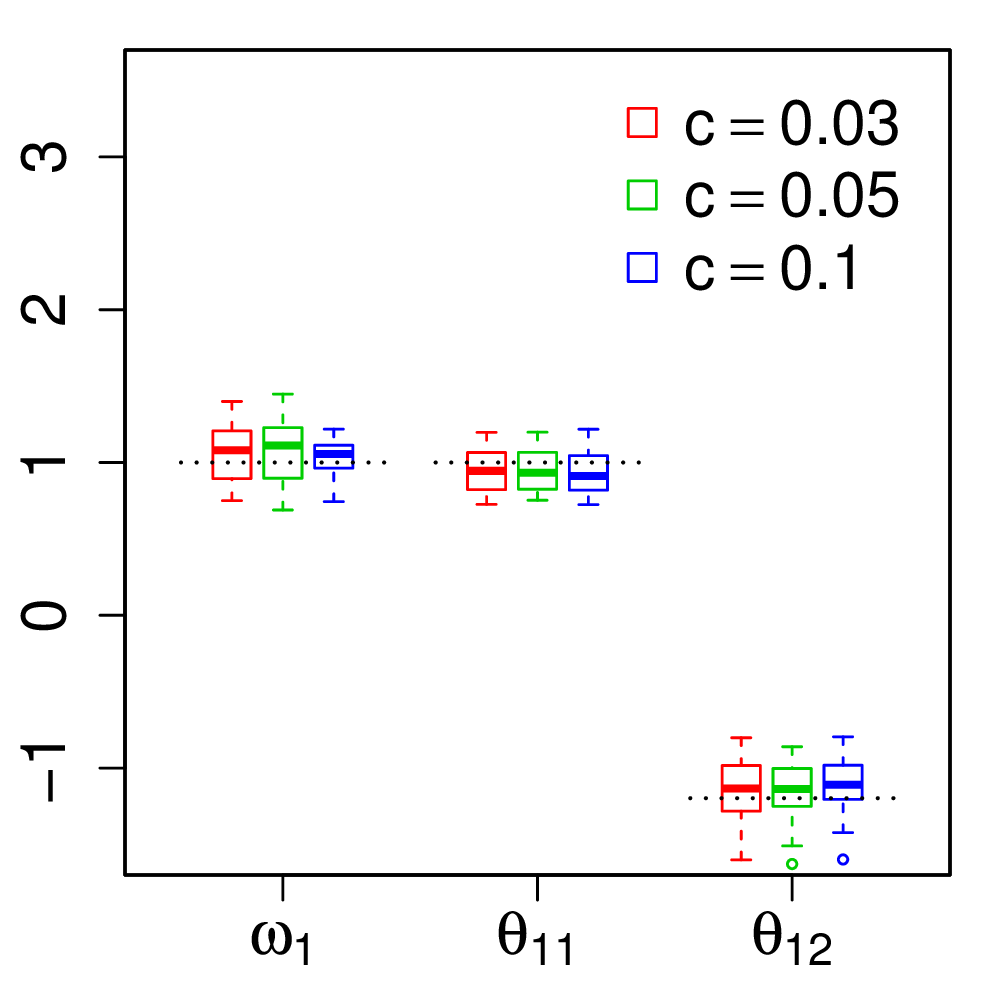}}
		\caption{Simulated datasets: The boxplots of $\hat{\omega}_1$, $\hat{\theta}_{11}$, and $\hat{\theta}_{12}$ under different choices of $c$ for each scenario: (a) high attraction, (b) low attraction, (c) low repulsion, and (d) high repulsion scenarios, with the black dashed lines indicating the true values.}
		\label{Figure 2}
	\end{figure}

	The proposed model contains one tunable parameter $c$, which defines the neighborhood for each point. A large value of $c$ quadratically increases the computational cost, while a small value may cause biased estimates. We suggest users choose a value of $0.1$ or less unless there is strong evidence in support of a larger value. Such evidence could be either subjective, such as an assessment from an experienced expert, or objective, such as MCF plots from the data (e.g. Figure \ref{Figure 1} (i)-(l)). For repulsion scenarios, it is found that the MCF curve converges right after $d$ passing over the true value of $c$. Thus, we could choose $c$ based on such an observation. However, for attraction scenarios, the curve tends to have a much bigger lag, especially for larger values of $\phi_{12}$ or $\phi_{21}$. In this case, we suggest users choose $c=0.1$. We also conducted a sensitivity analysis to the specification of $c$. We fit each of the $120$ simulated datasets generated from the homogeneous Poisson process ($30$ for each scenario, excluding the complete randomness one) into the proposed model with $c=0.03$, $0.05$, and $0.1$, respectively. Figure \ref{Figure 2} (a)-(d) show the boxplots of the three estimates $\hat{\omega}_1$, $\hat{\theta}_{11}$, and $\hat{\theta}_{12}$ under different values of $c$ for each scenario. As we can see, the model was quite robust to different choices of $c$.
	
	\section{Application}\label{application}
	In this section, we first investigate the performance of our methodology using three benchmark datasets in the \texttt{R} package \texttt{spatstat}, which is a major tool for spatial point pattern analysis. The proposed model is then applied to a large cohort of lung cancer pathology images, and it reveals novel potential imaging biomarkers for lung cancer prognosis.
	
	\subsection{\texttt{spatstat} Datasets}
	One of the basic data types offered by \texttt{spatstat} is multi-type point pattern data. We use two retinal cell datasets with marks on/off and one wood dataset with six species to quantify their attraction/repulsion characteristics by using the proposed model.
	
\begin{figure}
		\centering
		\subfloat[][Rescaled data]{\includegraphics[width=0.42\linewidth]{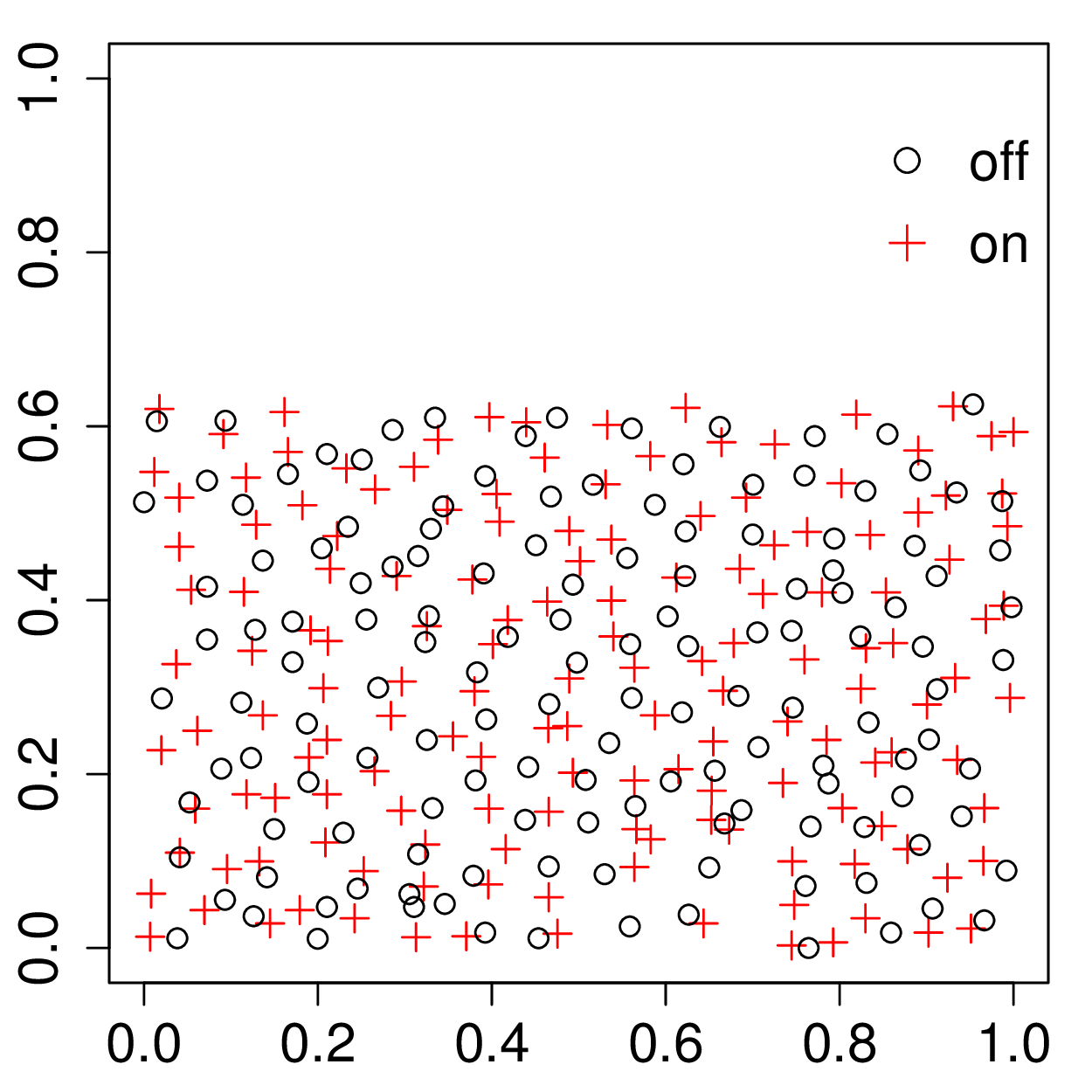}}
		\subfloat[][Mark connection functions]{\includegraphics[width=0.42\linewidth]{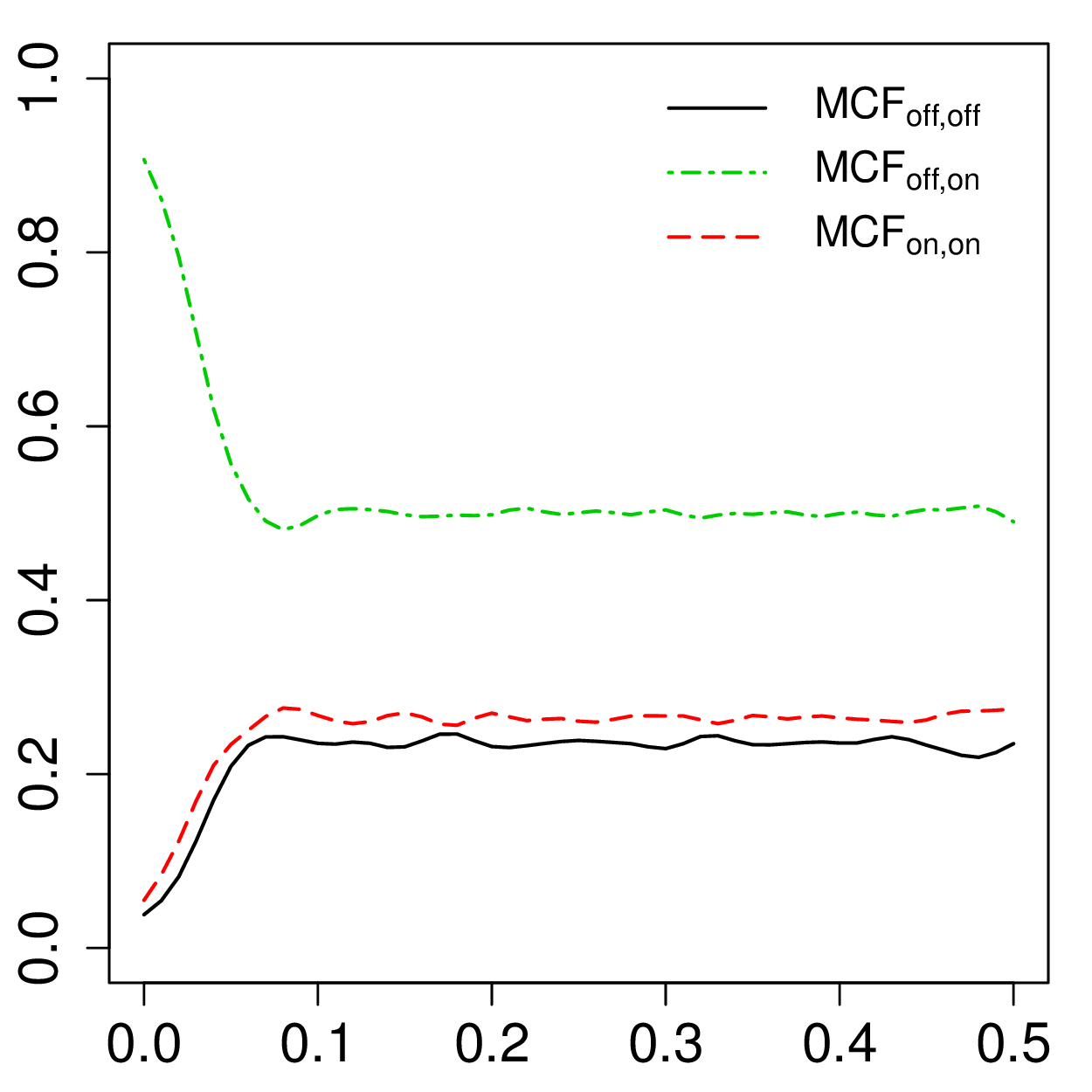}}\\\vspace{-15pt}
		\subfloat[][Estimated $\hat{\bm{\Phi}}$]{\includegraphics[width=0.42\linewidth]{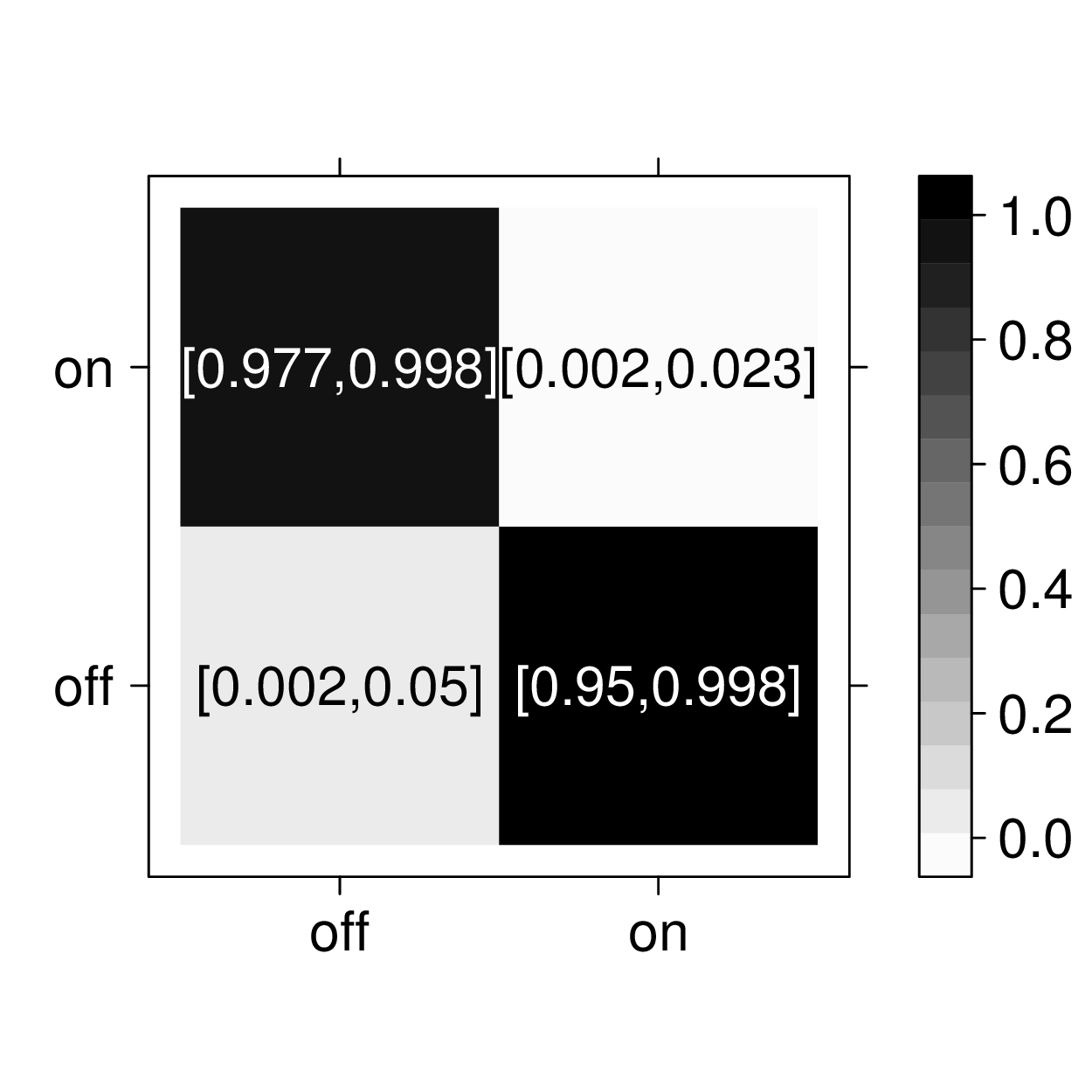}}
		\subfloat[][Mark interaction functions]{\includegraphics[width=0.42\linewidth]{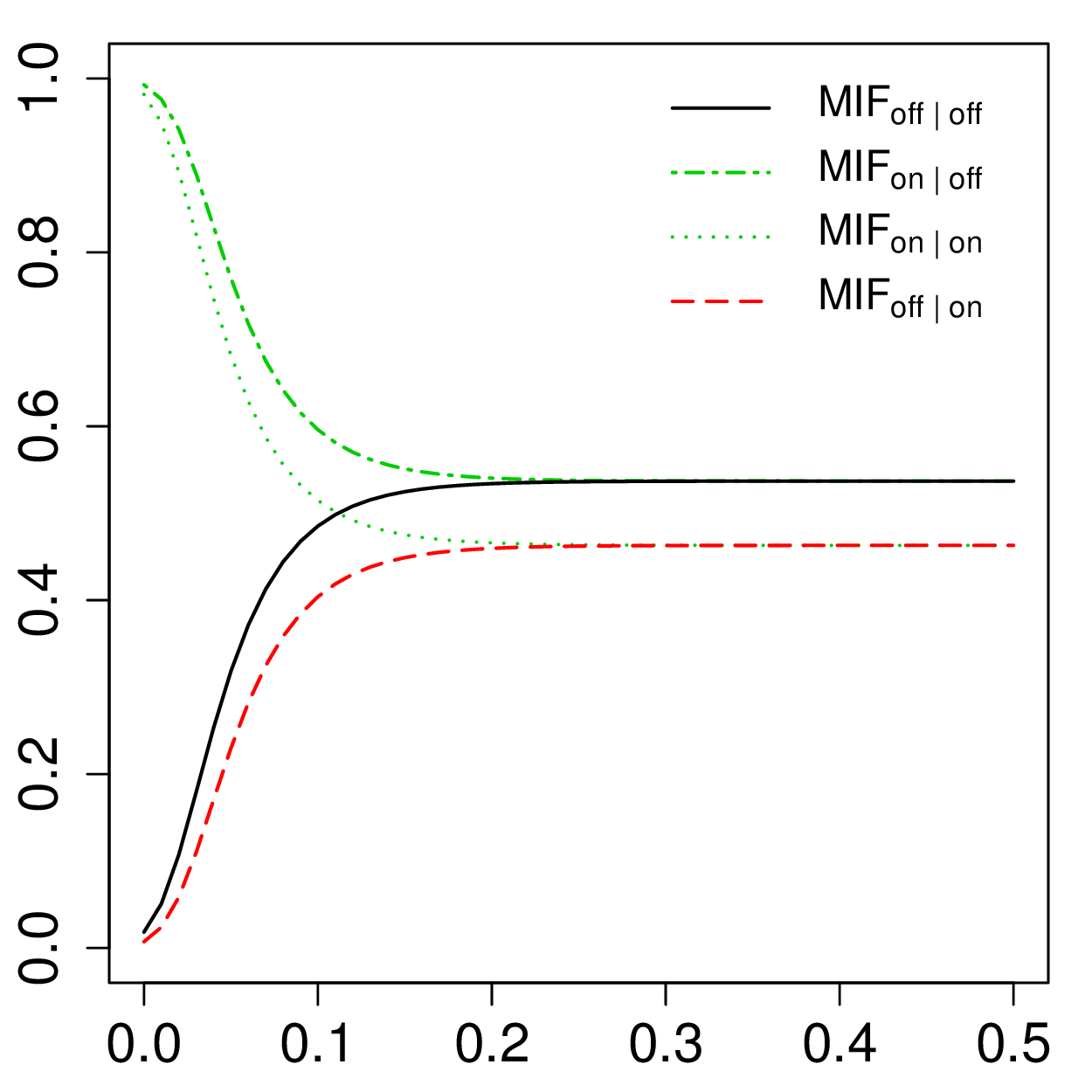}}
		\caption{\texttt{amacrine} dataset: (a) The rescaled marked point data, with a unit standing for approximate $1,000\mu$m; (b) The empirical mark connection function plots; (c) The levelplot of the estimated $\hat{\bm{\phi}}$, with the numbers in square brackets giving the $95\%$ credible interval; (d) The estimated mark interaction function plots.}
		\label{Figure 3}
	\end{figure}

\begin{figure}
		\centering
		\subfloat[][Rescaled data]{\includegraphics[width=0.42\linewidth]{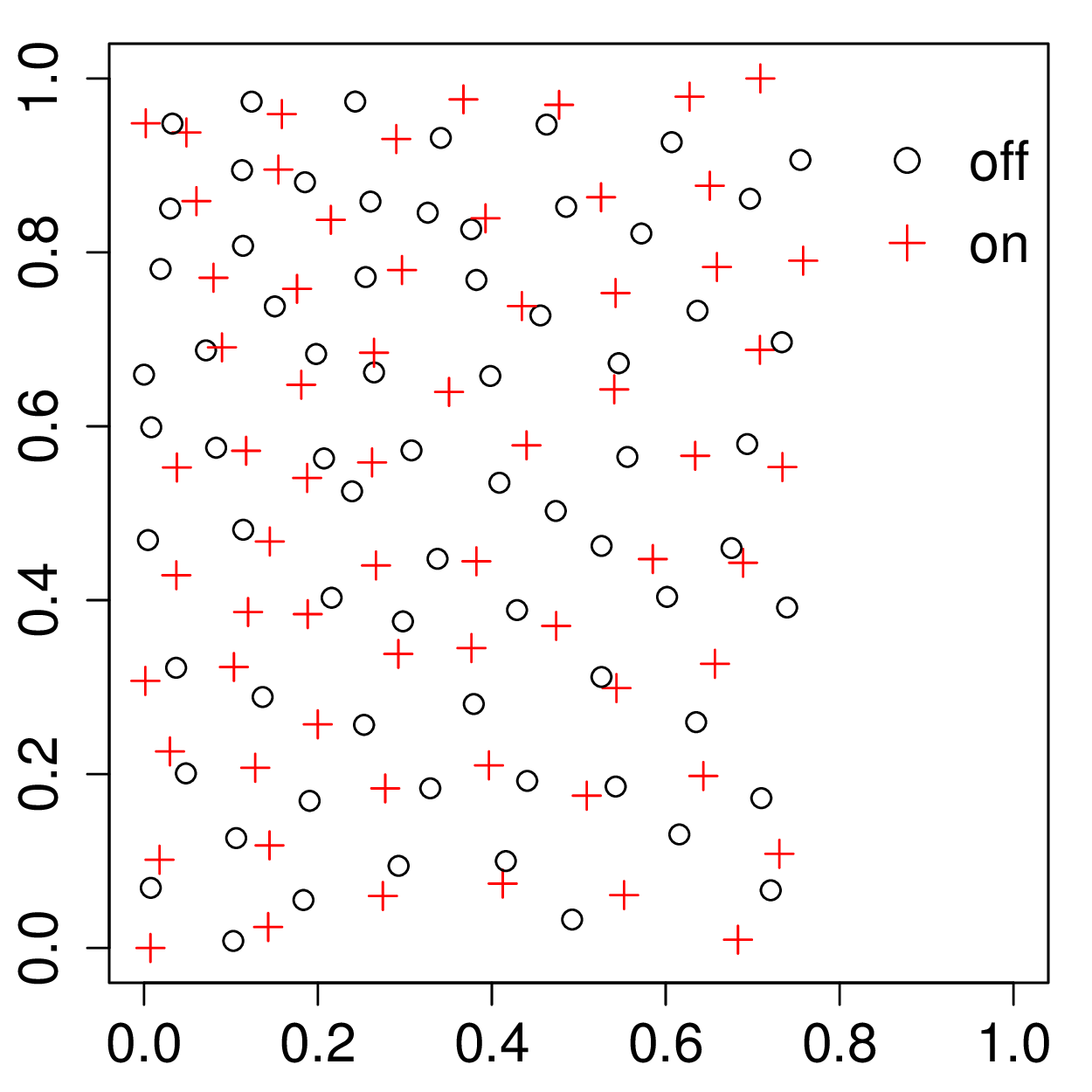}}
		\subfloat[][Mark connection functions]{\includegraphics[width=0.42\linewidth]{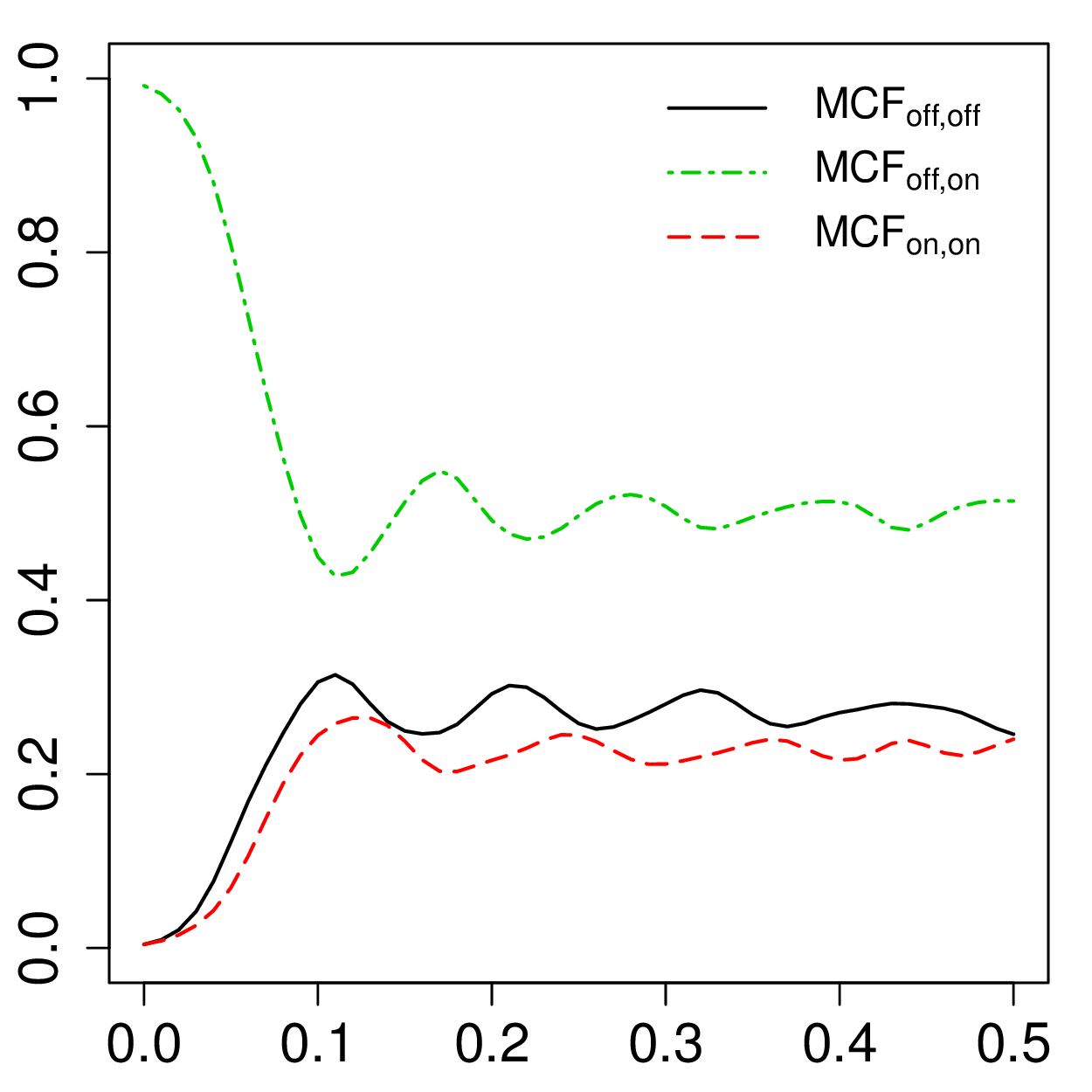}}\\\vspace{-15pt}
		\subfloat[][Estimated $\hat{\bm{\Phi}}$]{\includegraphics[width=0.42\linewidth]{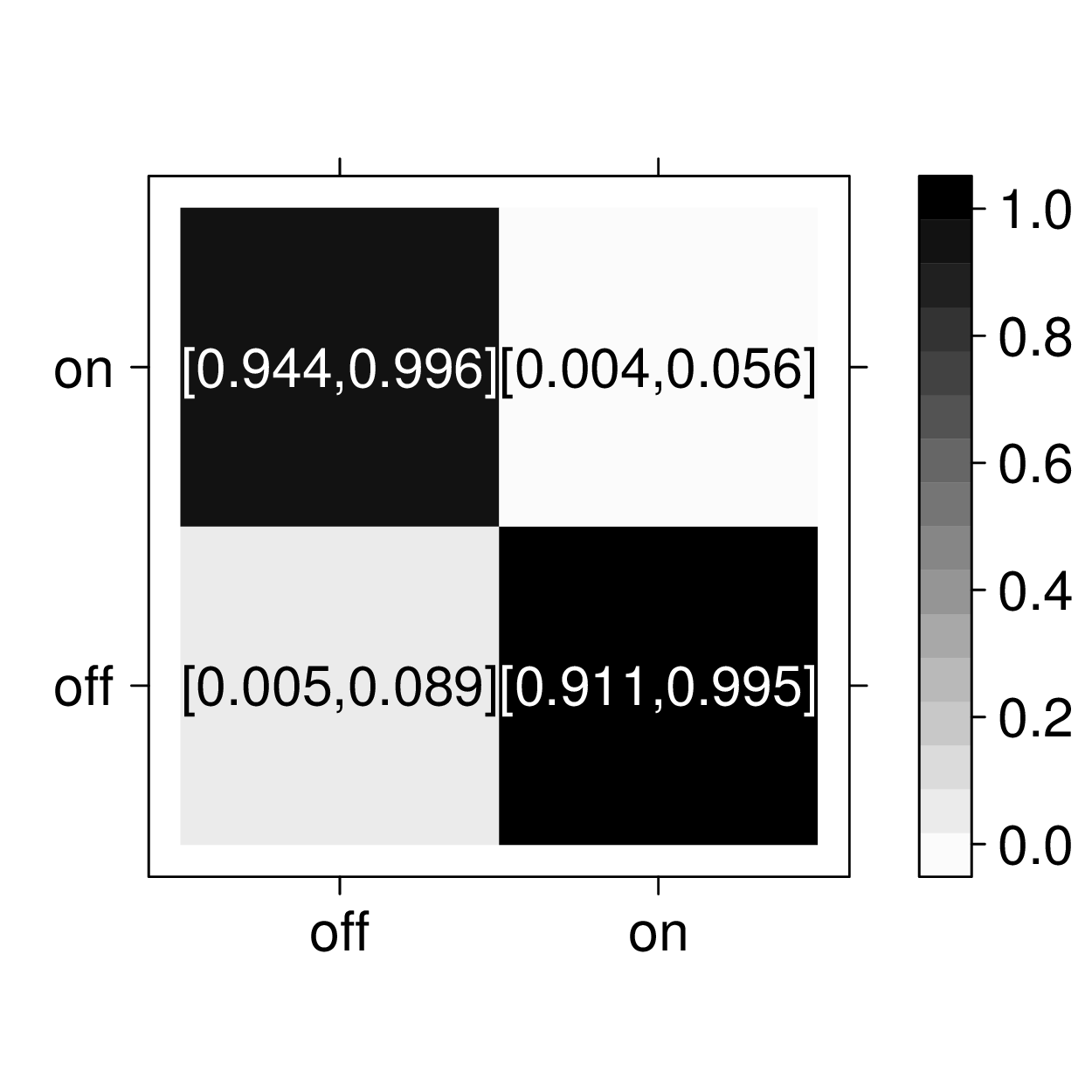}}
		\subfloat[][Mark interaction functions]{\includegraphics[width=0.42\linewidth]{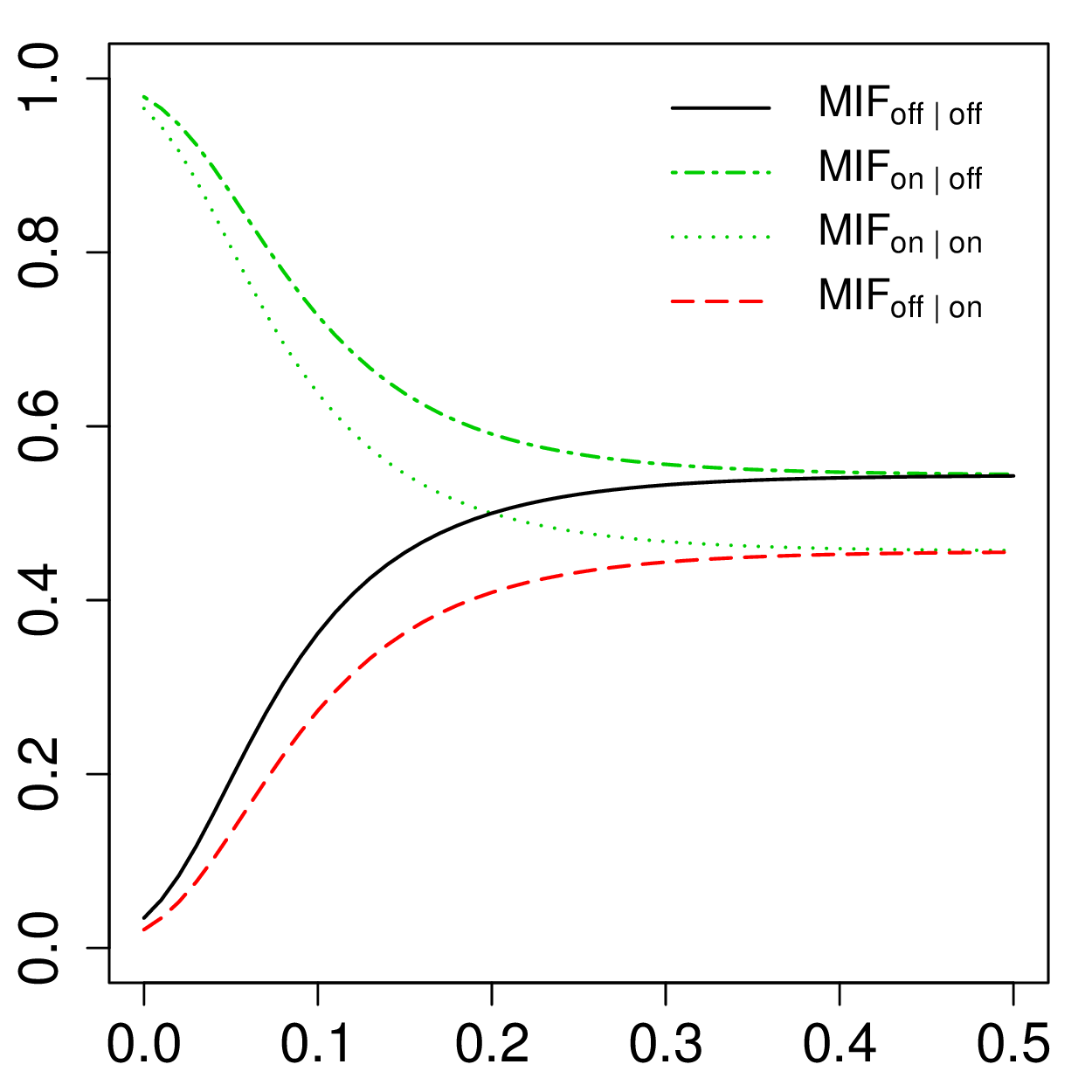}}
		\caption{\texttt{betacells} dataset: (a) The rescaled marked point data, with a unit standing for approximate $1,000\mu$m; (b) The empirical mark connection function plots; (c) The levelplot of the estimated $\hat{\bm{\phi}}$, with the numbers in square brackets giving the $95\%$ credible interval; (d) The estimated mark interaction function plots.}
		\label{Figure 4}
	\end{figure}

	Since 1970s, there has been considerable interest in studying the spatial pattern presented by particular types of mammalian retinal cell bodies \citep{Wassle1978,Wassle1978_2,Wassle1981,Wassle1981_2,Wassle1981_3,Hughes1981,Hughes1981_2,Peichl1981,Vaney1981,Rockhill2000}. One of the two commonly used examples is the \texttt{amacrine} cells dataset \citep{Diggle1986}, consisting of two types (i.e. on/off) of displaced amacrine cells within the retinal ganglion cell layer of a rabbit. The other is the \texttt{betacells} dataset \citep{Wassle1981}, composed of two types (i.e. on/off) beta cells that are associated with the resolution of fine details in the visual system of a cat. Figure \ref{Figure 3} (a) depicts how the two different types of amacrine cells distribute in a $1070\times600\mu$m rectangular region, where the $142$ circles ($\circ$) represent those cells processing ``light-off" information and the $152$ crosses (\textcolor{red}{$+$}) represent those cells processing ``light-on" information. Figure \ref{Figure 4} (a) shows the cell distribution map of the \texttt{betacells} dataset in an approximate $753\times1,000\mu$m rectangular window, where the $70$ circles ($\circ$) represent those ``off" beta cells and the $65$ crosses (\textcolor{red}{$+$}) represent those ``on" beta cells. Their mark connection function plots are shown in Figure \ref{Figure 3} (b) and Figure \ref{Figure 4} (b), respectively. Although both of the plots clearly indicate strong repulsion among cells with the same type and the interaction region radius around $0.1$, no quantities can be accurately estimated further. 
	
	For each dataset, we applied the proposed model with the same hyperparameter and algorithm settings as described in Section \ref{simulation} and the choice of $c=0.2$. We ran four independent MCMC chains with $50,000$ iterations, discarding the first half as burn-in. The Gelman and Rubin's convergence diagnostics \cite{Gelman1992} were used to inspect the convergence. Those statistics for all the model parameters were below $1.03$, ranging from $1.002$ to $1.029$, clearly suggesting that the MCMC chains were run for a sufficient number of iterations. Then, for each dataset, we pooled together the outputs from the four chains and report the results as below. For dataset \texttt{amacrine}, we obtained the decay $\hat{\lambda}=30.195$, the first-order intensity $\hat{\omega}_\text{off}=0.85$ corresponding to $\hat{\pi}_\text{off}=0.538$ and $\hat{\pi}_\text{on}=0.462$, and the second-order intensities $\hat{\theta}_\text{off,off}=0.35$ and $\hat{\theta}_\text{on,off}=\hat{\theta}_\text{off,on}=-4.024$ corresponding to $\hat{\phi}_\text{off,off}=0.012$, $\hat{\phi}_\text{on,off}=0.999$, $\hat{\phi}_\text{off,on}=0.993$, and $\hat{\phi}_\text{on,on}=0.007$. For dataset \texttt{betacells}, we obtained the decay $\hat{\lambda}=15.695$, the first-order intensity $\hat{\omega}_\text{off}=0.882$ corresponding to $\hat{\pi}_\text{off}=0.53$ and $\hat{\pi}_\text{on}=0.47$, and the second-order intensities $\hat{\theta}_\text{off,off}=0.65$ and $\hat{\theta}_\text{on,off}=\hat{\theta}_\text{off,on}=-3.104$ corresponding to $\hat{\phi}_\text{off,off}=0.023$, $\hat{\phi}_\text{on,off}=0.977$, $\hat{\phi}_\text{off,on}=0.984$, and $\hat{\phi}_\text{on,on}=0.016$. Figure \ref{Figure 3} (c) and Figure \ref{Figure 4} (c) also show the levelplot of the estimated $\hat{\bm{\phi}}$ and the $95\%$ credible interval for each $\hat{\phi}_{qq'}$. Figure \ref{Figure 3} (d) and Figure \ref{Figure 4} (d) plots the mark interaction functions according to the estimated model parameters. Our method, as well as other methods \citep{Diggle1986,Van1999}, suggest repulsion between the cells (e.g. most cells have a nearest neighbor of the opposite type). The message about oppositely labelled pairs between neighbor cells would strengthens the assumption that there are two separate channels for brightness and darkness as postulated by Hering in 1874. Indeed, we provide an accurate quantitative description $\bm{\omega}$ and $\bm{\Theta}$, along with the corresponding probability measurements $\bm{\pi}$ and $\bm{\Phi}$, which may benefit the development and retinal sampling efficiency.  
	
\begin{figure}
		\centering
		\subfloat[][Rescaled data]{\includegraphics[width=0.6\linewidth]{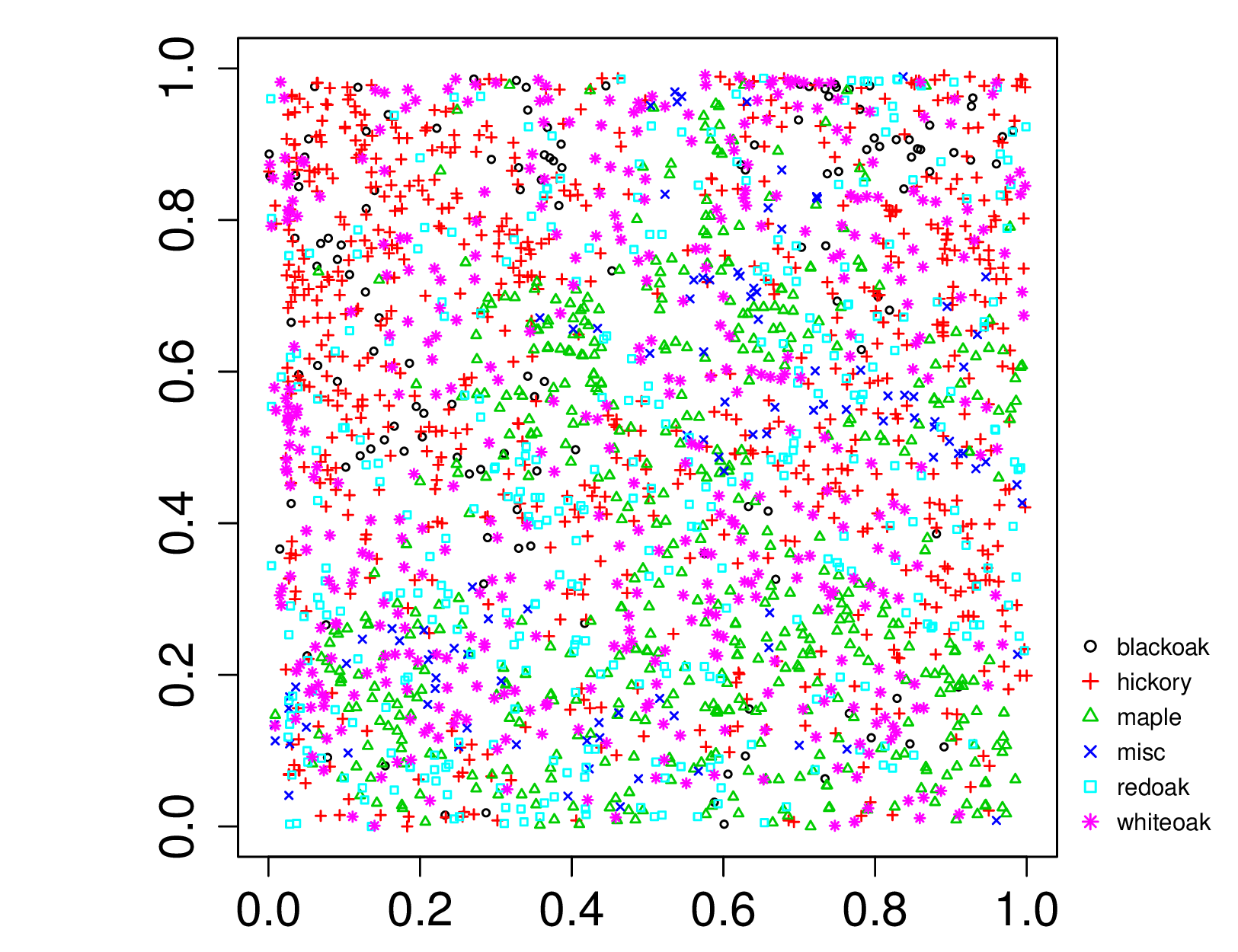}}\\\vspace{-15pt}
		\subfloat[][Mark connection functions]{\includegraphics[width=0.42\linewidth]{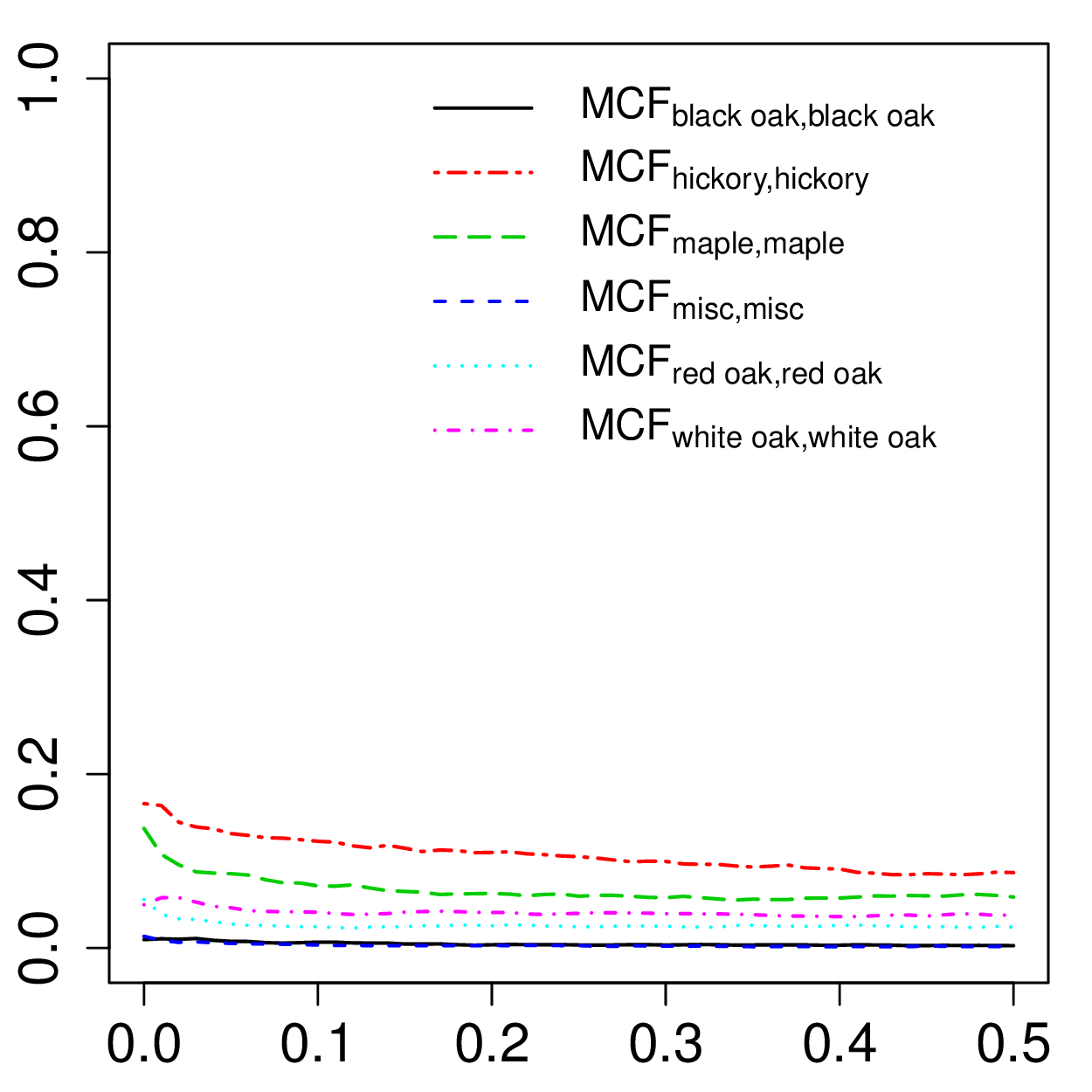}}
		%\subfloat[][Estimated $\hat{\bm{\Phi}}$]{\includegraphics[width=0.42\linewidth]{fig_5_3_3}}
		\subfloat[][Mark interaction functions]{\includegraphics[width=0.42\linewidth]{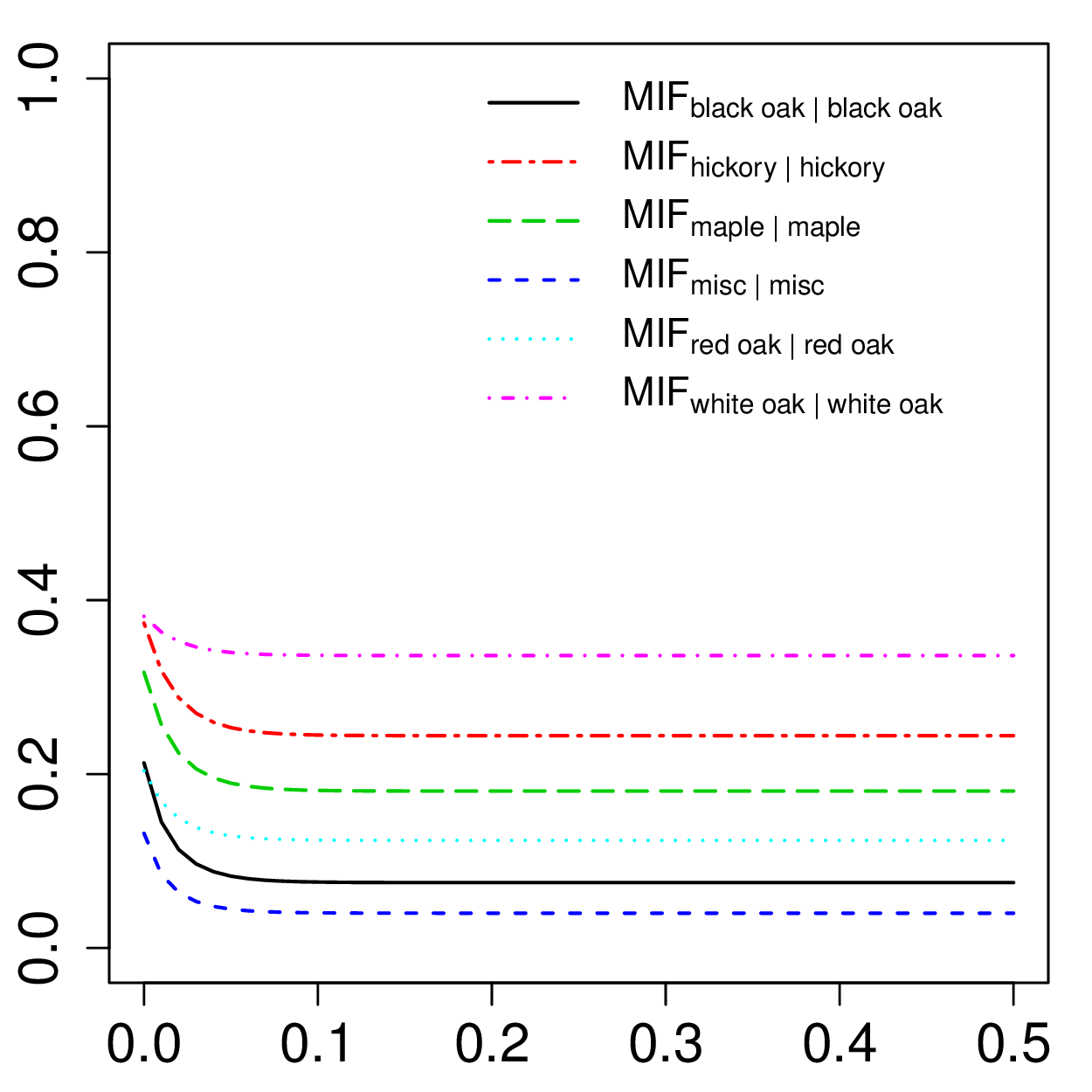}}
		\caption{\texttt{lansing} dataset: (a) The rescaled marked point data, with a unit standing for approximate $282$m ($\approx924$ft); (b) The empirical mark connection function plots (only the MCFs between the same mark are shown); (c) The estimated mark interaction function plots (only the MIFs between the same mark are shown).}
		\label{Figure 5}
	\end{figure}

	The third dataset in \texttt{spatstat} that was used to demonstrate the proposed model is the \texttt{lansing} dataset. It contains the locations and botanical classification of trees in Lansing Woods, Clinton County, Michigan, United States. \cite{Gerrard1969} investigated $n=2,251$ trees, including $135$ black oaks ($\circ$), $703$ hickories (\textcolor{red}{$+$}), $514$ maples (\textcolor{green}{$\bigtriangleup$}), $105$ miscellaneous trees (\textcolor{blue}{$\times$}), $346$ red oaks (\textcolor{cyan}{$\Box$}), and $448$  white oaks (\textcolor{magenta}{$\ast$}), over an area of $924\times924$ feet ($19.6$ acre). Figure \ref{Figure 5} (a) shows the rescaled multivariate spatial pattern that consists of $Q=6$ types of trees in the unit square, where each shape/color represents a tree category. Next, Figure \ref{Figure 5} (b) plots the corresponding mark connection functions between the same mark, which indicate exhibition of clustering among the trees with the same type.
	
\begin{figure}
	\centering
	\includegraphics[width=0.99\linewidth]{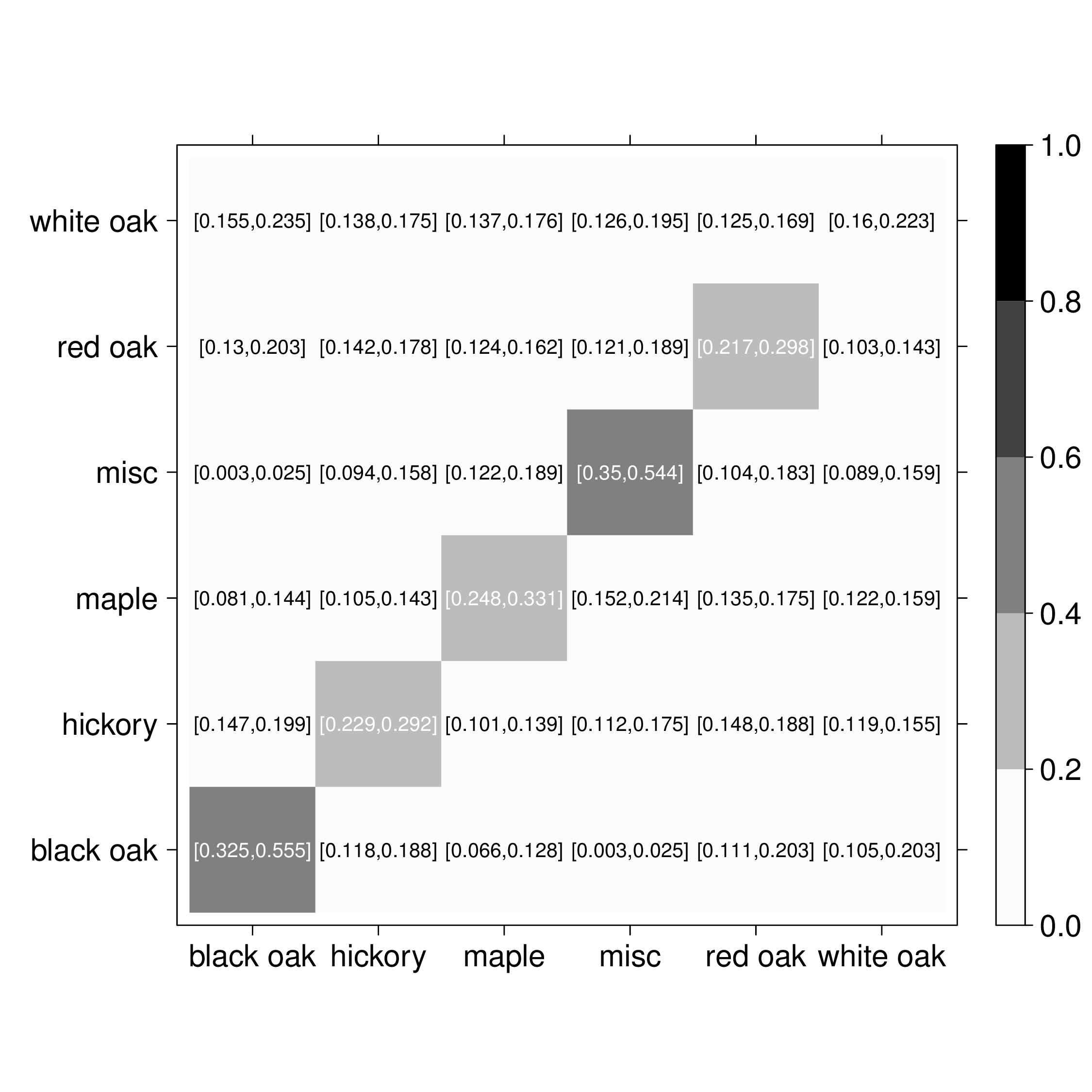}
	\caption{\texttt{lansing} dataset: The levelplot of the estimated $\hat{\bm{\phi}}$, with the numbers in square brackets giving the $95\%$ credible interval. }
	\label{Figure 55}
\end{figure}

	We applied the proposed model with the same hyperparameter and algorithm settings as described in Section \ref{simulation} and the choice of $c=0.1$. Again, we ran four independent MCMC chains with $50,000$ iterations and assessed the convergence by the Gelman and Rubin's convergence diagnostics. Those statistics for all parameters range from $1.003$ to $1.022$. Results we report here were obtained by pooling together the outputs from the four chains. We obtained the decay $\hat{\lambda}=49.764$, the first-order intensities $\hat{\omega}_\text{black oak}=2.514$, $\hat{\omega}_\text{hickory}=1.315$, $\hat{\omega}_\text{maple}=1.654$, $\hat{\omega}_\text{misc}=3.104$, and $\hat{\omega}_\text{red oak}=2.016$, corresponding to $\hat{\pi}_\text{black oak}=0.074$, $\hat{\pi}_\text{hickory}=0.247$, $\hat{\pi}_\text{maple}=0.176$, $\hat{\pi}_\text{misc}=0.041$, $\hat{\pi}_\text{red oak}=0.123$, and $\hat{\pi}_\text{white oak}=0.339$. The estimated $\bm{\Theta}$ is given as below
	\[
	\hat{\bm{\Theta}}=\bordermatrix{
		& \text{black oak} & \text{hickory} & \text{maple} & \text{misc} & \text{red oak} & \text{white oak}\cr
		\text{black oak}& $-0.066$ & $0.978$ & $1.449$ & $3.836$ &$0.970$& $0.997$\cr
		\text{hickory}& $0.978$& $0.570$ & $1.332$ & $1.166$& $1.003$ & $1.200$\cr
		\text{maple}& $1.449$ & $1.332$ & $0.495$ & $0.955$ & $1.108$& $1.202$\cr
		\text{misc} & $3.836$& $1.166$& $0.955$ & $-0.092$ & $1.044$& $1.187$\cr
		\text{red oak}& $0.970$ & $1.003$ & $1.108$ & $1.044$ & $0.535$ & $1.266$\cr
		\text{white oak}& $0.997$& $1.200$ & $1.202$ & $1.187$ & $1.266$ & $1.000$
	}\]
	, and the corresponding $\hat{\bm{\Phi}}$ is shown in Figure \ref{Figure 55} with the $95\%$ credible interval for each parameter. The pattern reveals that the first five types of trees exhibits clustering, especially for black oak and miscellaneous trees. This means if one species has a clump in an area, then no other species tends to form a clump in the same location. We also found white oak has the least $\hat{\phi}_{qq}$ value, which suggests its spatial pattern is more likely random. Those findings were also reported in \cite{Cox1976} and \cite{Cox1979}. In addition, our method outputs the mark interaction functions between the same mark, as shown in Figure \ref{Figure 5} (c), indicating there is no interaction between the same type trees beyond about $90$ feet.

\subsection{Case Study on Lung Cancer}
Lung cancer is the leading cause of death from cancer in both men and women. Non-small-cell lung cancer (NSCLC) accounts for about $85\%$ of deaths from lung cancer. Current guidelines for diagnosing and treating NSCLC are largely based on pathological examination of H\&E-stained tumor tissue section slides. We have developed a ConvPath pipeline (\href{https://qbrc.swmed.edu/projects/cnn/}{https://qbrc.swmed.edu/projects/cnn/}) to determine the locations and types of cells observed in the processed tumor pathology images. Specifically, the classifier, based on a convolutional neural network (CNN), was trained using a large cohort of lung cancer pathology images manually labelled by pathologists, and it can classify each cell by its $Q=3$ category: lymphocyte (a type of immune cell), stromal, or tumor cell. 

\begin{figure}
	\centering
	\subfloat[][Rescaled data]{\includegraphics[width=0.33\linewidth]{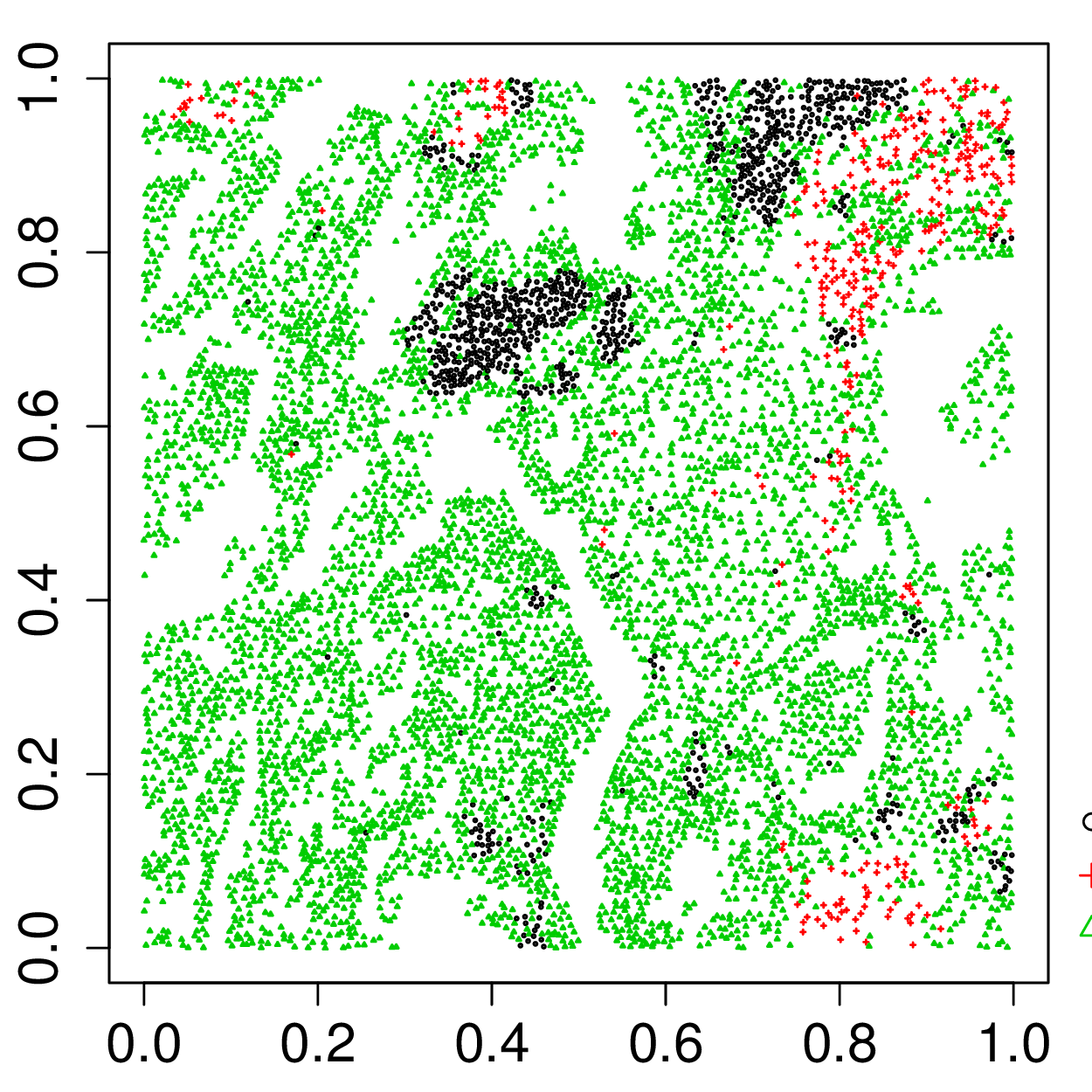}}
	\subfloat[][Mark connection functions]{\includegraphics[width=0.33\linewidth]{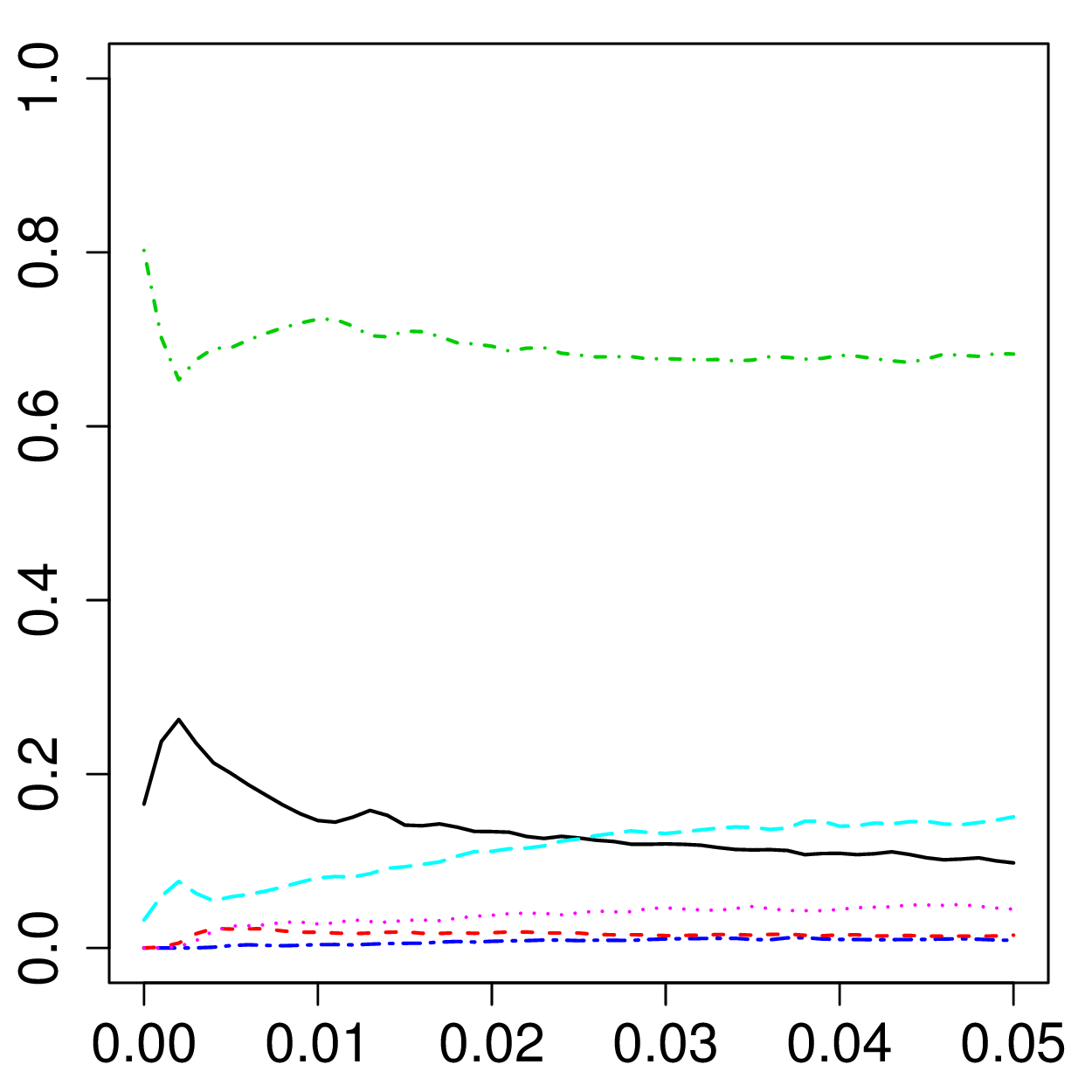}}
	\subfloat[][Mark interaction functions]{\includegraphics[width=0.33\linewidth]{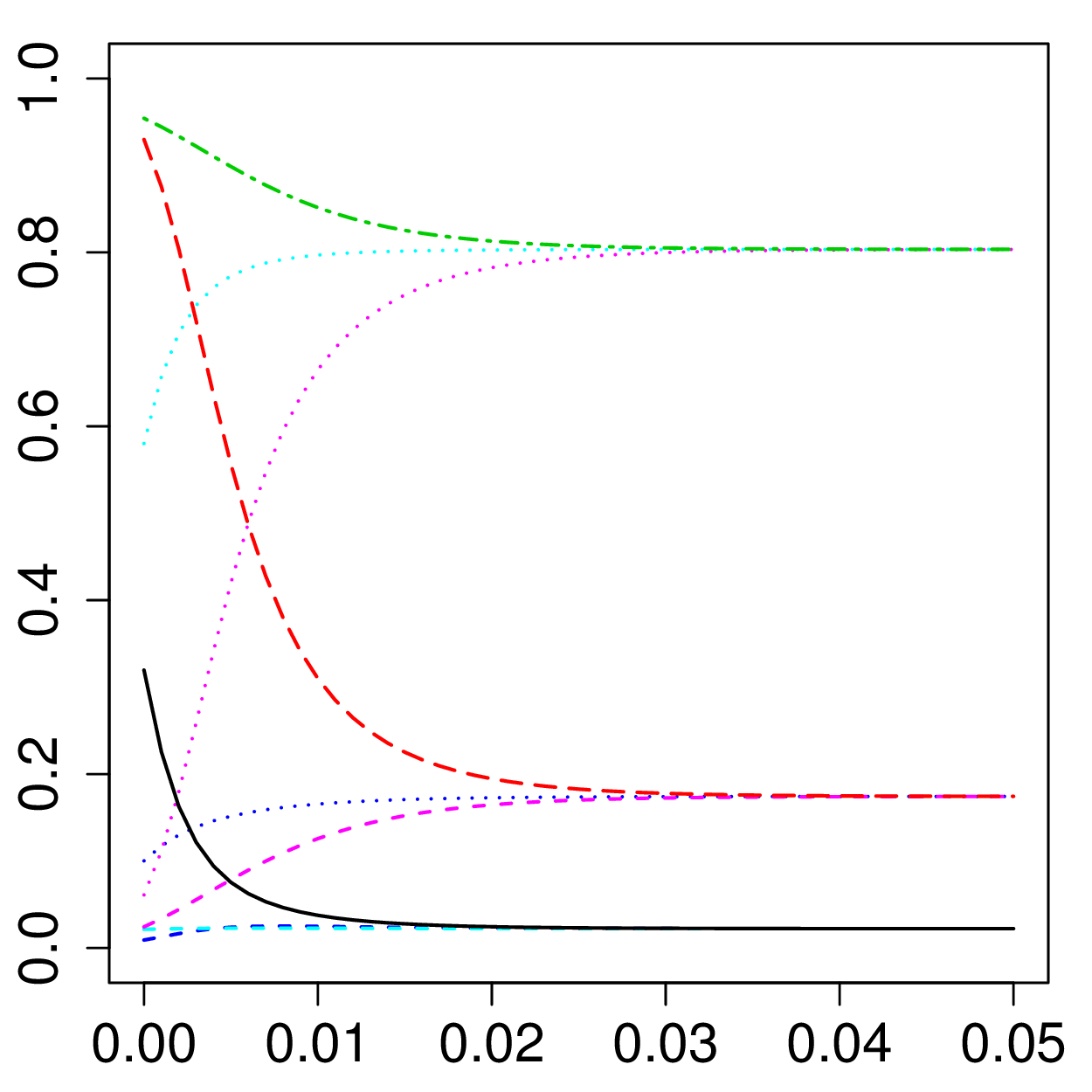}}\\
	\subfloat[][Rescaled data]{\includegraphics[width=0.33\linewidth]{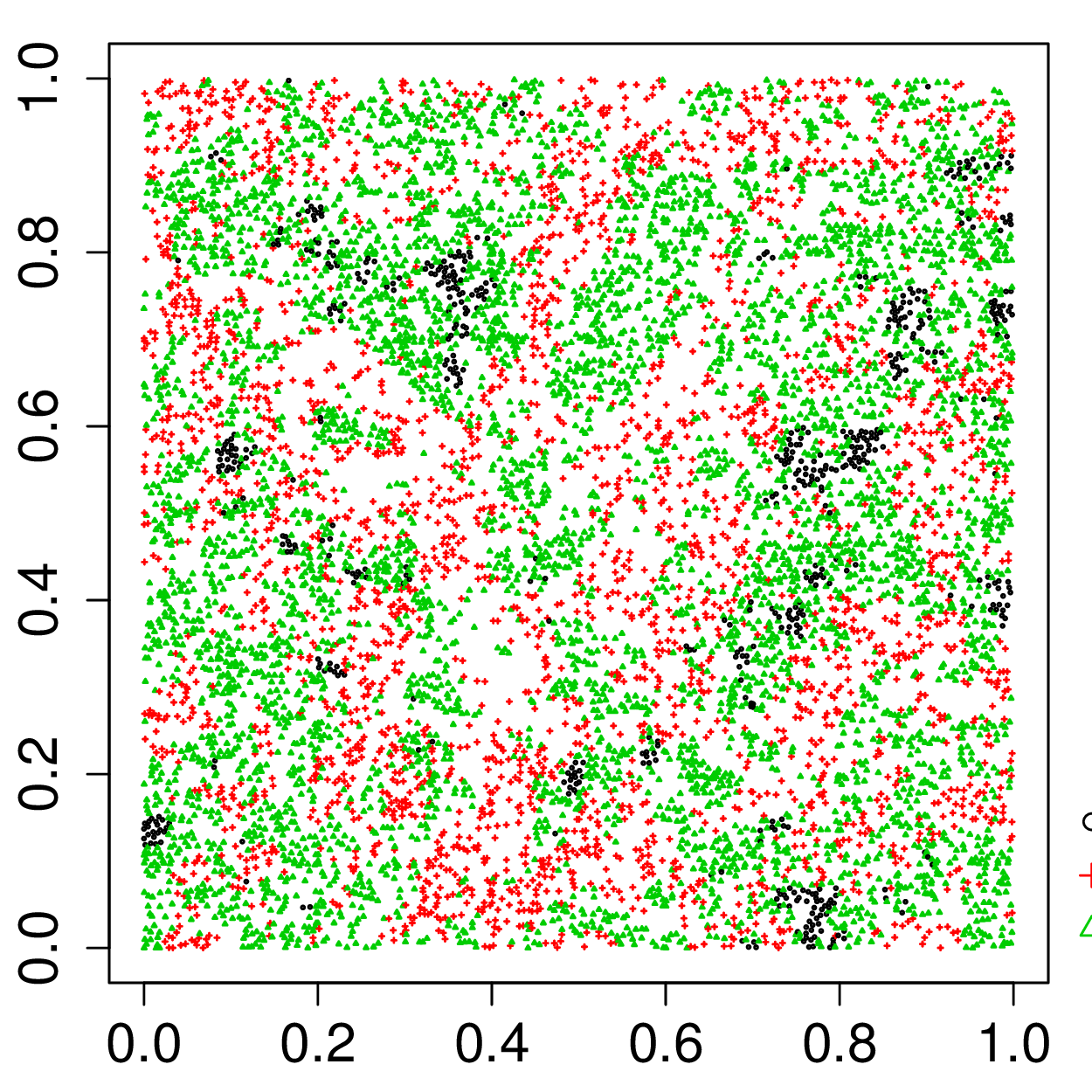}}
	\subfloat[][Mark connection functions]{\includegraphics[width=0.33\linewidth]{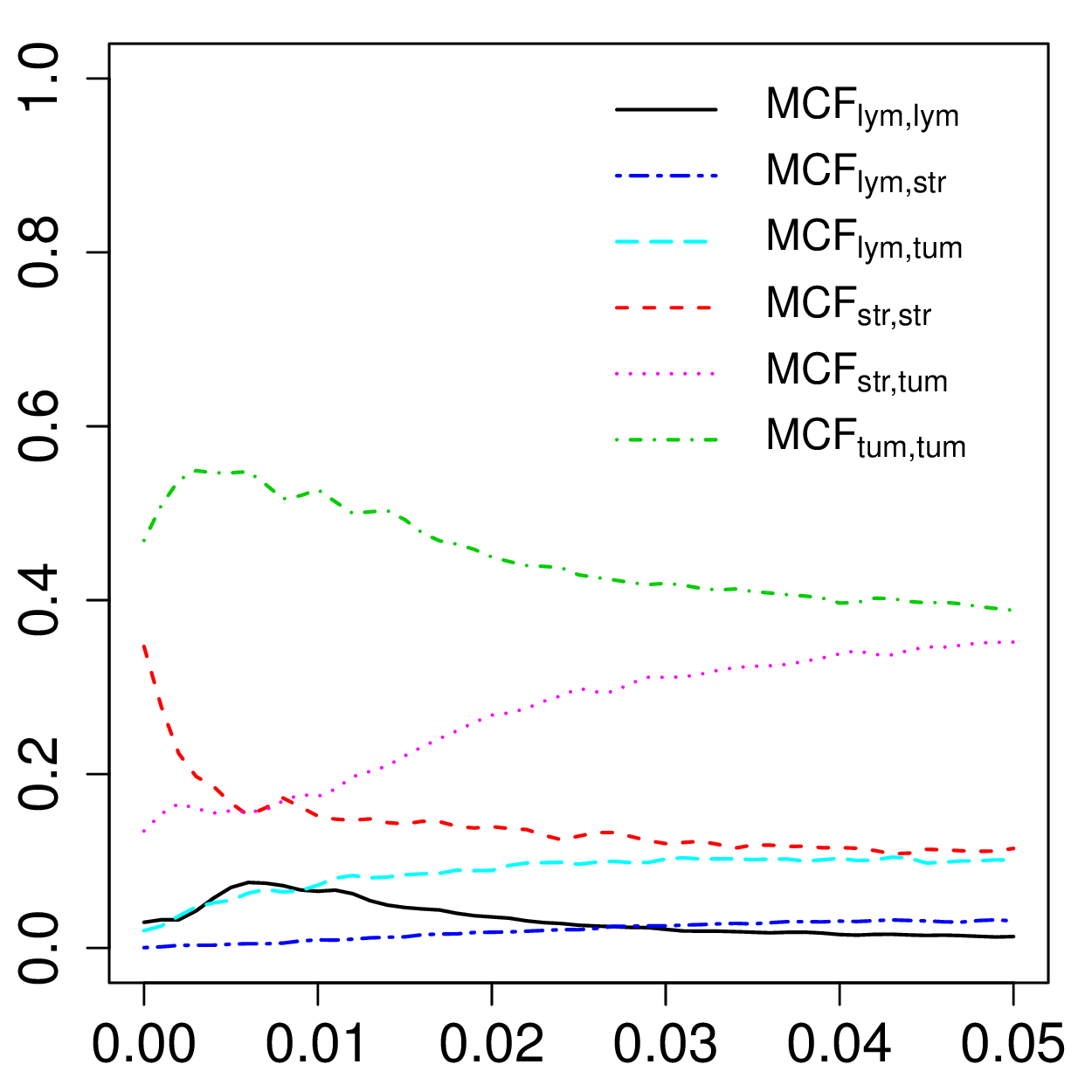}}
	\subfloat[][Mark interaction functions]{\includegraphics[width=0.33\linewidth]{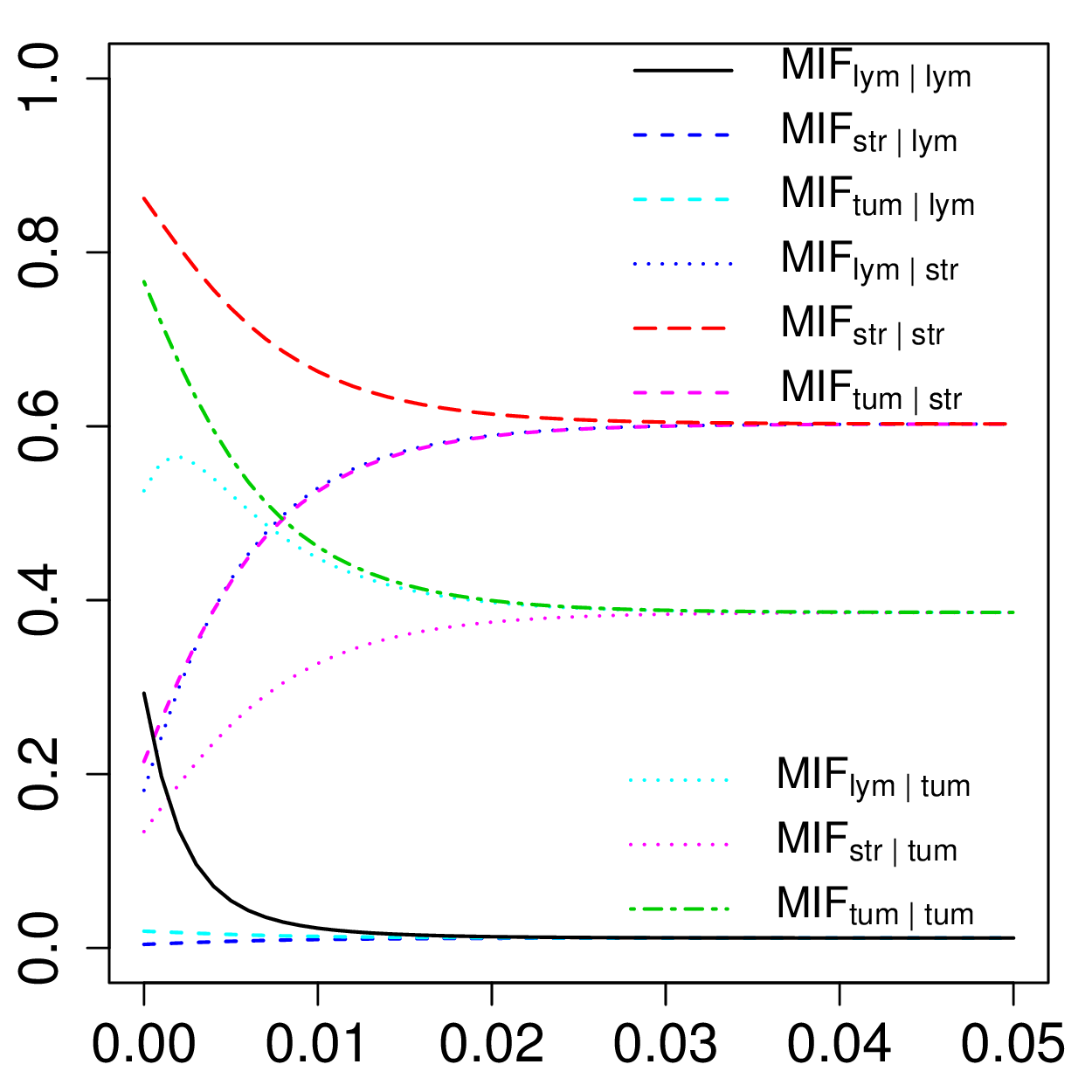}}\\
	\caption{Lung cancer case study: (a) and (d) Two examples of the rescaled marked point data from NLST dataset, where black, red, and green points represent lymphocyte, stromal, and green cells; (b) and (e) The empirical mark connection function plots; (c) and (f) The estimated mark interaction function plots. For the data shown in (a), $\hat{\lambda}=172.102$, $\hat{\pi}_\text{lym}=0.022$, $\hat{\pi}_\text{str}=0.173$, $\hat{\pi}_\text{tum}=0.805$, and $\hat{\phi}_\text{tum,str}=0.012$; For the data shown in (d), $\hat{\lambda}=169.268$, $\hat{\pi}_\text{lym}=0.011$, $\hat{\pi}_\text{str}=0.603$, $\hat{\pi}_\text{tum}=0.386$, and $\hat{\phi}_\text{tum,str}=0.162$}
	\label{Figure 6}
\end{figure}

\begin{figure}
	\centering
	\subfloat[][$\overline{\text{MCF}}_\text{lym,lym}(d)$]{\includegraphics[width=0.33\linewidth]{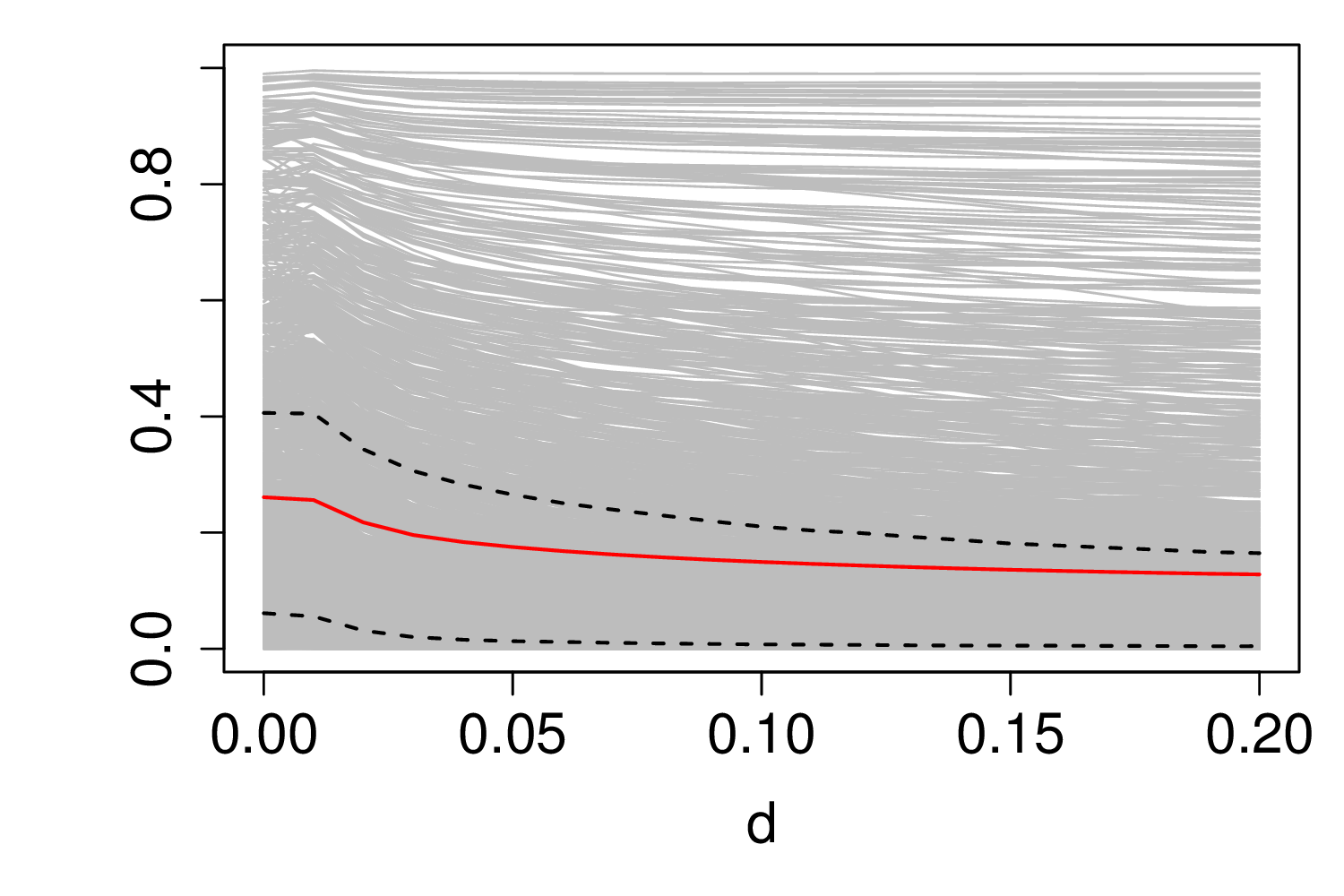}}
	\subfloat[][$\overline{\text{MCF}}_\text{str,str}(d)$]{\includegraphics[width=0.33\linewidth]{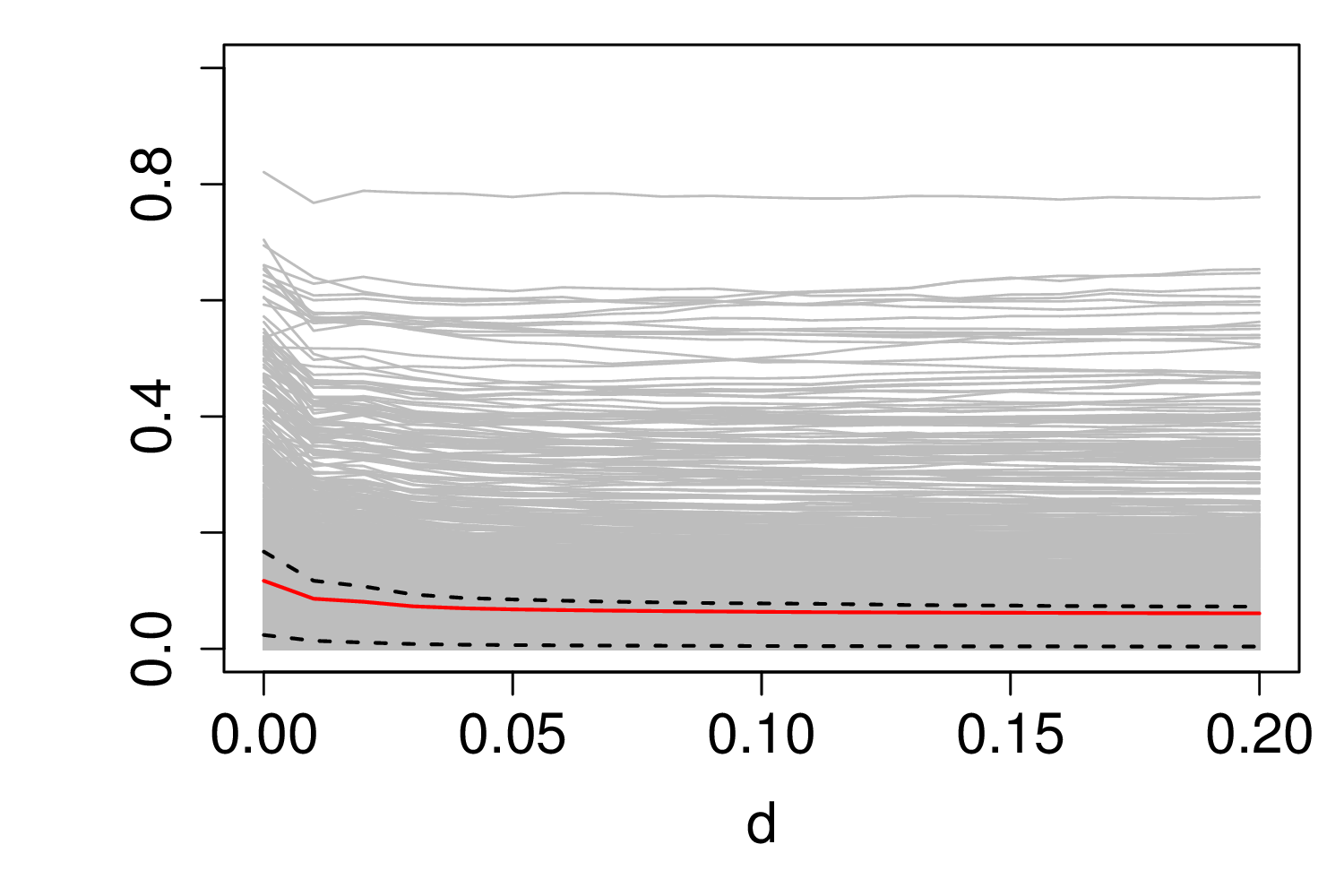}}
	\subfloat[][$\overline{\text{MCF}}_\text{tum,tum}(d)$]{\includegraphics[width=0.33\linewidth]{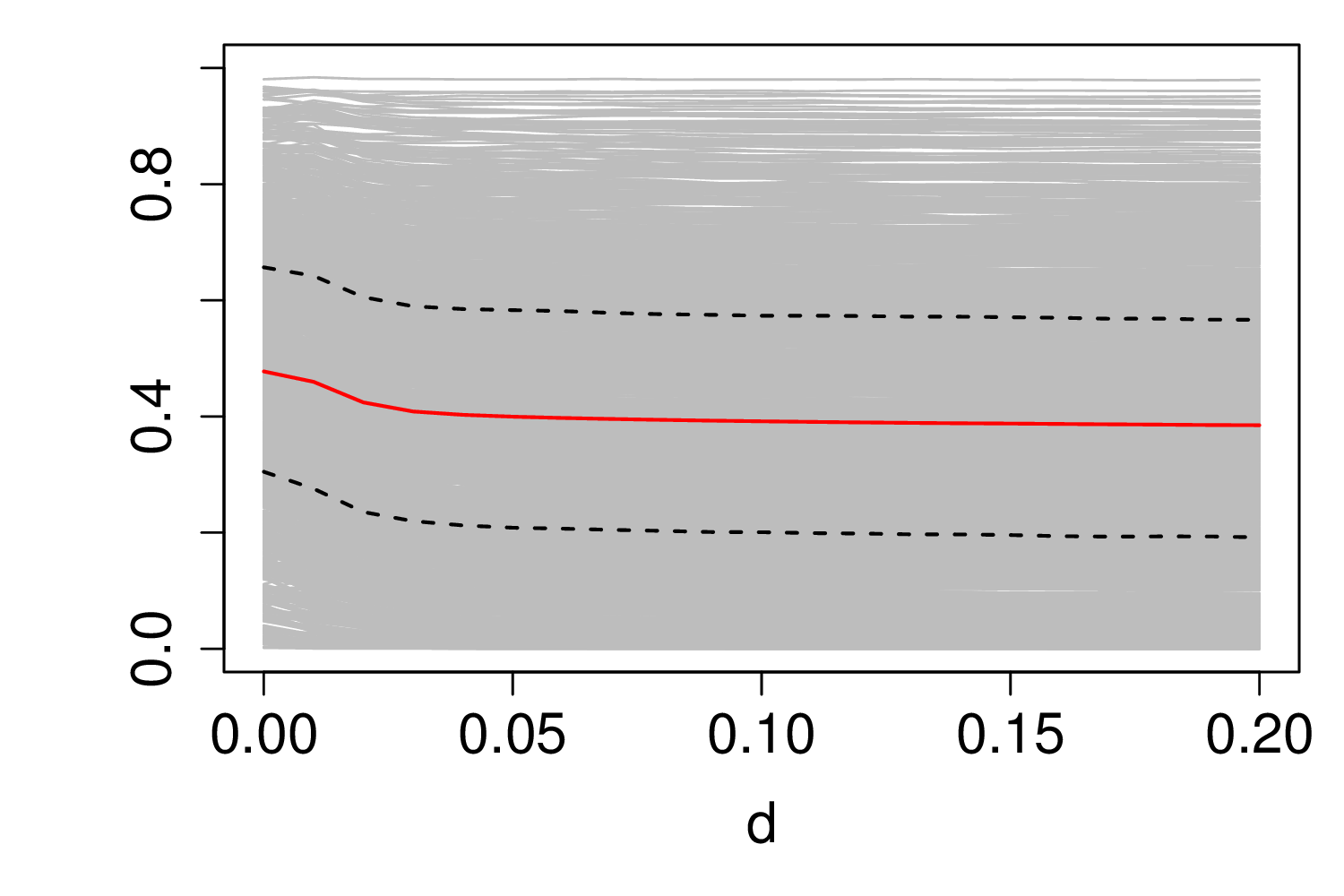}}\\
	\subfloat[][$\overline{\text{MCF}}_\text{lym,str}(d)$]{\includegraphics[width=0.33\linewidth]{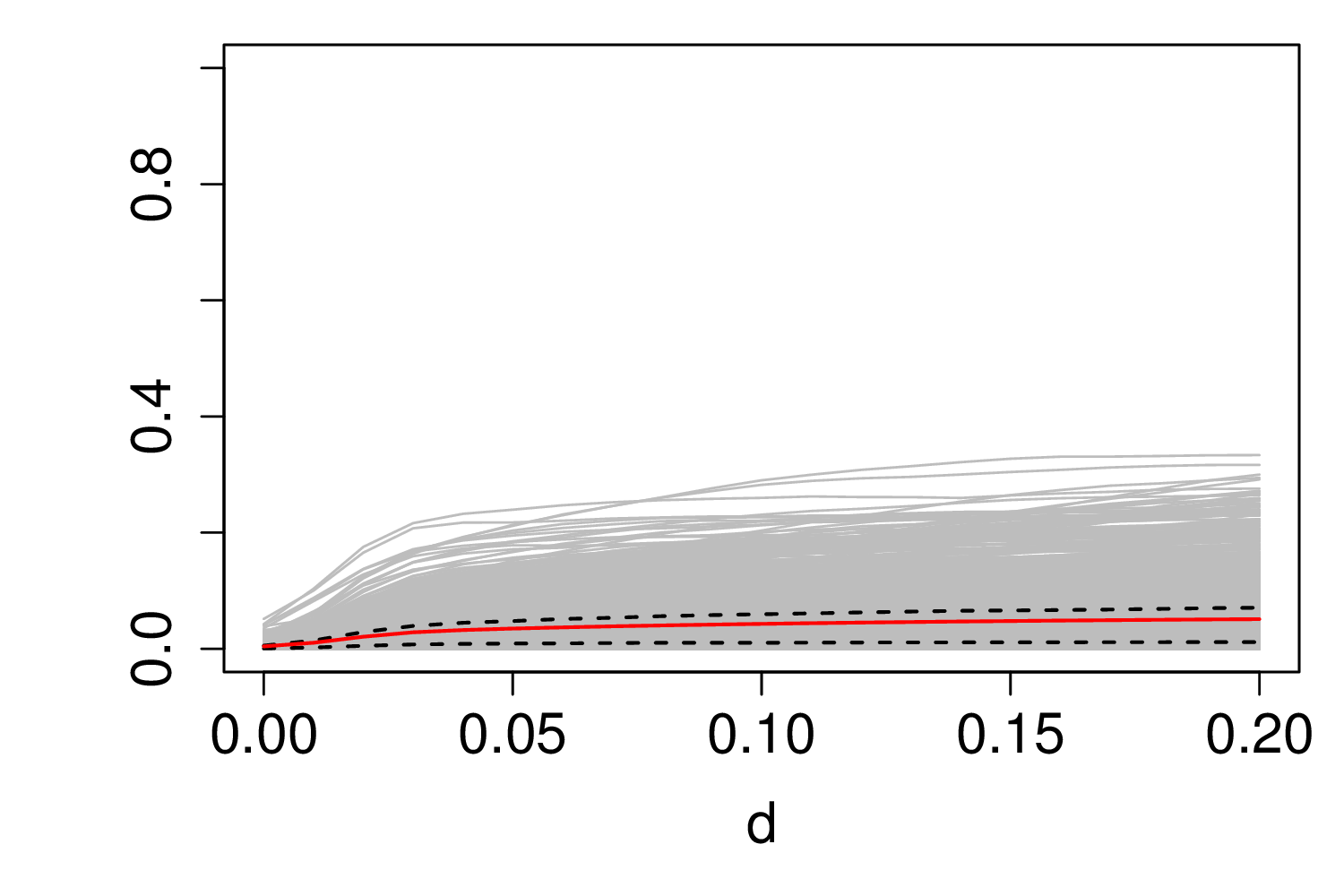}}
	\subfloat[][$\overline{\text{MCF}}_\text{lym,tum}(d)$]{\includegraphics[width=0.33\linewidth]{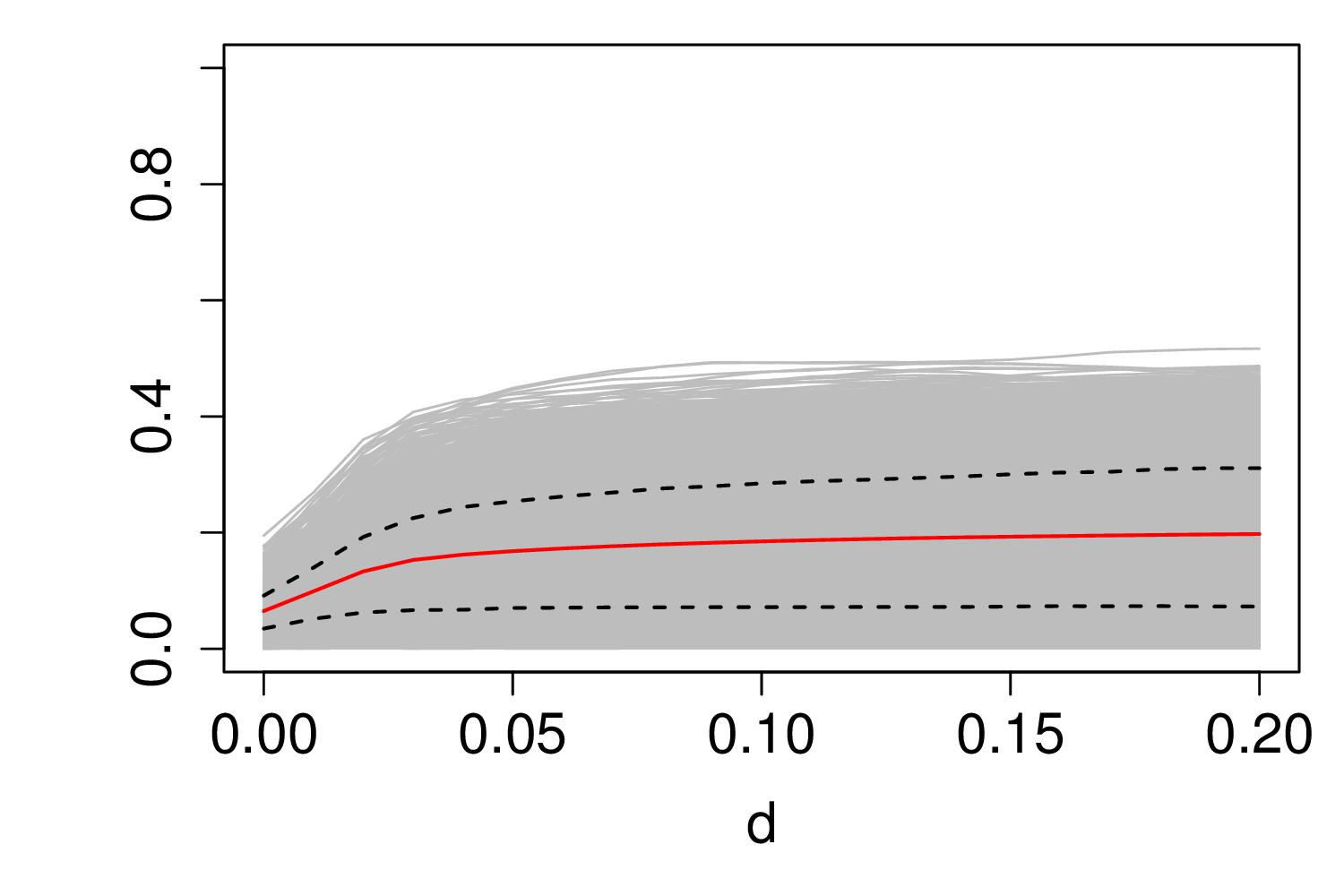}}
	\subfloat[][$\overline{\text{MCF}}_\text{str,tum}(d)$]{\includegraphics[width=0.3\linewidth]{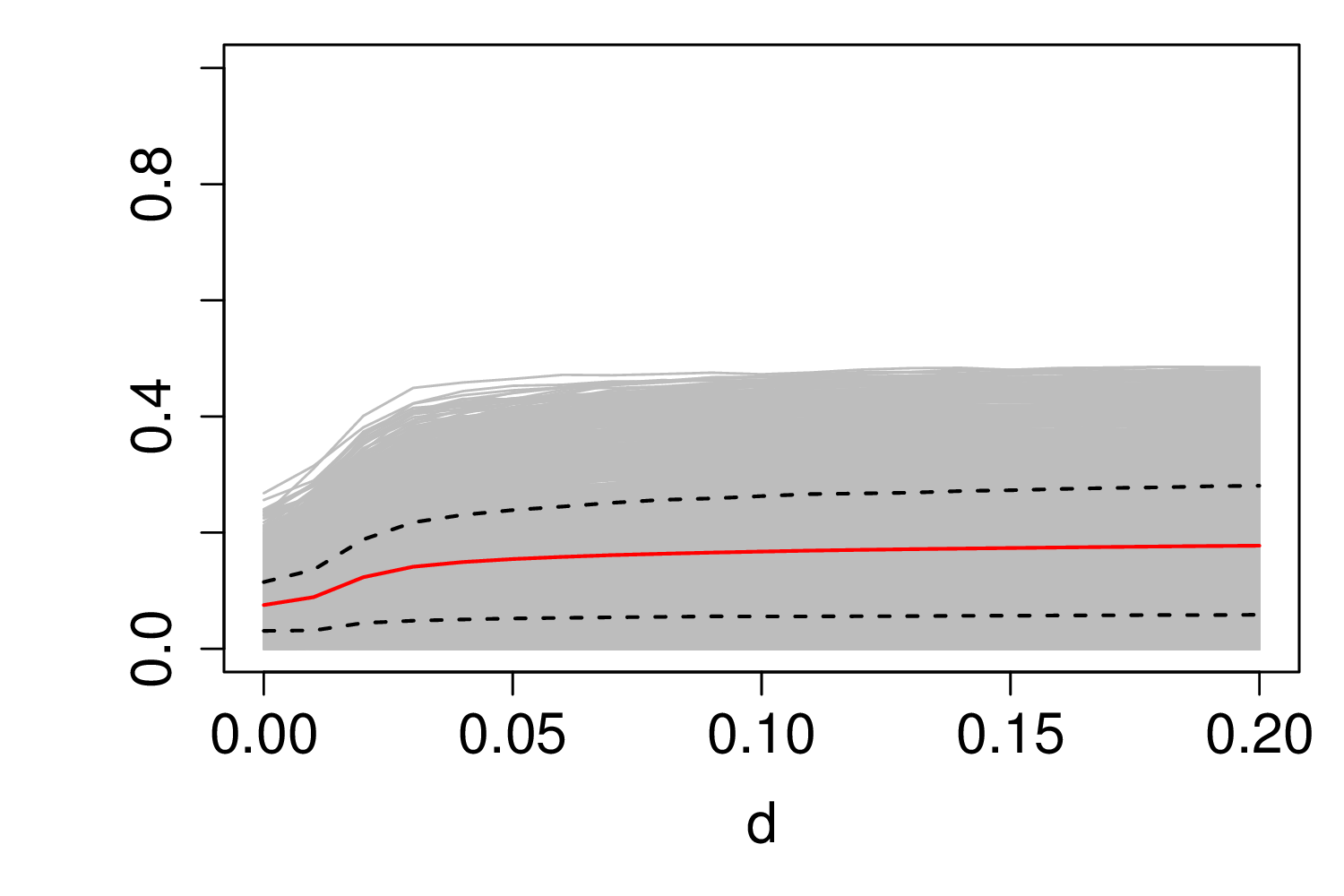}}
	\caption{Lung cancer case study: The empirical mark connection function plots between marks: (a) lymphocyte and lymphocyte, (b) stromal and stromal, (c) tumor and tumor, (d) lymphocyte and stromal, (e) lymphocyte and tumor, and (f) stromal and tumor. Each grey line represents one of the $1,585$ sample images. Red solid lines indicate the MCF means against the distance $d$.} 
	\label{Figure 7}
\end{figure}

In this case study, we used the pathology images from $188$ NSCLC patients in the National Lung Screening Trial (NLST). Each patient has one or more tissue slide(s) scanned at $40$x magnification. The median size of the slides is $24,244\times19,261$ pixels.  A lung cancer pathologist first determined and labeled the region of interest (ROI) within the tumor region(s) from each tissue slide using an annotation tool, ImageScope (Leica Biosystem). ROIs are regions of the slides containing the majority of the malignant tissues and are representative of the whole slide image. Then we randomly chose five square regions, each of which is in a $5000\times5000$ pixel window, per ROI as the sample images. The total number of sample images that we collected was $1,585$. For each sample image, the ConvPath software was used to identify cells from the sample images and classify each cell into one of three types, so that a corresponding spatial map of cells was generated and used as the input of our model. The number of cells in each sample image ranges from $n=2,876$ to $26,463$. Figure \ref{Figure 6} (a) and (b) show the examples of two sample images and Figure \ref{Figure 7} displays the mark connection functions of the whole datasets, which exhibits attraction (i.e. the cells with the same type tend to cluster). %The marked point pattern of most sample images exhibits attraction.

We applied the proposed model with the same hyperparameter and algorithm settings as described in Section \ref{simulation} and different choices of $c=0.02$, $c=0.05$, and $c=0.1$. We then computed the pairwise Pearson correlation coefficients between the estimated model parameters under different choices of $c$. These correlations indicated substantial agreement between any pair of settings, with values ranging from $0.967$ to $0.997$. Results we report below were obtained by using the estimated parameters under the choice of $c=0.1$.

With the estimated parameters in each sample image, we conducted a downstream analysis to investigate their associations with the other measurements of interest. Specifically, a Cox proportional hazards model \citep{Cox1992} was fitted to evaluate the association between the transformed model parameters $\hat{\bm{\pi}}$ and $\hat{\bm{\Phi}}$, and patient survival outcomes, after adjusting for other clinical information, such as age, gender and tobacco history. Multiple sample images from the same patient were modeled as correlated observations in the Cox regression model to compute a robust variance for each coefficient. The overall \textit{P}-value for the Cox model is $0.0021$ (Wald test), and the \textit{P}-value and coefficient for each individual variable are summarized in Table \ref{Table 5}. The results imply that a low interaction between stromal and tumor cells ($\phi_\text{tum, str}$) is associated with good prognosis in NSCLC patients (\textit{P}-value=$0.0073$). Interestingly, \cite{Beck2011} also discovered that the morphological features of the stroma in the tumor region are associated with patient survival in a systematic analysis of breast cancer. Besides, the abundance of the stromal cells itself (\textit{P}-value=$0.017$) is also a prognostic factor, while the underlying biological mechanism is currently unknown. The positive coefficient of the predictor $\phi_\text{tum,str}$ implies that a higher value may reveal a higher risk of death. Indeed, we obtained $\hat{\phi}_\text{tum,str}=0.012$ for the data shown in Figure \ref{Figure 6}(a) and it is from a patient who was still alive over $2,615$ days after the surgery, while the estimated value of ${\phi}_\text{tum,str}=0.162$ for the data shown in Figure \ref{Figure 6}(d) and it is from a patient died on the $1,246$th days after the surgery. These two images have distinctive patterns, as the former clearly shows the same type cells tend to clump in the same area, while the latter displays a case where stromal and tumor cells are thoroughly mixed together, indicating the spread of stromal cells into the tumor region. Although the high/low interaction between stromal and tumor cells can be easily seen by eyes in these two images, the patterns are much more subtle for many other images. Therefore, the proposed model can be used to predict the survival time when human visualization does not work.

\begin{figure}
	\begin{center}
		\includegraphics[width=0.3\linewidth]{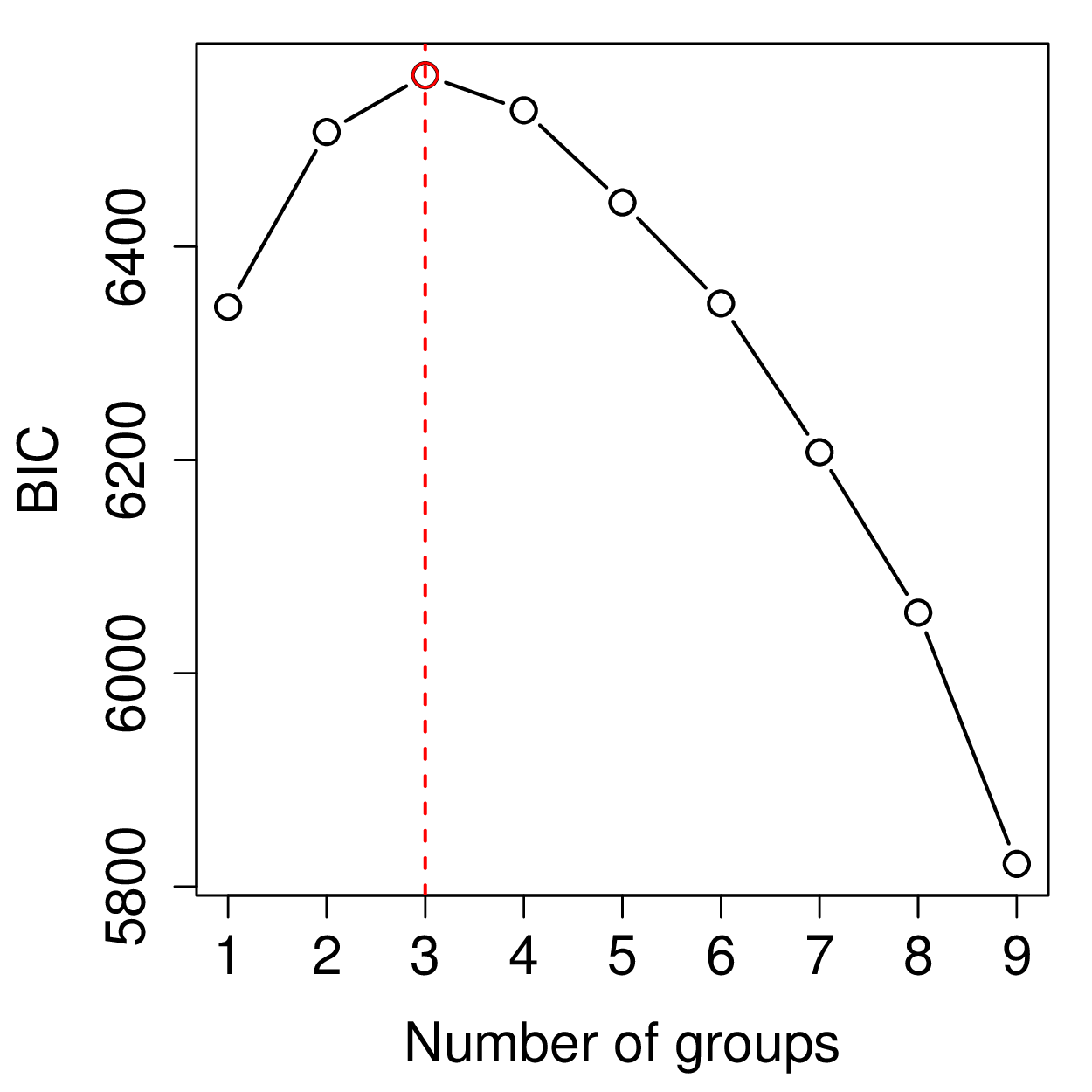}
		\includegraphics[width=0.3\linewidth]{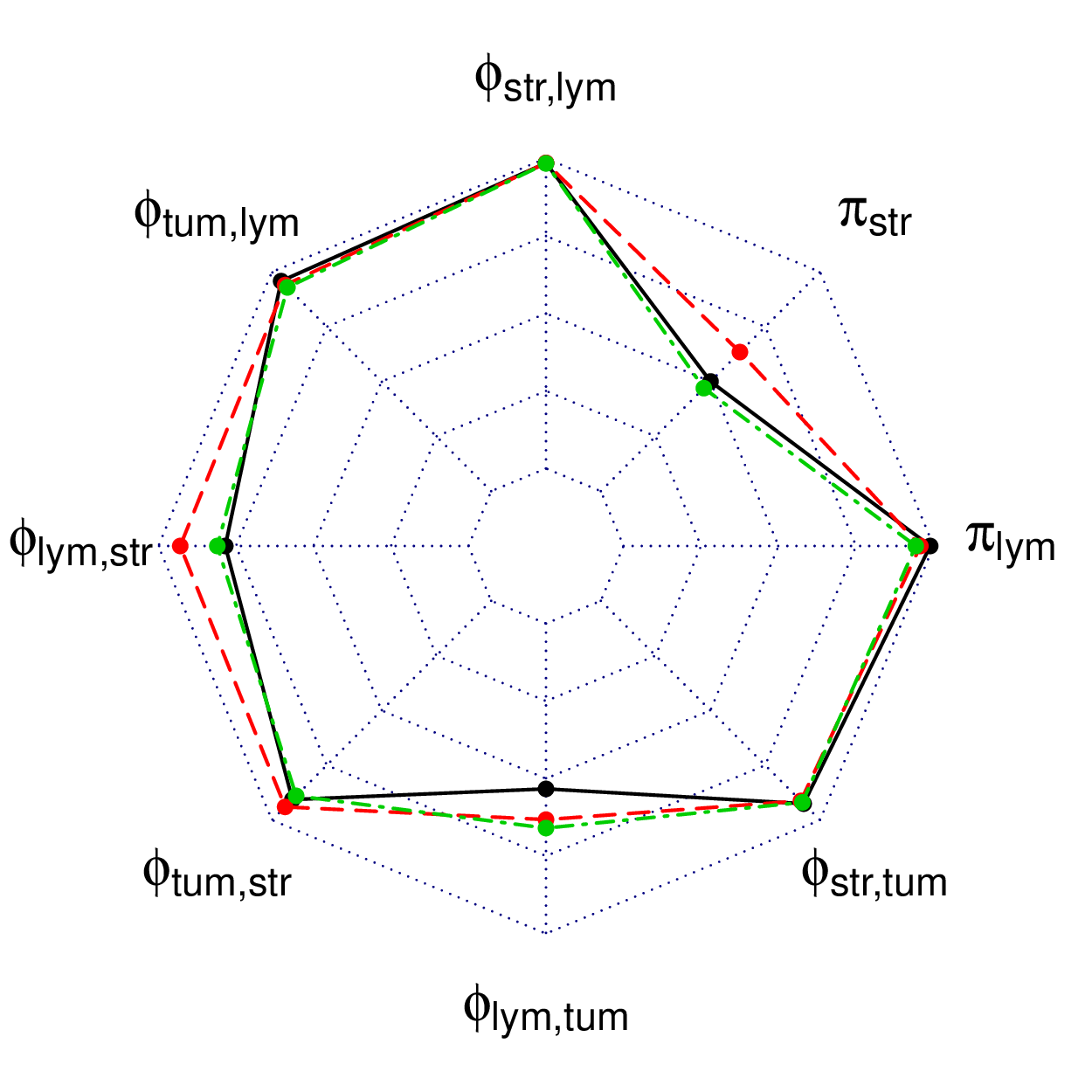}
		\includegraphics[width=0.3\linewidth]{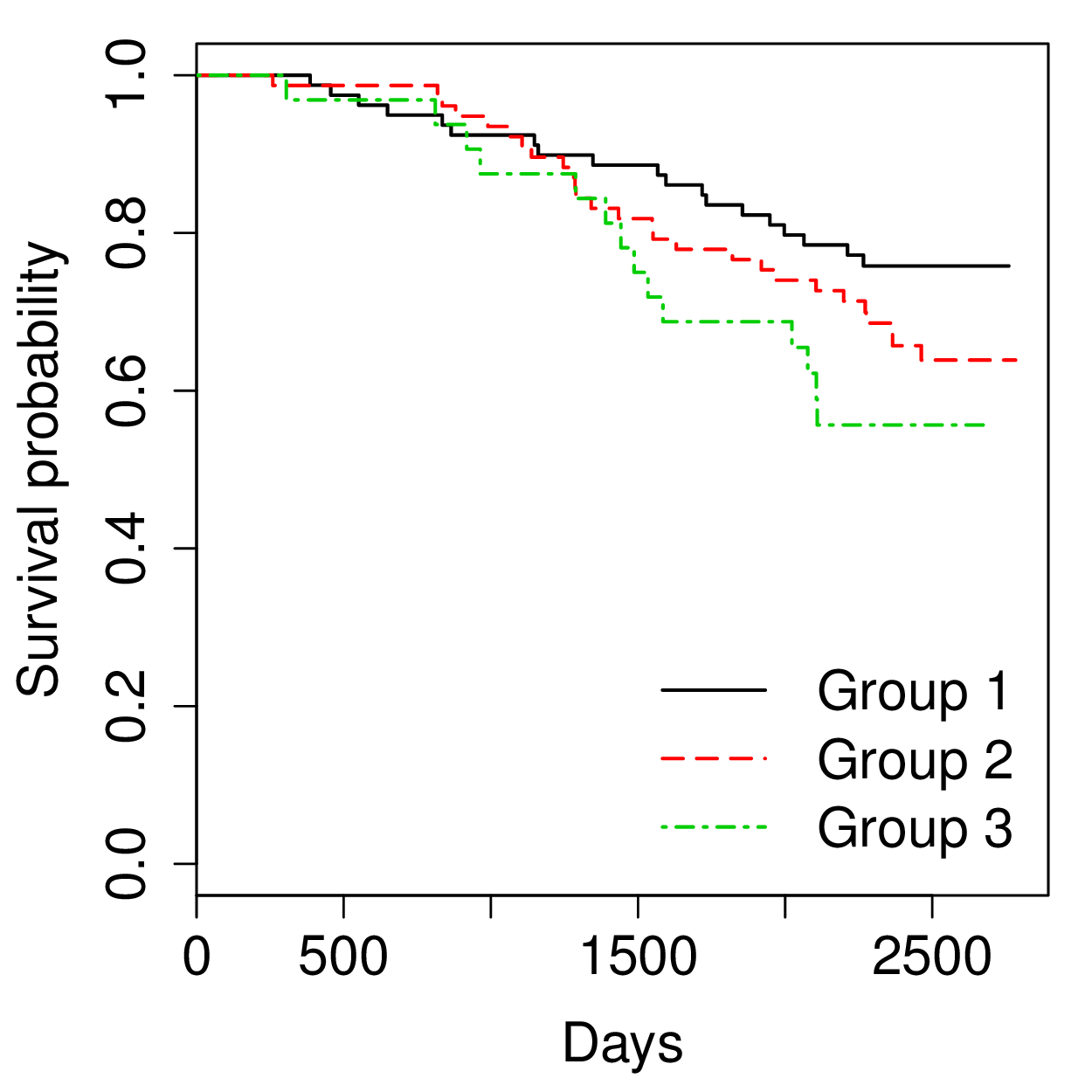}
	\end{center}
	\caption{Lung cancer case study: (a) The BIC plot of the model-based clustering on the patient-level parameters $(\hat{\phi}_\text{str,lym},\hat{\phi}_\text{tum,lym},\hat{\phi}_\text{lym,str},\hat{\phi}_\text{tum,str},\hat{\phi}_\text{lym,tum},\hat{\phi}_\text{str,tum},\hat{\pi}_\text{lym},\hat{\pi}_\text{str})$; (b) The radar chart of the averaged patient-level parameters of the three groups (shown in different colors), where the outer ring and the center have the values of $0$ and $1$, respectively; (c) The Kaplan-Meier plot for the three groups with patient survival.}\label{Figure 9}
\end{figure}

Furthermore, we performed a model-based clustering analysis on the features extracted by the model. First, each of the eight parameters $\hat{\phi}_\text{str,lym}$, $\hat{\phi}_\text{tum,lym}$, $\hat{\phi}_\text{lym,str}$, $\hat{\phi}_\text{tum,str}$, $\hat{\phi}_\text{lym,tum}$, $\hat{\phi}_\text{str,tum}$, $\hat{\pi}_\text{lym}$, $\hat{\pi}_\text{str}$ of multiple sample images from the same patient were averaged. We then used the multivariate Gaussian mixture model \citep{Fraley2002} to cluster patients using those $8$ parameter. This was done using \texttt{R} package \texttt{mclust}. To estimate the number of clusters that best represents the data as well as its covariance structure, we plotted the Bayesian information criterion (BIC) values against the number of clusters from $1$ to $9$, as shown in Figure \ref{Figure 9} (a). It shows that clustering patients into three groups achieves the best fit of the data measured by BIC, where the first (in black), second (in red), and third (in green) groups have $79$, $77$, and $32$ patients, respectively. Next, we visualized the means of these patient-level parameters for each group, shown as a radar chart in Figure \ref{Figure 9} (b), and plotted the Kaplan-Meier survival curve for each group in Figure \ref{Figure 9} (c). The patients from group 1 had higher survival probabilities, while the patients from the last group had the poor prognosis. The log-rank test shows that there are significant differences ($P$-value$=0.024$) among the survival curves of the three groups. The analysis, again, demonstrated that the proposed mark interaction features can be used as a potential biomarker for patient prognosis.
\begin{table}[h]
	\caption{Lung cancer case study: The $P$-values of the transformed model parameters by fitting a Cox regression model with survival time and vital status as responses, and $\hat{\phi}_{\text{str},\text{lym}}$, $\hat{\phi}_{\text{tum},\text{lym}}$, $\hat{\phi}_{\text{lym},\text{str}}$, $\hat{\phi}_{\text{tum},\text{str}}$, $\hat{\phi}_{\text{lym},\text{tum}}$, $\hat{\phi}_{\text{str},\text{tum}}$, $\hat{\pi}_\text{lym}$, $\hat{\pi}_\text{str}$, $\hat{\lambda}$, age, gender, and smoking history as predictors. The overall $P$-value corresponding to a Wald test for the model is $0.0024$.}\label{Table 5}
	\begin{center}
		\begin{tabular}{@{}lrcrr@{}}
			Predictor & Coefficient & $\exp$(Coef.) & SE & $P$-value\\\hline 						
			$\hat{\phi}_{\text{str},\text{lym}}$ & $13.22$ & $5.5\times10^{5}$ & $5.16$ & $0.15$\\	 				
			$\hat{\phi}_{\text{tum},\text{lym}}$ & $2.34$ & $10.39$ & $2.45$ & $0.62$\\			
			$\hat{\phi}_{\text{lym},\text{str}}$ & $-0.59$ & $0.55$ & $0.76$ & $0.68$\\	
			$\hat{\phi}_{\text{tum},\text{str}}$ & $8.83$ & $6.8\times10^3$ & $1.60$ & $\bm{0.0073}$\\
			$\hat{\phi}_{\text{lym},\text{tum}}$ & $-0.53$ & $0.59$ & $0.81$ & $0.70$\\
			$\hat{\phi}_{\text{str},\text{tum}}$ & $-6.08$ & $2.3\times10^{-3}$ & $1.88$ & ${0.12}$\\\hline
			$\hat{\pi}_\text{lym}$ & $3.44$ & $31.2$ & $1.49$ & $0.17$\\	
			$\hat{\pi}_\text{str}$ & $-3.21$ & $4.0\times10^{-2}$ & $0.72$ & $\bm{0.017}$\\\hline	
			$\hat{\lambda}$ & $-0.01$ & $0.99$ & $2.6\times10^{-3}$ & $0.19$\\\hline
			Age & $0.04$ & $1.04$ & $8.7\times10^{-3}$ & $0.17$\\
			Female vs. male & $-0.12$ & $0.89$ & $9.1\times10^{-2}$ & $0.67$\\
			Smoking vs. non-smoking & $0.07$ & $1.07$ & $8.9\times10^{-2}$ & $0.80$\\
		\end{tabular}
	\end{center}
\end{table}

By contrast, we fitted a similar Cox regression model by using the mark connection function features as predictors. Specifically, we first used $\text{MCF}_{\text{lym},\text{str}}(d)$, $\text{MCF}_{\text{lym},\text{tum}}(d)$, and $\text{MCF}_{\text{str},\text{tum}}(d)$, where $d=0.1$ for each sample image as covariates. The results are summarized in Table \ref{Table 6}. As we can see, there is no significant predictor and the overall \textit{P}-value for the Cox model is $0.47$ (Wald test). Then, we tried to vary the value of $d$ from $0$ to $0.2$, Figure \ref{Figure 7} shows the $P$-values of $\text{MCF}_{\text{lym},\text{str}}(d)$, $\text{MCF}_{\text{lym},\text{tum}}(d)$, and $\text{MCF}_{\text{str},\text{tum}}(d)$ against $d$. Again, we were unable to find any association between cell-cell interactions and clinical outcomes. The comparison demonstrates the advantage of modeling the pathology images via the proposed model over the traditional methods for characterizing spatial correlation.
\begin{table}[h]
	\caption{Lung cancer case study: The $P$-values of $\text{MCF}_{\text{lym},\text{str}}(0.1)$, $\text{MCF}_{\text{lym},\text{tum}}(0.1)$, and $\text{MCF}_{\text{str},\text{tum}}(0.1)$ by fitting a Cox regression model with survival time and vital status as responses, and $\text{MCF}_{\text{lym},\text{str}}(0.1)$, $\text{MCF}_{\text{lym},\text{tum}}(0.1)$, $\text{MCF}_{\text{str},\text{tum}}(0.1)$, proportion of lymphocyte cells, proportion of stromal cells, age, gender, and smoking history as predictors. The overall $P$-value corresponding to a Wald test for the model is $0.47$.}\label{Table 6}
	\begin{center}
		\begin{tabular}{@{}lrcrr@{}}
			Predictor & Coefficient & $\exp$(Coef.) & SE & $P$-value\\\hline 						
			$\text{MCF}_{\text{lym},\text{str}}(d=0.1)$ & $-2.92$ & $0.05$ & $1.38$ & $0.23$\\	 				
			$\text{MCF}_{\text{lym},\text{tum}}(d=0.1)$ & $-0.63$ & $0.53$ & $0.47$ & $0.54$\\			
			$\text{MCF}_{\text{str},\text{tum}}(d=0.1)$ & $0.12$ & $1.12$ & $0.79$ & $0.95$\\\hline
		    Prop. of lym cells & $0.75$ & $2.11$ & $0.25$ & $0.16$\\	
			Prop. of str cells & $-0.31$ & $0.73$ & $0.61$ & $0.79$\\\hline	
			Age & $0.03$ & $1.03$ & $0.01$ & $0.21$\\
			Female vs. male & $-0.17$ & $0.84$ & $0.09$ & $0.56$\\
			Smoking vs. non-smoking & $0.07$ & $1.07$ & $0.09$ & $0.81$\\
		\end{tabular}
	\end{center}
\end{table}
\begin{figure}
	\begin{center}
		\includegraphics[width=0.6\linewidth]{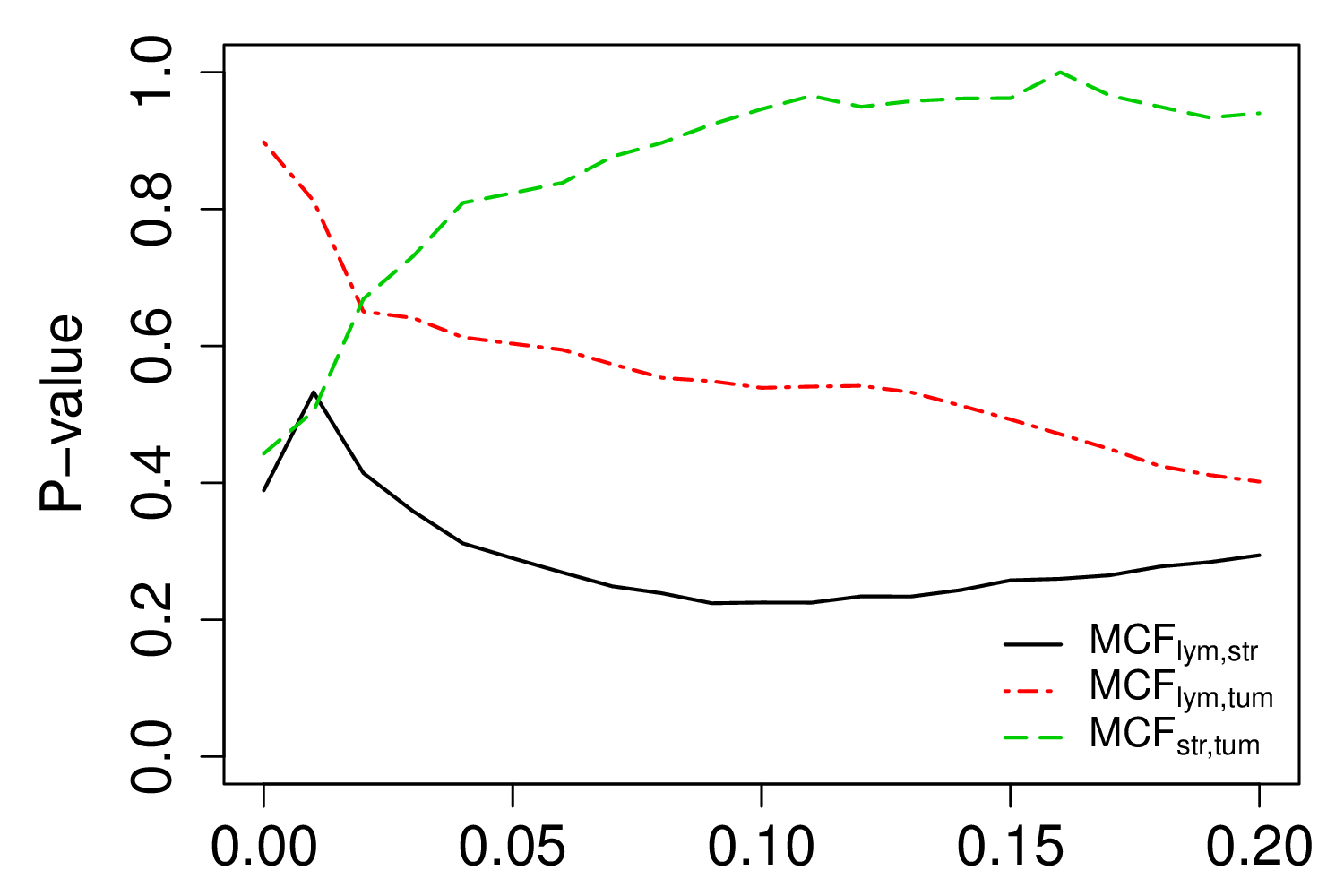}
	\end{center}
	\caption{Lung cancer case study: The $P$-values of $\text{MCF}_{\text{lym},\text{str}}(d)$, $\text{MCF}_{\text{lym},\text{tum}}(d)$, and $\text{MCF}_{\text{str},\text{tum}}(d)$ under different choices of $d$ by fitting Cox regression models with survival time and vital status as responses, and $\text{MCF}_{\text{lym},\text{str}}(d)$, $\text{MCF}_{\text{lym},\text{tum}}(d)$, $\text{MCF}_{\text{str},\text{tum}}(d)$, proportion of lymphocyte cells, proportion of stromal cells, gender, and smoking history as predictors.}\label{Figure 8}
\end{figure}

\section{Conclusion} \label{conclusion}
The major cell types in a malignant tissue of lung are tumor cells, stromal cells and infiltrating lymphocytes. The distribution of different types of cells and their interactions play a key role in tumor progression and metastasis. For example, stromal cells are connective tissue cells such as fibroblasts and pericytes, and their interaction with tumor cells is known to play a major role in cancer progression \citep{Wiseman2002}. Tumor-infiltrating lymphocytes have been associated with patient prognosis in multiple tumor types previously \citep{Huh2012,Brambilla2016}. Recent advances in deep learning methods have made possible the automatic identification and classification of cells at large scale. For example, the ConvPath pipeline could determine the location and cell type for thousands of cells. However, it is challenging to utilize the vast amount of information extracted digitally. In this study, we developed a rigorous statistical method to model the spatial interaction among different types of cells in tumor regions. We focused on modeling the spatial correlation of marks in a spatial pattern that arose from a pathology image study. A Bayesian framework was proposed in order to model how the mark in a pattern might have been formed given the points. The proposed model can utilize the spatial information of thousands of points from any point processes. The output of the model is the parameters that characterize the spatial pattern. After a certain transformation, the parameters are identifiable and interpretable, and most importantly, transferable for conducting an association study with other measurements of interest. Furthermore, this statistical methodology provides new insights into the biological mechanisms of cancer.

For the lung cancer pathology imaging data, our study shows the interaction strength between stromal and tumor cells is significantly associated with patient prognosis. This parameter can be easily measured using the proposed method and used as a potential biomarker for patient prognosis. This biomarker can be translated into real clinical tools at low cost because it is based only on tumor pathology slides, which are available in standard clinical care. 

Several extensions of our model are worth investigating. First, the proposed model can be extended to finite mixture models for inhomogeneous mark interactions. Second, the correlation among first- and second-order intensity parameters could be taken into account by modeling them as a multivariate normal distribution. Last but not least, the proposed model provides a good chance to investigate the performance of other approximate Bayesian computation methods. These could be future research directions.

	%\bigskip
	%\begin{center}
	%	{\large\bf SUPPLEMENTARY MATERIAL}
	%\end{center}
	
	%\begin{description}
		
	%	\item[Title:] Brief description. (file type)
		
	%	\item[R-package for  MYNEW routine:] R-package ÒMYNEWÓ containing code to perform the diagnostic methods described in the article. The package also contains all datasets used as examples in the article. (GNU zipped tar file)
		
	%	\item[HIV data set:] Data set used in the illustration of MYNEW method in Section~ 3.2. (.txt file)
		
	%\end{description}

	\bibliographystyle{natbib}
    \bibliography{JASA-template.bbl}
\end{document}